



\input harvmac.tex
\def\p{\partial}
\def\CN{{\cal N}}

\def\p{\partial}


\input epsf
\def\figin{\epsfcheck\figin}\def\figins{\epsfcheck\figins}
\def\epsfcheck{\ifx\epsfbox\UnDeFiNeD
\message{(NO epsf.tex, FIGURES WILL BE IGNORED)}
\gdef\figin##1{\vskip2in}\gdef\figins##1{\hskip.5in}
\else\message{(FIGURES WILL BE INCLUDED)}%
\gdef\figin##1{##1}\gdef\figins##1{##1}\fi}
\def\DefWarn#1{}
\def\figinsert{\goodbreak\midinsert}
\def\ifig#1#2#3{\DefWarn#1\xdef#1{fig.~\the\figno}
\writedef{#1\leftbracket fig.\noexpand~\the\figno}%
\figinsert\figin{\centerline{#3}}\medskip\centerline{\vbox{\baselineskip12pt
\advance\hsize by -1truein\noindent\footnotefont{\bf Fig.~\the\figno:} #2}}
\bigskip\endinsert\global\advance\figno by1}


\def\unlockat{\catcode`\@=11}
\def\lockat{\catcode`\@=12}

\unlockat

\def\newsec#1{\global\advance\secno by1\message{(\the\secno. #1)}
\global\subsecno=0\global\subsubsecno=0\eqnres@t\noindent
{\bf\the\secno. #1}
\writetoca{{\secsym} {#1}}\par\nobreak\medskip\nobreak}
\global\newcount\subsecno \global\subsecno=0
\def\subsec#1{\global\advance\subsecno
by1\message{(\secsym\the\subsecno. #1)}
\ifnum\lastpenalty>9000\else\bigbreak\fi\global\subsubsecno=0
\noindent{\it\secsym\the\subsecno. #1}
\writetoca{\string\quad {\secsym\the\subsecno.} {#1}}
\par\nobreak\medskip\nobreak}
\global\newcount\subsubsecno \global\subsubsecno=0
\def\subsubsec#1{\global\advance\subsubsecno by1
\message{(\secsym\the\subsecno.\the\subsubsecno. #1)}
\ifnum\lastpenalty>9000\else\bigbreak\fi
\noindent\quad{\secsym\the\subsecno.\the\subsubsecno.}{#1}
\writetoca{\string\qquad{\secsym\the\subsecno.\the\subsubsecno.}{#1}}
\par\nobreak\medskip\nobreak}

\def\subsubseclab#1{\DefWarn#1\xdef
#1{\noexpand\hyperref{}{subsubsection}%
{\secsym\the\subsecno.\the\subsubsecno}%
{\secsym\the\subsecno.\the\subsubsecno}}%
\writedef{#1\leftbracket#1}\wrlabeL{#1=#1}}
\lockat

\def\IL{\relax{\rm I\kern-.18em L}}
\def\IH{\relax{\rm I\kern-.18em H}}
\def\IR{\relax{\rm I\kern-.18em R}}
\def\IC{\relax\hbox{$\inbar\kern-.3em{\rm C}$}}
\def\IZ{\relax\ifmmode\mathchoice
{\hbox{\cmss Z\kern-.4em Z}}{\hbox{\cmss Z\kern-.4em Z}}
{\lower.9pt\hbox{\cmsss Z\kern-.4em Z}}
{\lower1.2pt\hbox{\cmsss Z\kern-.4em Z}}\else{\cmss Z\kern-.4em
Z}\fi}
\def\CM {{\cal M}}
\def\CN {{\cal N}}

\def\CP {{\cal P }}
\def\CK{{\cal K}}
\def\CL {{\cal L}}
\def\CV {{\cal V}}
\def\CO {{\cal O}}

\def\CE {{\cal E}}
\def\CG {{\cal G}}
\def\CH {{\cal H}}
\def\CC {{\cal C}}
\def\CB {{\cal B}}
\def\CS {{\cal S}}

\def\CM {{\cal M}}
\def\CN {{\cal N}}

\def\CO {{\cal O}}

\def\CP {{\cal P }}

\def\CE{{\cal E }}
\def\CV{{\cal V }}

\def\CS {{\cal S }}

\font\manual=manfnt \def\dbend{\lower3.5pt\hbox{\manual\char127}}

\def\IZ{\relax\ifmmode\mathchoice
{\hbox{\cmss Z\kern-.4em Z}}{\hbox{\cmss Z\kern-.4em Z}}
{\lower.9pt\hbox{\cmsss Z\kern-.4em Z}}
{\lower1.2pt\hbox{\cmsss Z\kern-.4em Z}}\else{\cmss Z\kern-.4em
Z}\fi}
\def\half {{1\over 2}}
\def\sdtimes{\mathbin{\hbox{\hskip2pt\vrule height 4.1pt depth -.3pt
width
.25pt
\hskip-2pt$\times$}}}
\def\p{\partial}

\def\CM {{\cal M}}
\def\CN {{\cal N}}

\def\CO {{\cal O}}

\def\CP {{\cal P }}

\def\CE{{\cal E }}
\def\CV{{\cal V }}

\def\CS {{\cal S }}


\def\Hom{{\rm Hom}}

\def\IZ{\relax\ifmmode\mathchoice
{\hbox{\cmss Z\kern-.4em Z}}{\hbox{\cmss Z\kern-.4em Z}}
{\lower.9pt\hbox{\cmsss Z\kern-.4em Z}}
{\lower1.2pt\hbox{\cmsss Z\kern-.4em Z}}\else{\cmss Z\kern-.4em
Z}\fi}
\def\IB{\relax{\rm I\kern-.18em B}}
\def\IC{{\relax\hbox{$\inbar\kern-.3em{\rm C}$}}}
\def\ID{\relax{\rm I\kern-.18em B}}
\def\IC{{\relax\hbox{$\inbar\kern-.3em{\rm C}$}}}
\def\ID{\relax{\rm I\kern-.18em D}}
\def\IE{\relax{\rm I\kern-.18em E}}
\def\IF{\relax{\rm I\kern-.18em F}}
\def\IG{\relax\hbox{$\inbar\kern-.3em{\rm G}$}}
\def\IGa{\relax\hbox{${\rm I}\kern-.18em\Gamma$}}
\def\IH{\relax{\rm I\kern-.18em H}}
\def\II{\relax{\rm I\kern-.18em I}}
\def\IK{\relax{\rm I\kern-.18em K}}
\def\IP{\relax{\rm I\kern-.18em P}}

\def\inbar{\,\vrule height1.5ex width.4pt depth0pt}

\def\p{\partial}

\font\cmss=cmss10 \font\cmsss=cmss10 at 7pt
\def\IR{\relax{\rm I\kern-.18em R}}

\def\sdtimes{\mathbin{\hbox{\hskip2pt\vrule
height 4.1pt depth -.3pt width .25pt\hskip-2pt$\times$}}}
\def\Tr{\rm Tr}


\def\boxit#1{\vbox{\hrule\hbox{\vrule\kern8pt
\vbox{\hbox{\kern8pt}\hbox{\vbox{#1}}\hbox{\kern8pt}}
\kern8pt\vrule}\hrule}}
\def\mathboxit#1{\vbox{\hrule\hbox{\vrule\kern8pt\vbox{\kern8pt
\hbox{$\displaystyle #1$}\kern8pt}\kern8pt\vrule}\hrule}}


\def\inbar{\,\vrule height1.5ex width.4pt depth0pt}

\def\p{\partial}

\font\cmss=cmss10 \font\cmsss=cmss10 at 7pt
\def\IR{\relax{\rm I\kern-.18em R}}

\def\sdtimes{\mathbin{\hbox{\hskip2pt\vrule
height 4.1pt depth -.3pt width .25pt\hskip-2pt$\times$}}}

\def\Tr{\rm Tr}

\def\CC{{\cal{C}}}
\def\CO{{\cal{O}}}

\def\Tr{{{\rm Tr}}}
\def\CP{{\cal P}}

\def\out{\iota^{*}}

\lref\quillen{D. G. Quillen, ``Rational homotopy theory", {\it Annals of Mathematics} \ 90 (1969), 205-295.}
\lref\mandell{M. A. Mandell, ``Cochains and homotopy type", math.AT/0311016.}
\lref\petrack{E. Getzler, J. D. S. Jones, and S. Petrack, ``Differential forms on loop space and the cyclic bar complex," {\it Topology} 30 (1991), 339-371.}
\lref\aspinwall{P. Aspinwall, ``A Note on the Equivalence of
 Vafa's and Douglas's Picture of Discrete Torsion,''
hep-th/0009045 }

\lref\AtiyahSegal{M.~Atiyah and G.~Segal,
   ``Twisted K-theory and cohomology,''
  arXiv:math.kt/0510674.
}

\lref\BanksAZ{
  T.~Banks,
   ``TASI lectures on matrix theory,''
  arXiv:hep-th/9911068.
}

 \lref\behrend{R.E.  Behrend, P.A. Pearce, V.B. Petkova, J-B Zuber,
``Boundary conditions in Rational Conformal Field Theories,''
hep-th/9908036}
\lref\bouwknegt{P. Bouwknegt and V. Mathai, ``D-Branes,
B-fields, and Twisted K-Theory,'' hep-th/0002023.}
\lref\bdlr{I. Brunner, M.R. Douglas, A. Lawrence,
and C. Romelsberger,  ``D-branes on the Quintic,''
hep-th/9906200}

\lref\brunnermoore{I. Brunner and G. Moore, unpublished. }
\lref\cardy{J. Cardy, ``Boundary conditions, fusion rules and the
Verlinde formula,'' Nucl. Phys. B324 (1989)581}
\lref\caldararu{A. Caldararu, ``The Mukai pairing, I: the Hochschild
structure,''math.AG/0308079; ``The Mukai pairing, II: the
Hochschild-Kostant-Rosenberg isomorphism,'' math.AG/0308080}

\lref\markarian{N. Markarian, ``Poincar\'e-Birkhoff-Witt Isomorphism
and Riemann Roch Theorem,'' unpublished. }
\lref\coste{A. Coste, T. Gannon, and P. Ruelle, ``Finite group
modular data,'' hep-th/0001158}
%

\lref\CostelloEI{
  K.~Costello,
   ``Topological conformal field theories and Calabi-Yau categories,''
  arXiv:math.qa/0412149.
}

 \lref\curtisreiner{C. Curtis and I. Reiner, {\it Representation
theory of finite groups and associative algebras}, AMS Chelsea,
2006}

\lref\dvvv{R.~Dijkgraaf, C.~Vafa, E.~P.~Verlinde and H.~L.~Verlinde,
   ``The Operator Algebra Of Orbifold Models,''
  Commun.\ Math.\ Phys.\  {\bf 123}, 485 (1989).
  }

\lref\DijkgraafPhD{R. Dijkgraaf, {\it A Geometrical Approach to
Two-Dimensional Conformal Field Theory}, PhD Thesis, 1989}

\lref\dvvtst{R. Dijkgraaf, H. Verlinde, and E. Verlinde,
``Notes on topological string theory and 2D quantum gravity,''
 Lectures given at Spring School on Strings and Quantum Gravity,
Trieste, Italy, Apr 24 - May 2, 1990 and
at Cargese Workshop on Random Surfaces,
 Quantum Gravity and Strings, Cargese, France, May 28 - Jun 1, 1990.
 }
\lref\douglas{M.R. Douglas, ... Douglas and Fiol}
\lref\dm{M.~R.~Douglas and G.~Moore,
``D-branes, Quivers, and ALE Instantons,'' hep-th/9603167.}
\lref\douglas{M. Douglas, ``D-branes, Categories and N=1 Supersymmetry,''
hep-th/0011017}
\lref\drozd{Yu. A. Drozd and V.V. Kirichenko,
{\it Finite Dimensional Algebras} Springer-Verlag 1994}
\lref\freedone{D. Freed,  ``Higher algebraic structures and
quantization'', Commun. Math. Phys., {\bf 159} (1994), 343--398;
hep-th/9212115.}
\lref\freedtwo{D. Freed, `` $K$-theory in quantum field theory,'' in
{\it  Current Developments in Mathematics 2001}, International
Press, Somerville, MA, pp. 41-87; math-ph/0206031.}
\lref\gp{E.~Gimon and J.~Polchinski,
``Consistency Conditions for Orientifolds and D-Manifolds,''
Phys.\ Rev.\  {\bf D54}, 1667 (1996)
hep-th/9601038.}
\lref\ginspargmoore{P. Ginsparg and G. Moore,
Lectures on 2D Gravity and 2D String Theory,
 hepth/9304011; in {\it Recent Directions in Particle Theory}, J. Harvey
and J. Polchinski,
eds., World Scientific, 1993.}

\lref\hofman{C. Hofman and W.K. Ma, ``Deformations of
topological open strings,'' hep-th/0006120}

\lref\HIV{K. Hori, A. Iqbal, and C. Vafa, ``D-Branes and Mirror
Symmetry,'' hep-th/0005247}

\lref\hori{K. Hori, ``Linear Models of Supersymmetric
D-Branes,'' hep-th/0012179}

\lref\kapustin{ A. Kapustin, ``$D$-branes in a Topologically
Nontrivial B-field,'' hep-th/9909089.}
\lref\koscorb{A.~Konechny and A.~Schwarz,
``Compactification of M(atrix) theory on noncommutative toroidal orbifolds,''
Nucl.\ Phys.\  {\bf B591}, 667 (2000),
hep-th/9912185;
``Moduli spaces of maximally supersymmetric solutions
on noncommutative  tori and noncommutative orbifolds,''
JHEP {\bf 0009}, 005 (2000), hep-th/0005174.}
\lref\koscrev{A.~Konechny and A.~Schwarz,
``Introduction to M(atrix) theory and noncommutative geometry,''
hep-th/0012145.}
\lref\lazaroiu{C.I. Lazaroiu,
`` Unitarity, D-brane dynamics and D-brane categories,''
hep-th/0102183; ``Generalized complexes and string field theory,''
hep-th/0102122;
``Instanton amplitudes in open-closed topological string theory,''
hep-th/0011257;
``On the structure of open-closed topological field theory in two dimensions,''
hep-th/0010269.}

\lref\abrams{L. Abrams, ``The quantum Euler class and the
quantum cohomology of the Grassmannians,''
q-alg/9712205}
\lref\sieberttian{B. Siebert and G. Tian, ``On the quantum
cohomology rings of Fano manifolds and a formula
of Vafa and Intriligator,'' alg-geom/9403010}
\lref\tianxu{G. Tian and G. Xu, ``On the semisimplicity of
the quantum cohomology algebras of complete intersections,''
alg-geom/9611035}
\lref\stienstra{J. Stienstra, ``Resonant hypergeometric systems
and Mirror symmetry,'' alg-geom/9711002}

\lref\martmoore{E. Martinec and G. Moore,
`` Noncommutative Solitons on Orbifolds,''hep-th/0101199 }
\lref\mtrxreviews{Reviews on matrix theory}
%

\lref\MooreNN{
  G.~W.~Moore,
   ``Some comments on branes, G-flux, and K-theory,''
  Int.\ J.\ Mod.\ Phys.\ A {\bf 16}, 936 (2001)
  [arXiv:hep-th/0012007].
}

\lref\stringstalk{http://feynman.physics.lsa.umich.edu/cgi-bin/s2ktalk.cgi?moore}

\lref\kitptalks{ See   http://online.itp.ucsb.edu/online/mp01}

 \lref\gmoorehomepage{Go to
http://www.physics.rutgers.edu/~gmoore/clay.html.}

\lref\stanfordnotes{G. Segal, Lectures at Stanford University and at
ITP Santa Barbara, August 1999. Available at
http://doug-pc.itp.ucsb.edu/online/geom99/  }

\lref\lewellen{D. Lewellen, ``Sewing constraints
for conformal field theories on surfaces with
boundaries,'' Nucl.Phys. {\bf B372} (1992) 654;
J.L. Cardy and D.C. Lewellen, ``Bulk and boundary
operators in conformal field theory,''
Phys.Lett. {\bf B259} (1991) 274.}

\lref\maclane{S. Maclane, {\it Categories for the working
mathematician} Springer-Verlag GTM 5, 1971}

\lref\matsuo{Y. Matsuo, ``Identity projector and
D-brane in string field theory,'' hep-th/0106027}

\lref\seibergwitten{N. Seiberg and E. Witten,
``String Theory and Noncommutative Geometry,''
hep-th/9908142,JHEP 9909 (1999) 032}
\lref\chassullivan{M. Chas, and D. Sullivan, `` Closed string
operators in topology leading to Lie bialgebras and higher string
algebra,'' {\it  The legacy of Niels Henrik Abel}, 771--784,
Springer, Berlin, 2004. }
\lref\catcoh{G. Segal, ``Categories and cohomology theories,"
Topology \ {\bf 13} (1974), 293}
\lref\sullivan{D. Sullivan, ``Infinitesimal computations in
topology,''  Inst. Hautes \'Etudes Sci. Publ. Math. No. 47,
269--331 (1978).}
\lref\sharpe{E. Sharpe, ``Quotient Stacks and String Orbifolds,''
hep-th/0103008;``String Orbifolds and Quotient Stacks,''
hep-th/010221;``
Stacks and D-Brane Bundles,''hep-th/0102197;
``Recent Developments in Discrete Torsion,''
hep-th/0008191,``Discrete Torsion,''hep-th/0008154}
%

\lref\TaylorVB{
  W.~Taylor,
   ``M(atrix) theory: Matrix quantum mechanics as a fundamental theory,''
  Rev.\ Mod.\ Phys.\  {\bf 73}, 419 (2001)
  [arXiv:hep-th/0101126].
}
\lref\turaev{V. Turaev, ``Homotopy field theory in
dimension 2 and group-algebras,'' math.QA/9910010}

\lref\vafa{
C. Vafa, ``Modular invariance and discrete torsion
on orbifolds,'' Nucl. Phys. {\bf B273}(1986) 592}
\lref\wittenab{E. Witten, ``Mirror manifolds and
topological field theory,'' hep-th/9112056;
in {\it Essays on Mirror Manifolds}, S.-T. Yau ed.
International Press, 1992}
\lref\wittencs{E. Witten, ``Chern-Simons gauge theory as a
string theory,'' hep-th/9207094}
\lref\wittenk{E. Witten, ``$D$-Branes And $K$-Theory,''
JHEP {\bf 9812}:019, 1998; hep-th/9810188.}
\lref\wittenstrings{E. Witten, ``Overview of K-theory applied to
strings,'' hep-th/0007175.}
\lref\zwiebach{
Gaberdiel and Zwiebach, hep-th/9705038;
Gaberdiel and Zwiebach, hep-th/9707051;
Zwiebach, hep-th/9705241.}

\lref\AlexeevskiRP{
  A.~Alexeevski and S.~Natanzon,
   ``Non-commutative extensions of two-dimensional topological field  theories
   and Hurwitz numbers for real algebraic curves,''
  arXiv:math.gt/0202164.
}
\lref\BrunnerDC{
  I.~Brunner, M.~Herbst, W.~Lerche and B.~Scheuner,
   ``Landau-Ginzburg realization of open string TFT,''
  arXiv:hep-th/0305133.
}
\lref\BrunnerEM{
  I.~Brunner and K.~Hori,
   ``Notes on orientifolds of rational conformal field theories,''
  JHEP {\bf 0407}, 023 (2004)
  [arXiv:hep-th/0208141].
}

\lref\kaufmann{R.M. Kaufmann, ``Orbifolding Frobenius algebras,''
math.AG/0107163}

\lref\LupercioAP{
  E.~Lupercio and B.~Uribe,
   ``Topological quantum field theories, strings, and orbifolds,''
  arXiv:hep-th/0605255.
}

\lref\QuinnKQ{
  F.~Quinn,
   ``Lectures on axiomatic topological quantum field theory,''
  %
{\it Given at Graduate Summer School on the Geometry and Topology of
Manifolds and Quantum Field Theory, Park City, Utah, 22 Jun - 20 Jul
1991} }

\lref\SawinRH{
  S.~Sawin,
   ``Direct sum decompositions and indecomposable TQFTs,''
  J.\ Math.\ Phys.\  {\bf 36}, 6673 (1995)
  [arXiv:q-alg/9505026].
}

\lref\HoriIC{
  K.~Hori and J.~Walcher,
   ``D-brane categories for orientifolds: The Landau-Ginzburg case,''
  arXiv:hep-th/0606179.
}

\rightline{RUNHETC-06-12}
 \Title{hep-th/0609042}
{\vbox{\centerline{D-branes and K-theory   }
\medskip
\centerline{in 2D topological field theory }}}

\smallskip
\centerline{Gregory W. Moore}
\smallskip
\centerline{\it Department of Physics, Rutgers University}
\centerline{\it Piscataway, New Jersey, 08855-0849}
\smallskip
\centerline{Graeme Segal}
\smallskip
\centerline{\it All Souls College }
\smallskip\centerline{\it Oxford University, Oxford, OX1 4AL, England }
\medskip

\bigskip
\noindent This expository paper describes sewing conditions in
two-dimensional open/closed topological field theory. We include a
description of the $G$-equivariant case, where $G$ is a finite
group. We determine the category of boundary conditions in the case
that the closed string algebra is semisimple. In this case we find
that  sewing constraints -- the most primitive form of worldsheet
locality -- already imply that D-branes are ($G$-twisted) vector
bundles on spacetime. We comment on extensions to cochain-valued
theories and various applications. Finally, we give uniform proofs
of all relevant sewing theorems using Morse theory.

\medskip

\Date{August 31, 2006}

\newsec{Introduction and Summary}

The theory of D-branes  has proven to be of great importance in the
development of string theory. In this paper we will focus on certain
mathematical structures central to the idea of D-branes. One of the
questions which motivated our work was: ``Given a closed string
background, what is the set of possible D-branes?'' This is a rather
complicated question. One might at first be tempted to declare that
D-branes simply correspond to conformally invariant boundary
conditions for the open string. This viewpoint is not very useful
because there are too many such boundary conditions, and in general
they have no geometrical description. It also neglects important
restrictions imposed by sewing consistency conditions.

In this paper we address the above problem in the drastically
simpler case of two-dimensional {\it topological } field theory
(TFT), where the whole content of the theory is encoded in a
finite-dimensional commutative Frobenius algebra. We shall find that
describing the
 sewing conditions, and their solutions  for
2d topological open and closed TFT, is a tractable but not entirely
trivial problem. We also extend our results to the equivariant case,
where we are given a finite group $G$, and the worldsheets are
surfaces equipped with $G$-bundles.  This is relevant to the
classification of D-branes in orbifolds.

Another one of our primary motivations has been the desire to
understand the relation between D-branes and K-theory in the
simplest possible terms. This relation is often justified by
considerations of anomaly cancellation or of brane-antibrane
annihilation. Our analysis shows that the relation is, in some
sense, more primitive, and follows simply from sewing constraints.

We hope the present work will be of some pedagogical interest
in explaining the structure of boundary conformal field
theory and its connections to K-theory in the simplest  context.
There are also, however, some potential applications of our results.
One  ambitious goal is to classify
the boundary conditions in topologically twisted
nonlinear sigma models and their allied topological
string theories. Here we have some suggestive results, but they
are far from a complete theory.

Our main concrete results are the following two theorems.
To state
the first we must point out that a semisimple Frobenius
algebra\foot{For basic material on Frobenius algebras see, for
example, \curtisreiner, ch. 9 or \drozd .}
 $\CC$ is automatically the algebra of complex-valued functions
 on a finite set $X={\rm Spec}(\CC)$ --- the ``space-time" ---
 which is equipped with a ``volume-form" or ``dilaton field"
 $\theta$ which assigns a measure $\theta_x$ to each point $x \in X$.

\bigskip

\noindent{\bf Theorem A}\ \  For a semisimple 2-dimensional TFT
corresponding to a finite space-time $(X, \theta)$ the  choice of a
maximal category of D-branes fixes a choice of a square-root of
$\theta_x$ for each point $x$ of $X$. The category of boundary
conditions is equivalent to the category Vect$(X)$ of
finite-dimensional complex vector bundles on $X$. The correspondence
is, however, not canonical, but is arbitrary up to composition with
an equivalence Vect$(X)\to {\rm Vect}(X)$ given by tensoring each
vector bundle with a fixed line bundle (i.e. one which does not
depend on the particular D-brane).

Conversely, if we are given a semisimple Frobenius
category $\CB$, then it is the category of boundary
conditions for a canonical 2-dimensional TFT corresponding
to the commutative Frobenius algebra which is the ring of
endomorphisms of the identity functor of $\CB$.
\bigskip

We shall explain in the next section the sense in which the boundary
conditions form a category. The theorem will be proved in section 3.
In \S3.4\ we shall describe an analogue of the theorem for spin
theories.

\bigskip

The second theorem relates to ``$G$-equivariant" or ``$G$-gauged"
TFTs, where $G$ is a finite group. Turaev has shown that in
dimension 2 a semisimple $G$-equivariant TFT corresponds to a finite
space-time $X$ on which the group $G$ acts in a given way, and which
is equipped with a $G$-invariant dilaton field $\theta$ and as well
as a ``B-field" $B$ representing an element of the equivariant
cohomology group $H^3_G(X;\IZ)$.
\bigskip

\noindent{\bf Theorem B}\ \  For a semisimple $G$-equivariant TFT
corresponding to a finite space time $(X,\theta , B)$ the choice of
a maximal category of D-branes fixes a $G$-invariant choice of
square roots  $\sqrt\theta_x$ as before, and then the category is
equivalent to the category of finite-dimensional $B$-twisted
$G$-vector-bundles on $X$, up to an overall tensoring with a
$G$-line-bundle.

In this case the category of D-branes is equivalent to that of the
``orbifold" theory obtained from the gauged theory by integrating
over the gauge fields, and it does not remember the equivariant
theory from which the orbifold theory arose. There is, however, a
natural enrichment which does remember the equivariant theory.

\bigskip

This will be explained and proved in section 7.

\bigskip

The restriction to the semisimple case in our results seems at first
a damaging weakness. We believe, however, that it is this case that
reveals the essential structure of the theory. To go beyond it, the
appropriate objects of study, in our view, are
cochain-complex-valued TFTs rather than non-semisimple TFTs in the
usual sense. (This is analogous to the fact that in ordinary algebra
the duality theory of non-projective modules is best studied in the
derived category.) We have said something about this line of
development in sections 2 and 6, explaining how the category of
boundary conditions is naturally an $A_{\infty}$ category in the
sense of Fukaya, Kontsevich, and others.

Let us comment on one important conceptual aspect of the results of
this paper. In the  Matrix theory approach to nonperturbative string
theory \BanksAZ\TaylorVB\ {\it open} string field theory, or rather
its low-energy Yang-Mills theory, is taken to be the fundamental
starting point for the formulation of the entire string theory. In
particular, spacetime, and the closed strings, are regarded as
derived concepts. A similar philosophy lies at the root of the
AdS/CFT correspondence.  In this paper we begin our discussion with
the viewpoint that the spacetime and closed strings are fundamental
and then ask what category of boundary conditions is compatible with
that background. However, in the semisimple case, we find that one
could equally well start with the Frobenius category of boundary
conditions and derive the closed strings and the spacetime. Thus,
our treatment is in harmony with the philosophy of Matrix theory.
Indeed, it is possible to obtain the closed string algebra from the
open string algebra by taking the {\it center} of the open string
algebra $Z(\CO) \cong \CC$. (In general the Cardy condition only
shows that $\iota_*(\CC)$ maps into the center.) A more
sophisticated version of this idea is that the closed string algebra
is obtained from the category of boundary conditions by considering
the endomorphisms of the identity functor. All this is discussed in
\S3.3, and justifies the important point that there is a converse
statement to Theorem A.

A closely related point is that in open string field theory there
are different open string algebras $\CO_{aa}$ for the different
boundary conditions $a$. For boundary conditions with maximal
support, however, they are Morita equivalent via the bimodules
$\CO_{ab}$.  For some purposes it might seem more elegant to start
with a single algebra. (Indeed, Witten has suggested in
\wittenstrings\ that one should use something analogous to
stabilization of $C^*$ algebras, namely one should replace the
string field algebra by $\CO_{aa}\otimes \CK$ where $\CK$ is the
algebra of compact operators.) In our framework, the single algebra
is replaced by the category of boundary conditions. If one believes
that a stringy spacetime is a non-commutative space, our framework
is in good agreement with Kontsevich's approach to non-commutative
geometry, according to which a non-commutative space {\it is} a
linear category --- essentially the category of modules for the
ring, if the space is defined by a ring. For commutative rings the
category of modules determines the ring, but in the non-commutative
case the ring is determined only up to Morita equivalence. We
discuss this further in section \S3.

Finally, we comment briefly on some related literature. There is a
rather large literature on 2d TFT and it is impossible to give
comprehensive references. Here we just indicate some closely related
works.  The 2d closed sewing theorem is a very old result implicit
in the earliest papers in string theory. The algebraic formulation
was perhaps first formulated by  Friedan. Accounts have been given
in \DijkgraafPhD\SawinRH\QuinnKQ\ and in the Stanford lectures by
Segal \stanfordnotes. Sewing constraints in 2D open and closed
string theory were first investigated in \lewellen. Extensions to
unorientable worldsheets were described in \brunnermoore
\AlexeevskiRP\BrunnerEM\HoriIC.  Our work-- which is primarily
intended as a pedagogical exposition -- was first described at
Strings 2000 \stringstalk\  and summarized briefly in \MooreNN. It
was described more completely in lectures at the KITP in 2001 and at
the 2002 Clay School \kitptalks. In \gmoorehomepage\ one can find
alternative (more computational) proofs and examples to those we
give below, together with better quality pictures. \foot{Please note
that the arrows on some morphisms in figures 4, 35-38, 47-48 have
the wrong orientation.}  Some of our work was independently obtained
in the papers of C. Lazaroiu \lazaroiu\ although the emphasis in
these papers is on applications to disk instanton corrections in low
energy supergravity. Regarding $G$-equivariant theories, there is a
very large literature on D-branes and orbifolds not reflected in the
above references. In the context of 2D TFT two relevant references
are \kaufmann\LupercioAP. Alternative discussions on the meaning of
B-fields in orbifolds (in TFT) can be found in
\freedone\freedtwo\sharpe. Our treatment of cochain-level theories
and $A_\infty$ algebras  has been developed considerably further by
Costello \CostelloEI.

\bigskip
\centerline{\bf Acknowledgments}

We would like to thank Ilka Brunner, Robbert Dijkgraaf, Dan Freed,
Kentaro Hori and Anton Kapustin for some useful and clarifying
discussions. GM would like to thank Phil Candelas and Xenia de la
Ossa for hospitality at Oxford during the course of this work. We
thank K. Rabe for assistance with the figures.  GM and GS thank the
ITP for hospitality during the initiation of this work (August 1998)
and during the writing of the manuscript (March 2001). We also thank
the Aspen Center for Physics where some of this work was done.
 GM would like to thank the IAS and the Monell foundation for
hospitality during the completion of this work. The work of GM is
also supported by DOE grant DE-FG02-96ER40949.

\bigskip

\newsec{The sewing theorem}

\subsec{Definition of open and closed 2D TFT}

Roughly speaking, a quantum field theory is a functor from a
geometric category to a linear category. The simplest example is a
{\it topological field theory}, where we choose the geometric
category to be the category whose objects are closed, oriented
$(d-1)$-manifolds, and whose morphisms are oriented cobordisms (two
such cobordisms being identified if they are diffeomorphic by a
diffeomorphism which is the identity on the incoming and outgoing
boundaries). The linear category in this case is simply the category
of complex vector spaces and linear maps, and the only property we
require
of the functor is that (on objects and morphisms) it takes
disjoint unions to tensor products. The case $d=2$ is of course
especially well known and understood.

There are several natural ways to generalize the geometric category.
One may, for example, consider
manifolds equipped with some structure
such as a Riemannian metric. (We shall discuss some examples in the
following.) The focus of this paper is on a different kind of
generalization where the objects of the geometric category are
oriented $(d-1)$-manifolds with boundary, and each boundary
component is labelled with an element of a fixed set $\CB_0$ called
the set of {\it boundary conditions}. In this case a cobordism from
$Y_0$ to $Y_1$ means a $d$-manifold $X$ whose boundary consists of
three parts $\partial X = Y_0 \cup Y_1 \cup \partial _{\rm cstr}X$,
where the ``constrained boundary" $\partial _{\rm cstr}X$ is a
cobordism from $\partial Y_0$ to $\partial Y_1$. Furthermore, we
require the connected components of $\partial _{\rm cstr}X$ to be
labelled with elements of $\CB_0$ in agreement with the labelling of
$\partial Y_0$ and $\partial Y_1$.

Thus when $d=2$  the objects of the geometric category  are disjoint
unions of circles and oriented intervals with labelled ends. A
functor from this category to complex vector spaces which takes
disjoint unions to tensor products will be called an {\it open and
closed topological field theory}: such theories will give us a
``baby'' model of the theory of D-branes. We shall always write
$\CC$ for the vector space associated to the standard circle $S^1$,
and $\CO_{ab}$ for the vector space associated to the interval
$[0,1]$ with ends labelled by $a,b \in \CB_0$.

\ifig\FIGONE{Basic cobordism on open strings. }
{\epsfxsize2.0in\epsfbox{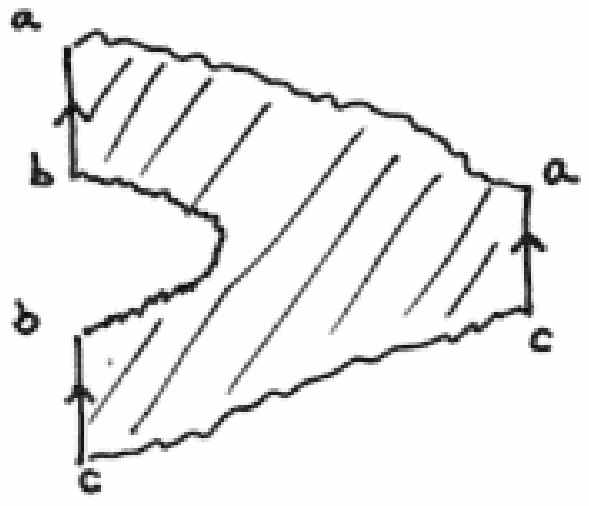}}

The cobordism \FIGONE\  gives us a linear map $\CO_{ab}\otimes
\CO_{bc}\to \CO_{ac}$, or equivalently a bilinear map
\eqn\compn{ \CO_{ab}\times \CO_{bc}\to \CO_{ac}, } which we think of
as a composition law. In fact we
have a $\IC$-linear {\it category}
$\CB$ whose objects are the elements of $\CB_0$, and whose set of
morphisms from $b$ to $a$ is the vector space $\CO_{ab}$, with
composition of morphisms given by \compn\ .   (To say that $\CB$ is
a category means no more than that the composition \compn\ is
associative in the obvious sense, and that there is an identity
element $1_a \in \CO_{aa}$ for each $a\in\CB_0$; we shall explain
presently why these properties hold.)

For any open and closed TFT we have a
map $e:\CC \to \CC$ defined by
the cylindrical cobordism $S^1\times [0,1]$, and a map
$e_{ab}:\CO_{ab} \to \CO_{ab}$ defined by the square $[0,1]\times
[0,1]$. Clearly $e^2 =e$ and $e_{ab} ^2 = e_{ab}$. If all these maps
are identity maps we say the theory is {\it reduced}. There is no
loss in restricting ourselves to reduced theories, and we shall do
so from now on.

\subsec{Algebraic characterization  }

The most general 2D open and closed TFT, formulated as in
the previous section, is given by the following algebraic data:

1. $(\CC, \theta_\CC, 1_{\CC} )$ is a commutative Frobenius algebra.

2a. $\CO_{ab}$ is a collection of vector spaces for $a,b\in \CB_0$
with an associative bilinear product
\eqn\assoc{
\CO_{ab} \otimes \CO_{bc} \rightarrow \CO_{ac}
}

2b. The $\CO_{aa}$ have nondegenerate traces
\eqn\trcs{
\theta_a: \CO_{aa} \rightarrow \IC
}
In particular, each $\CO_{aa}$ is a not-necessarily commutative
Frobenius algebra.

2c. Moreover,
\eqn\dualpair{
\eqalign{
\CO_{ab}\otimes \CO_{ba} & \rightarrow \CO_{aa}~~
 {\buildrel \theta_a \over \rightarrow}~~ \IC \cr
\CO_{ba}\otimes \CO_{ab} & \rightarrow \CO_{bb}~~
 {\buildrel \theta_b \over \rightarrow} ~~ \IC \cr}
}
are perfect pairings with
\eqn\cyclic{
\theta_a(\psi_1\psi_2) = \theta_b(\psi_2\psi_1)
}
for $\psi_1\in \CO_{ab}, \psi_2\in\CO_{ba}$.

3. There are linear maps:
\eqn\linmaps{
\eqalign{
\iota_a : \CC & \rightarrow \CO_{aa} \cr
\iota^a : \CO_{aa} & \rightarrow \CC \cr}
}

such that

3a. $\iota_a$ is an algebra homomorphism
\eqn\centerh{
\eqalign{
\iota_a(\phi_1\phi_2) & = \iota_a(\phi_1) \iota_a(\phi_2) \cr}
}

3b. The identity is preserved

\eqn\ident{
\iota_a(1_\CC) = 1_a
}

3c. Moreover, $\iota_a$ is  central in the sense that

\eqn\center{
\iota_a(\phi) \psi  =  \psi \iota_b(\phi)
}
for all $\phi\in \CC$ and $\psi\in \CO_{ab}$

3d. $\iota_a$ and $\iota^a$ are adjoints:
\eqn\adjoint{
\theta_\CC(\iota^a(\psi)\phi) = \theta_a(\psi \iota_a(\phi))
}
for all $\psi\in \CO_{aa}$.

3e. The ``Cardy conditions.''\foot{These are actually
generalization of the conditions stated
by Cardy. One recovers his conditions by taking the
trace. Of course, the factorization of the double twist
diagram in the closed string channel is an observation
going back to the earliest days of string theory. }
 Define $\pi_b^{~~a}: \CO_{aa} \rightarrow
\CO_{bb}$ as follows.  Since
$\CO_{ab}$ and $\CO_{ba}$ are in duality (using $\theta_a$ or
$\theta_b$),  if we let  $\psi_\mu$ be a basis for $\CO_{ba}$
then there is a dual basis  $\psi^\mu$
for $\CO_{ab}$. Then we define
\eqn\dblttw{
\pi_b^{~~a}(\psi) = \sum_\mu \psi_\mu
\psi\psi^\mu ,
}
and we have the ``Cardy condition'':
\eqn\cardycon{
\pi_b^{~~a} = \iota_b\circ \iota^a .
}

\ifig\FIGTWO{Four diagrams defining the Frobenius structure in a
closed 2d TFT. It is often more convenient to represent the
morphisms by the planar diagrams. In this case our convention is
that a circle oriented so that the right hand points into the
surface is an ingoing circle. }
{\epsfxsize2.0in\epsfbox{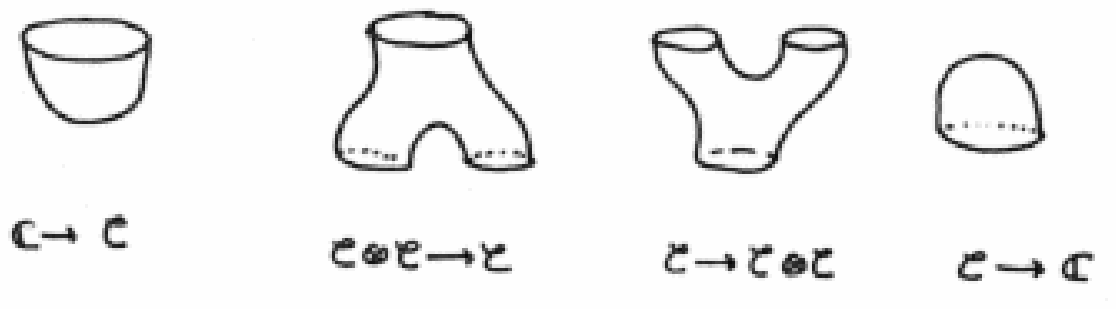}}

\ifig\FIGTHREE{Associativity, commutativity, and unit constraints in
the closed case. The unit constraint requires the natural assumption
that the cylinder correspond to the identity map $\CC \to \CC$. }
{\epsfxsize2.0in\epsfbox{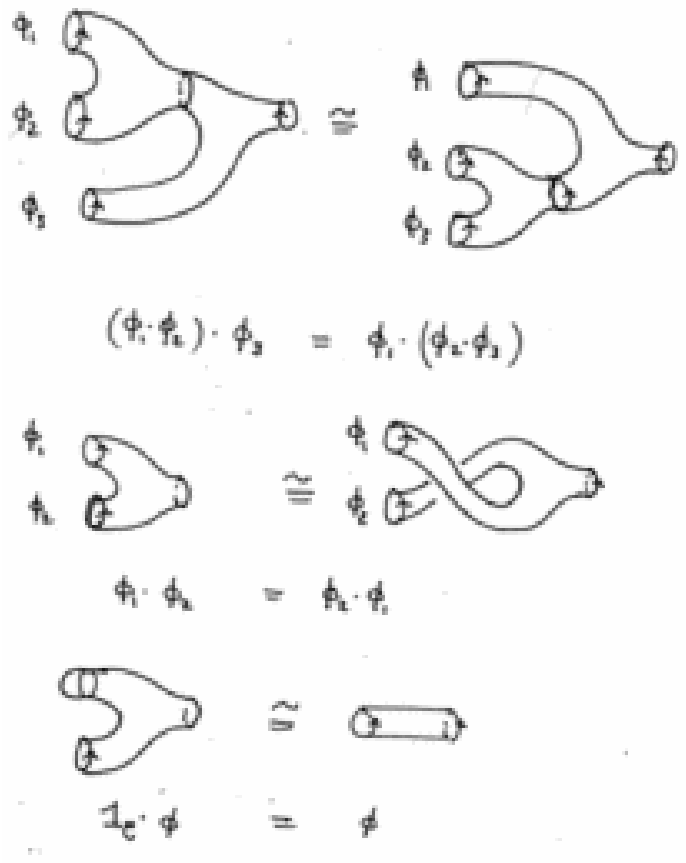}}

\subsec{Pictorial representation}

Let us explain the pictorial basis for these algebraic conditions.
The case of a closed 2d TFT is very well-known. The data of the
Frobenius structure is provided by the diagrams in \FIGTWO. The
consistency conditions follow from \FIGTHREE.

\ifig\FIGFOUR{Basic data for the open theory. Constrained boundaries
are denoted with wiggly lines, and carry a boundary condition
$a,b,c,\dots \in \CB_0.$. } {\epsfxsize2.0in\epsfbox{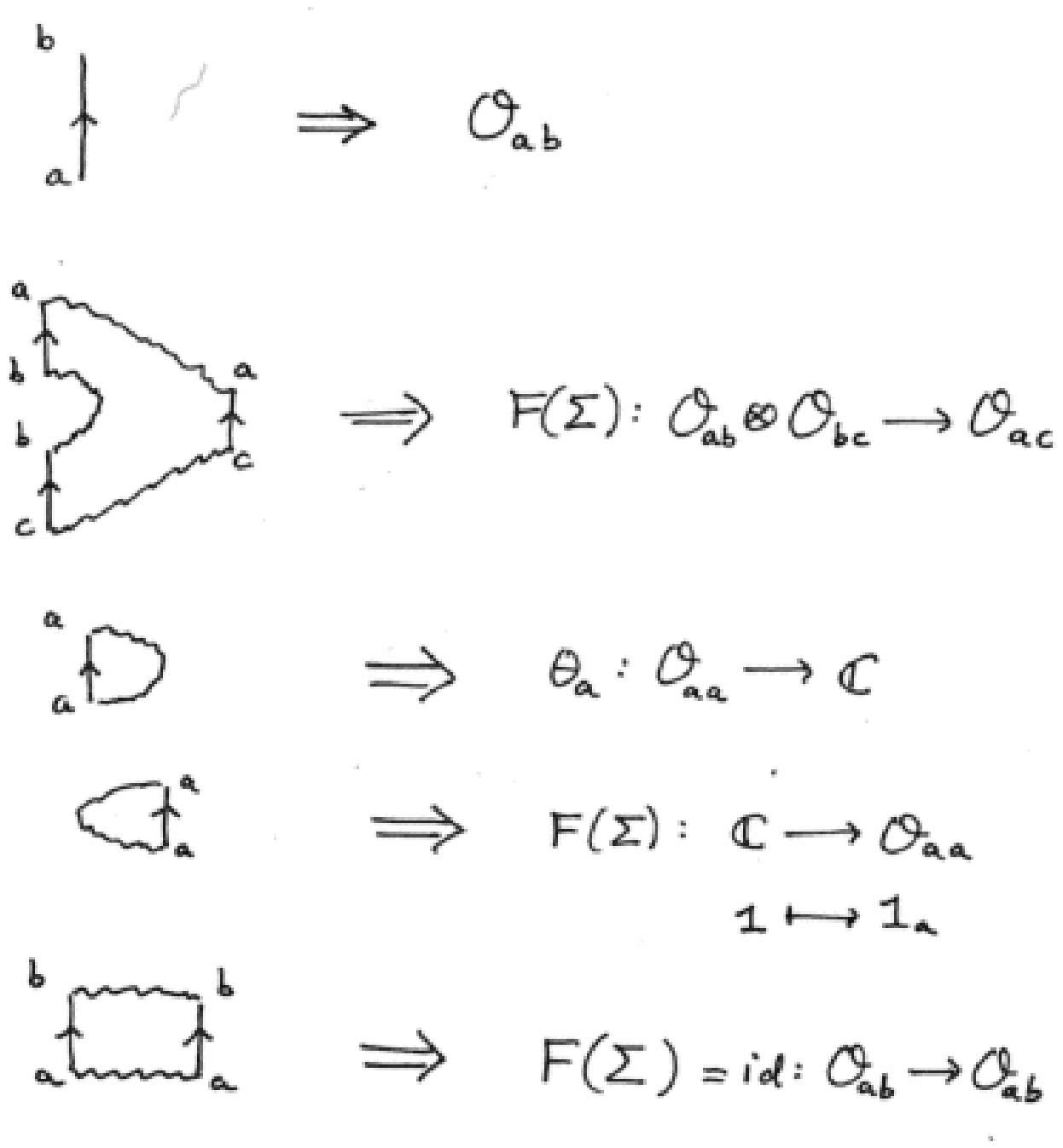}}

\ifig\FIGFIVE{Assuming that the strip corresponds to the identity
morphism we must have perfect pairings in \dualpair. }
{\epsfxsize2.0in\epsfbox{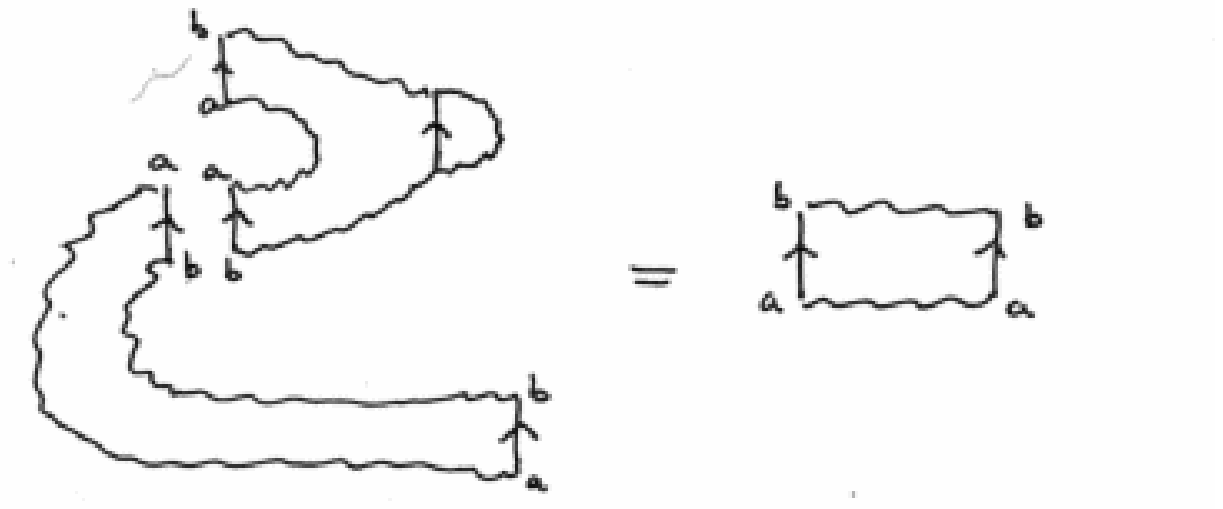}}

\ifig\FIGSIX{Two ways of representing open to closed and closed to
open transitions. } {\epsfxsize2.0in\epsfbox{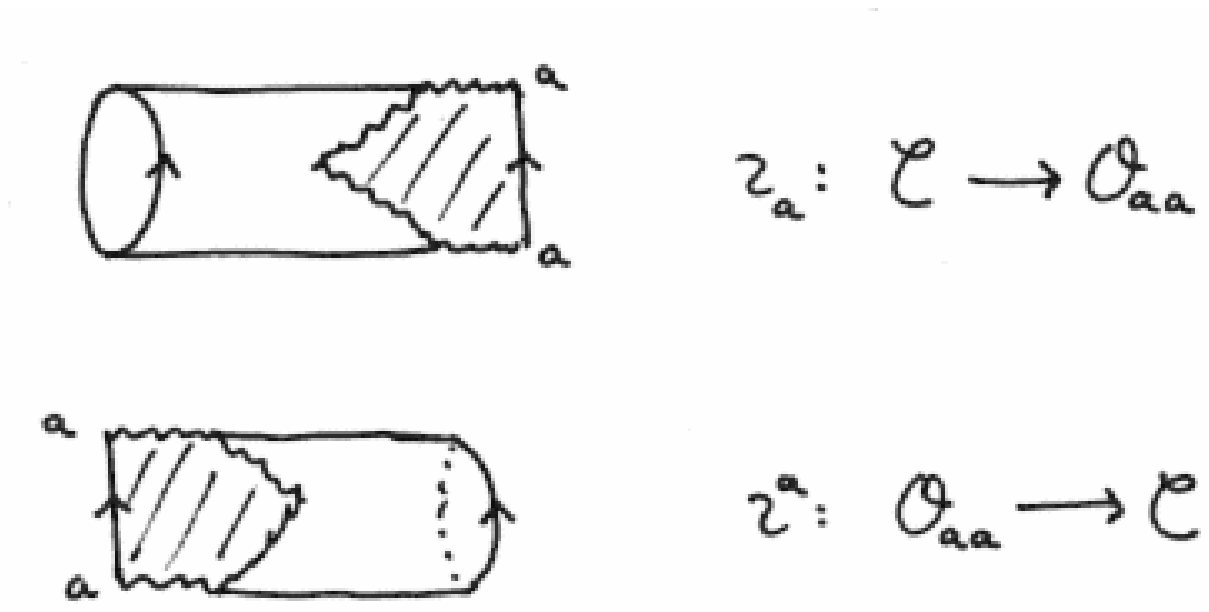}}

\ifig\FIGSEVEN{$\iota_a$ is a homomorphism. }
{\epsfxsize2.0in\epsfbox{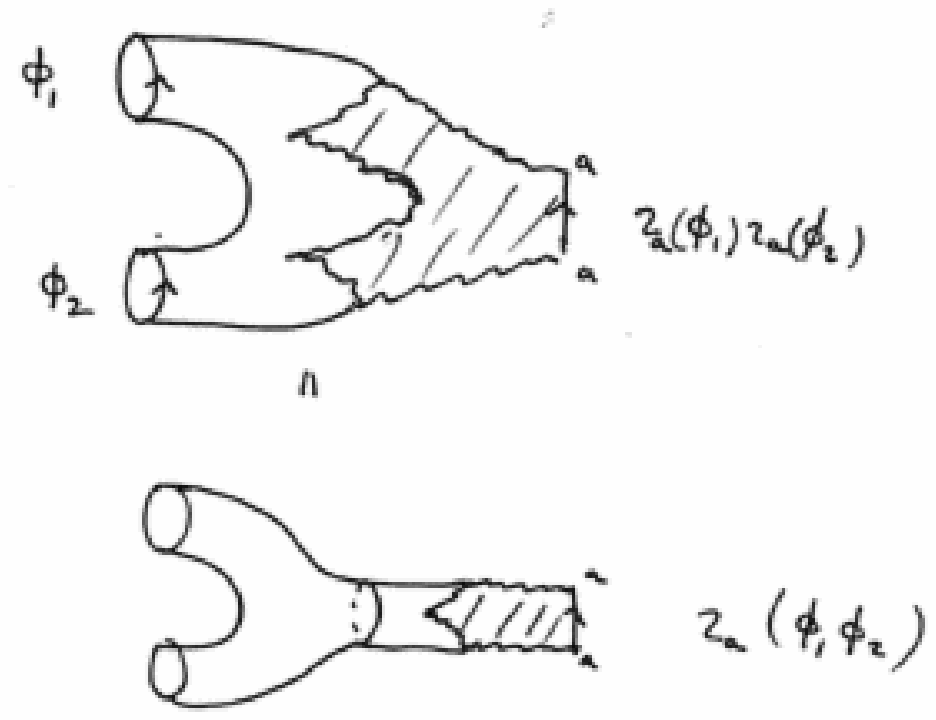}}

\ifig\FIGEIGHT{$\iota_a$ preserves the identity. }
{\epsfxsize2.0in\epsfbox{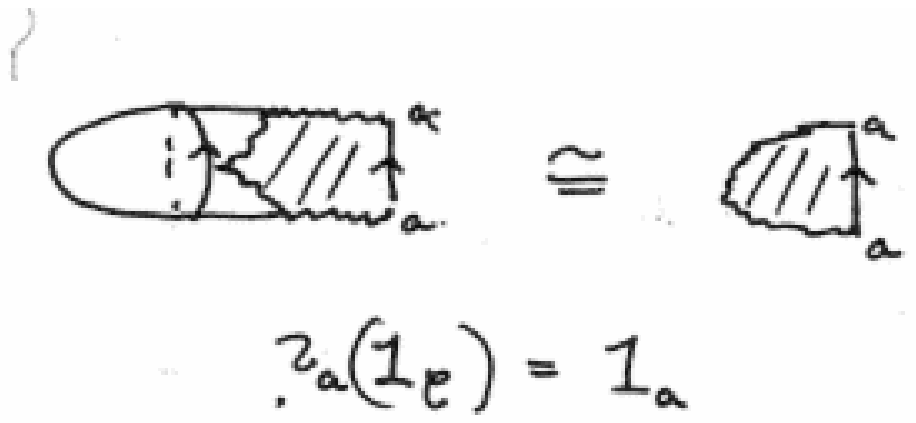}}

\ifig\FIGNINE{$\iota_a$ maps into the center of $\CO_{aa}$. }
{\epsfxsize2.0in\epsfbox{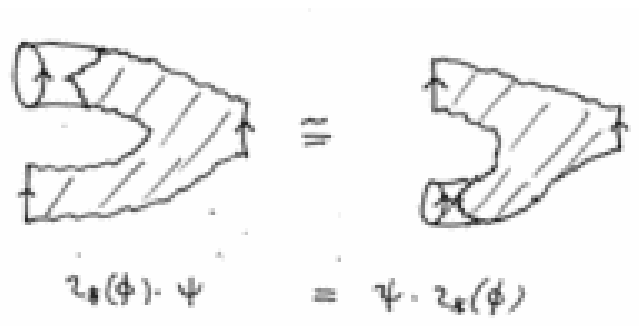}}

\ifig\FIGTEN{$\iota^a$ is the adjoint of $\iota_a$. }
{\epsfxsize2.0in\epsfbox{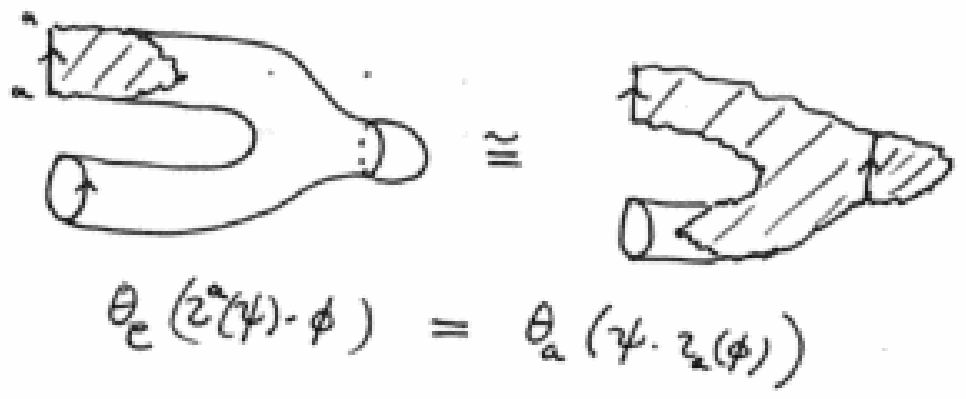}}

\ifig\FIGELEVEN{The double-twist diagram defines the map
$\pi_b^{~a}:\CO_{aa} \to \CO_{bb}$. }
{\epsfxsize2.0in\epsfbox{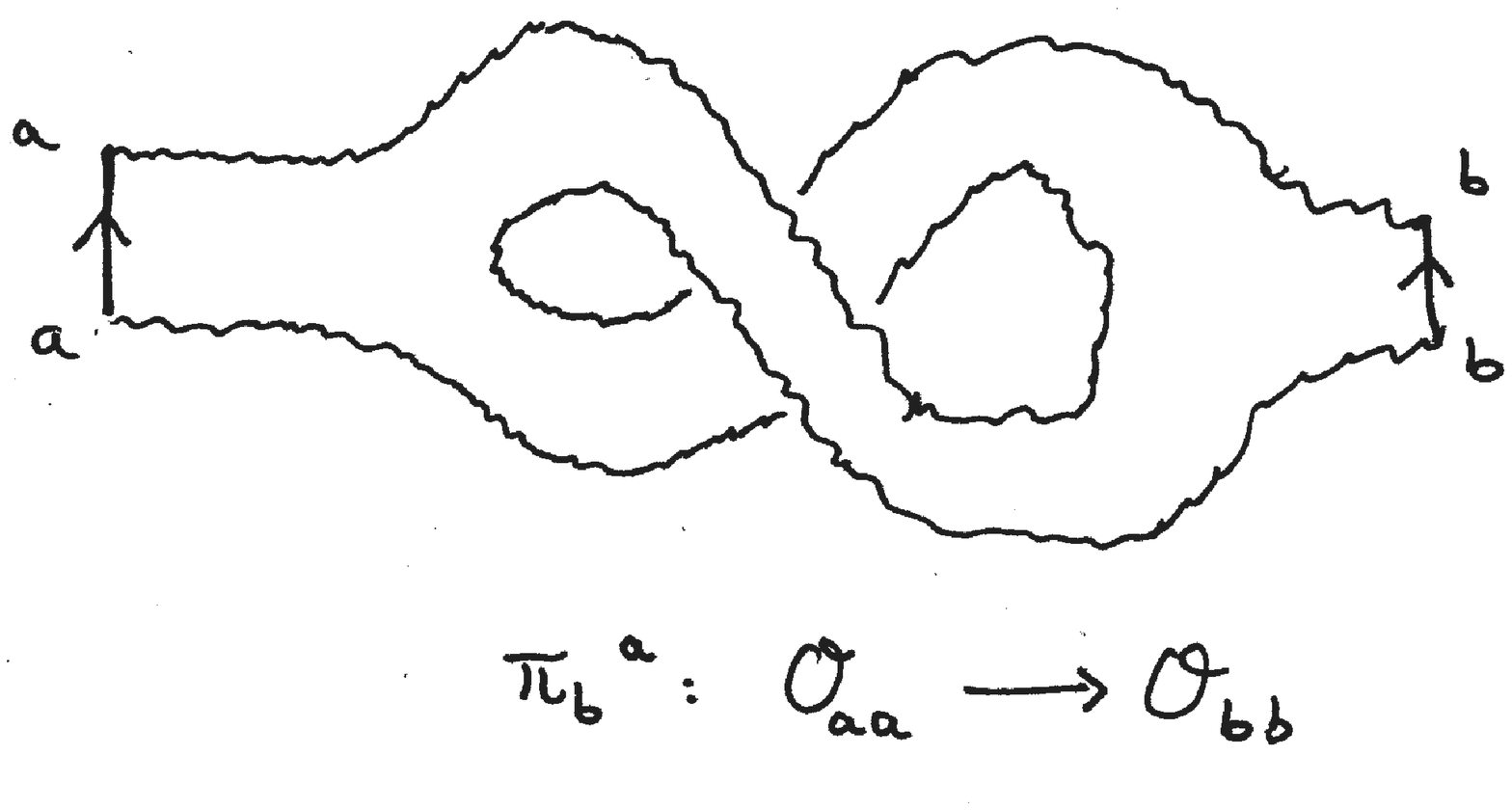}}

\ifig\FIGTWELVE{The (generalized) Cardy-condition expressing
factorization of the double-twist diagram in the closed string
channel. } {\epsfxsize2.0in\epsfbox{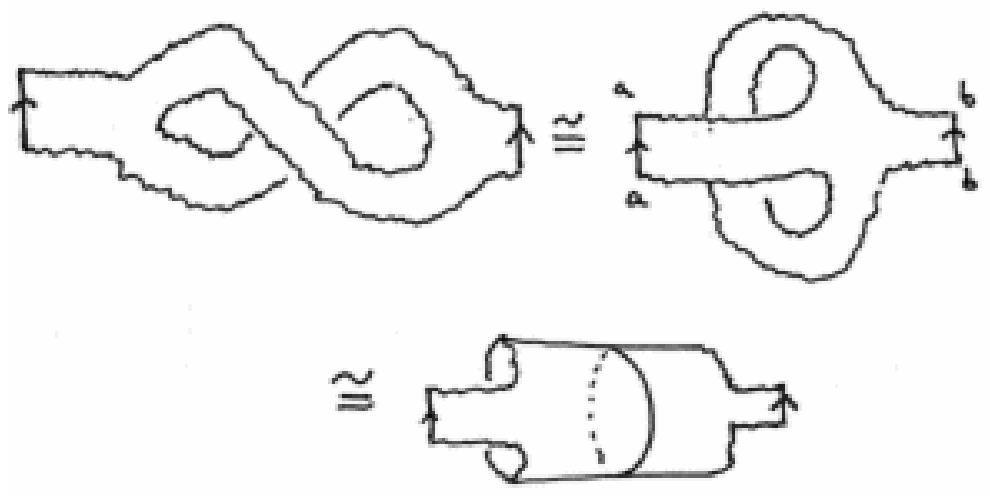}}

 In the open case, entirely
analogous considerations lead to the construction of a
non-necessarily commutative Frobenius algebra in the open sector.
The basic data are summarized in \FIGFOUR. The fact that \dualpair\
are dual pairings follows from \FIGFIVE.  The essential new
ingredient in the open/closed theory are the open to closed and
closed to open transitions. In 2d TFT these are the maps $\iota_a,
\iota^a$. They are represented by \FIGSIX. There are five  new
consistency conditions associated with the open/closed transitions.
These are illustrated in \FIGSEVEN\ to \FIGTWELVE.

\subsec{Sewing theorem}

Geometrically, any oriented surface can be decomposed
into a composition of morphisms
corresponding to the basic data defining the
Frobenius structure. However, a given surface
can be decomposed in many different ways. The
above sewing axioms follow from consistency of
these decompositions. The sewing theorem guarantees
that there are no further relations on the algebraic
data imposed by consistency of sewing.

\bigskip
\noindent
{\bf Theorem 1}  Conditions 1,2,3 above are the only conditions on the
algebraic data coming from cutting the morphisms in all possible
ways.
\bigskip

The proof is in appendix A.

\subsec{The category of boundary conditions}

 The category $\CB$ of boundary conditions of an open
 and closed TFT is an additive category. We can always
 adjoin new objects to it in various ways. For example, we
 may as well assume that it possesses direct sums, as we can define
 for any two objects $a$ and $b$  a new object $a\oplus b$ by
\eqn\addobj{
\CO_{a\oplus b, c} := \CO_{ac} \oplus \CO_{bc}
}
\eqn\addobjj{
\CO_{c,a\oplus b} := \CO_{ca} \oplus \CO_{cb},
}
and hence

\eqn\linstruct{
\CO_{a\oplus b,a\oplus b } := \pmatrix{\CO_{aa} & \CO_{ab}\cr
\CO_{ba}& \CO_{bb} \cr},
}
with the obvious composition laws, and
\eqn\newtr{
\theta_{a\oplus b}:\CO_{a\oplus b ,a\oplus b}\to \IC
}
given by

\eqn\newelttr{
 \theta_{a\oplus b}\pmatrix{\psi_{aa} & \psi_{ab}\cr
\psi_{ba}& \psi_{bb} \cr}=\theta_a(\psi_{aa})+\theta_b(\psi_{bb}).
}
The new object is the direct sum of $a$ and $b$ in the enlarged
category of boundary conditions. If there was already a direct sum
of $a$ and $b$ in the category $\CB$ then the new object will be
canonically isomorphic to it. In the opposite direction, if we have
a boundary condition $a$ and a projection $p \in \CO_{aa}$ (i.e. an
element such that $p^2=p$) then we may as well assume there is a
boundary condition $b = {\rm image}(p)$ such that for any $c$ we
have $\CO_{cb} = \{f \in \CO_{ab}: pf = f \}$ and $\CO_{bc} = \{f
\in \CO_{ba}: fp = f \}$. Then we shall have $a \cong {\rm image}(p)
\oplus {\rm image}(1-p)$.\foot{A linear category in which
idempotents split in this way is often called {\it Karoubian}.}

\bigskip

One very special property that the category $\CB$ possesses
is that
for any two objects $a$ and $b$ the space $\CO_{ab}$ of morphisms is
canonically dual to $\CO_{ba}$, by a pairing which factorizes
through the composition in either order. It is natural to call  a
category with this property a {\it Frobenius} category, or perhaps a
{\it Calabi-Yau} category.\foot{The latter terminology comes from
the case of coherent sheaves on a compact K\"{a}hler manifold, where
for two sheaves $E$ and $F$ the dual of the morphism space
Ext$(E;F)$ is in general Ext$(F;E\otimes \omega)$, where $\omega$ is
the canonical bundle. This coincides with Ext$(F;E)$ only when
$\omega$ is trivial, i.e. in the Calabi-Yau case. We shall discuss
this example further in \S 6.} It is a strong restriction on the
category: for example the category of finitely generated modules
over a finite dimensional algebra does not have the property unless
the algebra is semisimple.

\bigskip

\noindent{\it Example}\ \ \ Probably the simplest example of
an open
and closed theory of the type we are studying is one associated to a
finite group $G$. The category $\CB$ is the category of finite
dimensional complex representations $M$ of $G$, and the trace
$\theta_M :\CO_{MM}= {\rm End}(M) \to \IC$ takes $\psi :M\to M$ to
$1/\vert G \vert \ {\rm trace}(\psi)$. The closed algebra $\CC$ is
the center of the group-algebra $\IC[G]$, which maps to each
End$(M)$ in the obvious way. The trace $\theta _{\CC}:\CC \to \IC$
takes a central element $\sum \lambda _g g$ of the group-algebra to
$\lambda_1/\vert G \vert $.

In this example the partition function of the theory on a
surface
$\Sigma$ with constrained boundary circles $C_1, \ldots ,C_k$
labelled $M_1,\ldots,M_k$ is the weighted sum over the isomorphism
classes of principal $G$-bundles $P$ on $\Sigma$ of
$$\chi_{M_1}(h_P(C_1))\ldots \chi_{M_k}(h_P(C_k)), $$
where $\chi_M :G\to\IC$ is the character of a representation $M$,
and $h_P(C)$ denotes the holonomy of $P$ around a boundary
circle
$C$. Each bundle $P$ is weighted by the reciprocal of the order of
its group of automorphisms.

\bigskip

Returning to the general theory, we can now ask three basic
questions.

(i) If we are given a ``closed" TFT, can we enlarge it to
an open and
closed theory, and, if so, is the enlargement unique?

(ii)  If we are given the category $\CB$ of
boundary conditions of an open and closed theory,
together with the linear maps $\theta_a:\CO_{aa}\to\IC$
which define the Frobenius structure, can we reconstruct
the whole theory, i.e. can we find the closed Frobenius algebra $\CC$?

(iii) Is an arbitrary Frobenius category the category of boundary
conditions for some closed theory?

\bigskip

For the first question to be well-posed, we should assume that the
category of boundary conditions is {\it maximal}, in the sense that
if $\CB'$ is an enlargement of it then any object of $\CB'$ is
isomorphic to an object of $\CB$. Even so, we shall see that there
are subtleties which prevent any of these question from having a
simple affirmative answer.

\subsec{Generalizations}

We can obtain many interesting generalizations of the
above structure by modifying either the geometrical
or the linear category.

The most general target category we can consider is a symmetric
tensor category: clearly we need a tensor
product, and the axiom $\CH
_{Y_1 \sqcup Y_2}\cong \CH_{Y_1}\otimes \CH _{Y_2}$ only makes sense
if there is an involutory canonical isomorphism $\CH_{Y_1}\otimes
\CH_{Y_2}\cong \CH_{Y_2}\otimes \CH_{Y_1}$

A very common choice in physics is the
category of {\it super vector
spaces}, i.e. vector spaces $V$ with a mod 2 grading $V=V^0 \oplus
V^1$, where the canonical isomorphism $V\otimes W \cong W\otimes V$
is $ v\otimes w \mapsto (-1)^{\deg v \deg w}w\otimes v.$ One can
also consider the category of $\IZ$-graded vector spaces, with the
same sign convention for the tensor product.

In either case the closed string algebra is a graded-commutative
algebra $\CC$ with a
trace $\theta :\CC \rightarrow \IC$. In
principle the trace should have degree zero, but in fact the
commonly encountered theories have a {\it grading anomaly} which
makes the trace have degree $-n$ for some integer $n$.\foot{It is
easy to see that, up to an overall translation of the grading, the
most general anomaly assigns an operator of degree $\half
n(i-o-\chi)$ to a cobordism with Euler number $\chi$ and $i$
incoming and $o$ outgoing boundary circles.}  The formulae \cyclic,
\center, and \dblttw\  must be replaced by their graded-commutative
analogues. In particular if we choose a basis $\psi_\mu$ and its
dual $\psi^\mu$ so that
\eqn\oneway{
\theta_{\CC}(\psi^\mu \psi_{~\nu}) = \delta^{\mu}_{~\nu}
}
then
\eqn\dblttwii{
\pi_b^{~~a}(\psi) = \sum_\mu (-1)^{\deg \psi_\mu \deg \psi } \psi_\mu
\psi\psi^\mu
}

We can also obtain interesting structures by changing
the geometrical category of manifolds and cobordisms
by equipping them with extra stucture.

\bigskip
\noindent
{\it Example 1}\ \ \ \  We define {\it topological-spin theories}
by replacing  ``manifolds''
with ``manifolds with spin-structure.''

A {\it spin structure} on a surface means a double covering of its
space of non-zero tangent vectors which is non-trivial on each
individual tangent space. On an oriented 1-dimensional manifold $S$
it means a double covering of the space of positively-oriented
tangent vectors. For purposes of gluing it is useful to note that
this is the same thing as a spin structure on a ribbon neighbourhood
of $S$ in an orientable surface. Each spin structure has an
automorphism which interchanges its sheets, and this will induce an
involution $T$ on any vector space which is naturally associated to
a 1-manifold with spin structure, giving the vector space a mod 2
grading by its $\pm 1$-eigenspaces. We define a topological-spin
theory as a functor from the cobordism category of manifolds with
spin structures to the category of super vector spaces with its
graded tensor structure. The functor is required to take disjoint
unions to super tensor products, and we also require the
automorphism of the spin structure of a 1-manifold to induce the
grading automorphism $T=(-1)^{\rm degree}$ of the super vector
space. We shall see presently that this choice of the supersymmetry
of the tensor product rather than the naive symmetry which ignores
the grading is forced on us by the geometry of spin structures if we
want to allow the possibility of a semisimple category of boundary
conditions. There are two non-isomorphic circles with spin
structure: $S^1_{ns}$, with the M\"obius or ``Neveu-Schwarz"
structure, and $S^1_r$, with the trivial or ``Ramond" structure. A
topological-spin theory gives us state-spaces $\CC_{ns}$,
respectively $\CC_{r}$ corresponding to $S^1_{ns},S^1_{r}$.

There are four annuli with spin structures, for, alongside the
cylinders $A^+_{ns,r} = S^1_{ns,r} \times [0,1]$ which induce the
identity maps of $\CC_{ns,r}$ there are also cylinders $A^-_{ns,r}$
which connect $S^1_{ns,r}$ to itself while interchanging the sheets.
These cylinders $A^-_{ns,r}$ induce the grading automorphism on the
state spaces. But because $A^-_{ns} \cong A^+_{ns}$ by an
isomorphism which is the identity on the boundary circles --- the
Dehn twist which ``rotates one end of the cylinder by $2\pi$" ---
the grading on $\CC_{ns}$ must be purely even. The space $\CC_{r}$
can have both even and odd components. The situation is a little
more complicated for ``U-shaped" cobordisms, i.e. cylinders with two
incoming or two outgoing boundary circles. If the boundaries are
$S^1_{ns}$ there is only one possibility, but if the boundaries are
$S^1_{r}$ there are two, corresponding to $A_-^{ns,r}$. The
complication is that there seems no special reason to prefer either
of the spin structures as ``positive". We shall simply {\it choose}
one --- let us call it $P$
--- with incoming boundary $S^1_r \sqcup S^1_r$, and use $P$ to
define a pairing $\CC_r \otimes \CC_r \to \IC$. We then choose a
preferred cobordism $Q$ in the other direction so that when we sew
its right-hand outgoing $S^1_r$ to the left-hand incoming one of $P$
the resulting S-bend is the ``trivial" cylinder $A^+_r$. We shall
need to know, however, that the closed torus formed by the
composition $P \circ Q$ has an {\it even} spin structure. Note that
Frobenius structure $\theta$ on $\CC$ restricts to $0$ on $\CC_{r}$.

There is a unique spin structure on the pair-of-pants cobordism of
fig.2 which restricts to $S^1_{ns}$ on each boundary circle, and it
makes $\CC_{ns}$ into a commutative Frobenius algebra in the usual
way. If one incoming circle is $S^1_{ns}$ and the other is $S^1_{r}$
then the outgoing circle is $S^1_r$, and there are two possible spin
structures, but the one obtained by removing a disc from the
cylinder $A^+_r$ is preferred: it makes $\CC_r$ into a graded module
over $\CC_{ns}$. The chosen U-shaped cobordism $P$, with two
incoming circles $S^1_r$, can be punctured to give us a pair of
pants with an outgoing $S^1_{ns}$, and it induces a  graded bilinear
map $\CC_r \times \CC_r \to \CC_{ns}$ which, composing with the
trace on $\CC_{ns}$, gives a non-degenerate inner product on
$\CC_r$. At this point the choice of symmetry of the tensor product
becomes important. For the diffeomorphism of the pair of pants which
shows us in the usual case that the Frobenius algebra is
commutative, when we lift it to the spin structure, induces the
identity on one incoming circle but reverses the sheets over the
other incoming circle, and this proves that the cobordsism must have
the same output when we change the input from $S(\phi_1 \otimes
\phi_2)$ to $T(\phi_1) \otimes \phi_2$, where $T$ is the grading
involution and $S:\CC_r \otimes \CC_r \to \CC_r \otimes \CC_r$ is
the symmetry of the tensor category. If we take $S$ to be the
identity, this shows that the product on the graded vector space
$\CC_r$ is graded-symmetric with the usual sign; but if $S$ is the
graded symmetry then we see that {\it the product on $\CC_r$ is
symmetric in the naive sense}. (We must bear in mind here that if
$\psi_1$ and $\psi_2$ do not have the same parity then their product
is in any case zero, as we have seen that $\CC_+$ is purely even.)

\bigskip

There is an analogue for spin theories of the theorem which tells us
that a two-dimensional topological field theory ``is" a commutative
Frobenius algebra. It asserts that a spin-topological theory ``is" a
Frobenius algebra $\CC =  (\CC_{ns} \oplus \CC_{r}, \theta_{\CC})$
with the properties just mentioned, and with the following
additional property. Let $\{\phi_k\}$ be a basis for $\CC_{ns}$,
with dual basis $\{\phi^k\}$ such that $\theta_{\CC}(\phi_k
\phi^m)=\delta^m_k$, and let $\beta_k$ and $\beta^k$ be similar dual
bases for $\CC_r$. Then the Euler elements $\chi_{ns} := \sum \phi_k
\phi^k$ and $\chi_r = \sum \beta_k \beta^k$ are independent of the
choices of bases, and the condition we need on the algebra $\CC$ is
that $\chi_{ns}=\chi_{r}$. In particular, this condition implies
that the vector spaces $\CC_{ns}$ and $\CC_r$ have the same
dimension. \foot{Thus, in a sense, the theory has ``spacetime
supersymmetry.''}  In fact, the Euler elements can be obtained from
cutting a hole out of the torus. There are actually four spin
structures on the torus. The output state is necessarily in
$\CC_{ns}$. The Euler elements for the three even spin structures
are equal to $\chi_e=\chi_{ns}=\chi_r$. There is in addition an
Euler element $\chi_o$ corresponding to the odd spin structure, it
is given by $\chi_o = \sum (-1)^{\deg \beta_k} \beta_k\beta^k$.

We shall omit the proof that the general spin theory is what we have
just described,  but it is almost identical with the proof we shall
give in the appendix of the theorem of Turaev about $G$-equivariant
theories in the simple case when the group $G$ is $\IZ/2$. Indeed a
spin theory is very similar to --- but not the same as --- a
$\IZ/2$-equivariant theory, which is the structure obtained when the
surfaces are equipped with principal $\IZ/2$-bundles (i.e. double
coverings) rather than spin structures. We shall discuss equivariant
theories in \S7. (One difference is that in the equivariant case the
$\IZ/2$ action is nontrivial in the sector $\CC_1$ and trivial in
$\CC_g$, precisely the opposite of what we have found in the spin
case.) Comparing with the equivariant theory, the surprising result
that the product on $\CC_r$ is naive-symmetric can be understood as
twisted-anticommutativity.

\bigskip

It seems reasonable to call a spin theory {\it semisimple} if the
algebra $\CC_{ns}$ is semisimple, i.e. is the algebra of functions
on a finite set $X$. Then $\CC_r$ is the space of sections of a
vector bundle $E$ on $X$, and it follows from the condition
$\chi_{ns} = \chi_r $ that the fibre at each point must have
dimension $1$. Thus the whole structure is determined by the
Frobenius algebra $\CC_{ns}$ together with the binary choice of the
grading of the fibre of the line bundle $E$ at each point.

We can now see that if we had used the graded symmetry in defining
the tensor category we should have forced the grading of $\CC_r$ to
be purely even. For on the odd part the inner product would have had
to be skew, and that is impossible on a 1-dimensional space. And if
both $\CC_{ns}$ and $\CC_r$ are purely even then the theory is in
fact completely independent of the spin structures on the surfaces.

\bigskip

A concrete example of a two-dimensional topological-spin theory is
given by $\CC = \IC \oplus \IC \eta$ where $\eta^2=1$ and $\eta$ is
odd. The Euler elements are $\chi_e=1$ and $\chi_o=-1$. It follows
that the partition function of a closed  surface with spin structure
is $\pm 1$ according as the spin structure is even or odd. (To prove
this it is useful to compute the Arf invariant of the quadratic
refinement of the intersection product associated to the spin
structure and to note that it is multiplicative for adding handles.)

\bigskip

The most common theories defined on surfaces with spin structure are
not topological: they are 2-dimensional conformal field theories
with $\CN=1$ supersymmetry. The general features of the structure
are still as we have described, but it should be noticed that if the
theory is not topological one does not expect the grading on $\CC
_{ns}$ to be purely even: states can change sign on rotation by
$2\pi$. If a surface $\Sigma$ has a conformal structure then a
double covering of the non-zero tangent vectors is the complement of
the zero-section in a two-dimensional real vector bundle $L$ on
$\Sigma$ which is called the {\it spin bundle}. The covering map
then extends to a symmetric pairing of vector bundles $L\otimes L
\to T\Sigma$, which if we regard $L$ and $T\Sigma$ as complex line
bundles in the natural way, induces an isomorphism $L\otimes _{\IC}L
\cong T\Sigma$. An $\CN = 1$ superconformal field theory is a
conformal-spin theory with an additional map
\eqn\scft{ \Gamma (S;L)\otimes \CH_{S,L}  \to\CH_{S,L} }
\eqn\scftelt{ (\sigma ,\psi)  \mapsto G_{\sigma}\psi }
such that $G_{\sigma}$ is real-linear in the section $\sigma$ of $L$
and satisfies $G_{\sigma}^2 = D_{\sigma ^2}$, where $D_{\sigma ^2}$
is the Virasoro action of the vector field $\sigma ^2$. Furthermore,
when we have a cobordism $(\Sigma ,L)$ from $(S_0,L_0)$ to
$(S_1,L_1)$ and a holomorphic section $\sigma$ of $L$ which
restricts to $\sigma_i$ on $S_i$ we have the intertwining property
\eqn\scftcob{ G_{\sigma_1}\circ U_{\Sigma,L}=U_{\Sigma,L}\circ
G_{\sigma_0}. }

\bigskip

\noindent {\it Example 2}\ \ \   We define {\it topological-spin$^c$
theories},
which model 2d theories with $\CN=2$ supersymmetry, by
replacing ``manifolds" with ``manifolds with spin$^c$-structure".

A {\it spin}$^c$-{\it structure} on a surface with a
conformal
structure is a pair of holomorphic line bundles $L_1,L_2$ with an
isomorphism $L_1\otimes L_2 \cong T\Sigma$ of holomorphic line
bundles. A spin structure is the particular case when $L_1 =L_2$. An
$\CN = 2$ superconformal theory assigns a vector space
$\CH_{S;L_1,L_2}$ to each 1-manifold $S$ with spin$^c$-structure,
and an operator
\eqn\scfop{
U_{S_0;L_1,L_2}:\CH_{S_0;L_1,L_2}\to\CH_{S_1;L_1,L_2}
}
to each spin$^c$-cobordism from $S_0$ to $S_1$.
To explain the rest
of the structure we need to define the $\CN =2$ Lie superalgebra
associated to a spin$^c$ 1-manifold $(S;L_1,L_2)$. Let $\CG = {\rm
Aut}(L_1)$ denote the group of bundle isomorphisms $L_1\to L_1$
which cover diffeomorphisms of $S$. (We can identify this group with
Aut$(L_2)$.) Its Lie algebra Lie$(\CG)$ is an extension of Vect$(S)$
by $\Omega ^0 (S)$. Let $\Lambda ^0_{S;L_1,L_2}$ denote the complex
Lie algebra obtained from Lie$(\CG)$ by complexifying Vect$(S)$.
This is the even part of a Lie superalgebra whose odd part is
$\Lambda^1_{S;L_1,L_2} = \Gamma (L_1)\oplus \Gamma(L_2)$. The
bracket $\Lambda^1 \otimes \Lambda^1 \to \Lambda^0$ is completely
determined by the property that elements of $\Gamma(L_1)$ and of
$\Gamma(L_2)$ anticommute among themselves, while the composite
\eqn\sla{
\Gamma(L_1)\otimes\Gamma(L_2)\to\Lambda^1\to{\rm Vect}_{\IC}(S)
}
takes $(\lambda_1,\lambda_2)$ to $\lambda_1\lambda_2\in\Gamma(TS).$

In an $\CN =2$ theory we require the superalgebra
$\Lambda(S;L_1,L_2)$ to act on the vector space $\CH_{S;L_1,L_2}$,
compatibly with the action of the group $\CG$, and with a similar
intertwining property with the cobordism operators to that of the
$\CN = 1$ case. For an $\CN=2$ theory the state space always has an
action of the circle group coming from its embedding in $\CG$ as the
group of fibrewise multiplications on $L_1$ and $L_2$. Equivalently,
the state space is always $\IZ$-graded.

An $\CN = 2$ theory always gives rise to
two ordinary conformal field theories by
equipping a surface $\Sigma$ with the spin$^c$
structures $(\IC,T\Sigma)$ and $(T\Sigma,\IC)$.
These are called the ``$A$-model" and the ``$B$-model"
associated to the $\CN =2$ theory. In each case the state
spaces are cochain complexes in which the differential is the
action of the constant section 1 of the trivial component of the spin$^c$-structure.

\bigskip

\noindent
{\it Cochain level theories}

\bigskip

The most important ``generalization," however, of the open and
closed topological field theory we have described is the one of
which it is intended to be  a toy model. In closed string theory the
central object is the vector space $\CC = \CC_{S^1}$ of states of a
single parametrized string. This has an integer grading by the
``ghost number", and an operator $Q:\CC \to \CC$ called the ``BRST
operator" which raises the ghost number by 1 and satisfies $Q^2=0$.
In other words, $\CC$ is a {\it cochain complex}. If we think of the
string as moving in a space-time $M$ then $\CC$ is roughly the space
of differential forms defined along the orbits of the action of the
reparametrization group Diff$^+(S^1)$ on the free loop space $\CL
M$. (More precisely, square-integrable forms of semi-infinite
degree.) Similarly, the space $\CC$ of a topologically-twisted $N=2$
supersymmetric theory, as just described, is a cochain complex which
models the space of semi-infinite differential forms on the loop
space of a K\"{a}hler manifold --- in this case, {\it all}
square-integrable differential forms, not just those along the
orbits of Diff$^+(S^1)$. In both kinds of example, a cobordism
$\Sigma$ from $p$ circles to $q$ circles gives an operator
$U_{\Sigma,\mu}:\CC^{\otimes p}\to \CC^{\otimes q}$ which depends on
a conformal structure $\mu$ on $\Sigma$. This operator is a cochain
map, but its crucial feature is that changing the conformal
structure $\mu$ on $\Sigma$ changes the operator $U_{\Sigma,\mu}$
only by a cochain-homotopy. The cohomology $H(\CC)={\rm ker}(Q)/{\rm
im}(Q)$ --- the ``space of physical states" in conventional string
theory --- is therefore the state space of a topological field
theory. (In the usual string theory situation the
topological field
theory we obtain is not very interesting, for the BRST cohomology is
concentrated in one or two degrees, and there is a ``grading
anomaly" which means that the operator associated to a cobordism
$\Sigma$ changes the degree by a multiple of the Euler number
$\chi(\Sigma)$. In the case of the $N=2$ supersymmetric models,
however, there is no grading anomaly, and the full structure is
visible.)


A good way to describe how the operator $U_{\Sigma,\mu}$ varies with
$\mu$ is as follows.

If $\CM_{\Sigma}$ is the moduli space of conformal structures on the
cobordism $\Sigma$, modulo diffeomorphisms of $\Sigma$ which are the
identity on the boundary circles, then we have a cochain map

\eqn\cochncomp{
U_{\Sigma}:\CC^{\otimes p}\to \Omega (\CM_{\Sigma};\CC^{\otimes q})
}
where the right-hand side is the de Rham complex of forms
on
$\CM_{\Sigma}$ with values in $\CC^{\otimes q}$. The operator
$U_{\Sigma,\mu}$ is obtained from $U_{\Sigma}$ by restricting from
$\CM_{\Sigma}$ to $\{\mu \}$. The composition property when two
cobordisms $\Sigma_1$ and $\Sigma_2$ are concatenated is that the
diagram

\eqn\cpn{
\matrix{\CC^{\otimes p} & \longrightarrow &
\Omega(\CM_{\Sigma_1};\CC^{\otimes q}) \cr
\downarrow & & \downarrow \cr
\Omega(\CM_{\Sigma_2 \circ \Sigma_1};\CC^{\otimes r}) &
 \longrightarrow & \Omega(\CM_{\Sigma_1}\times
 \CM_{\Sigma_2};\CC^{\otimes r})=\Omega(\CM_{\Sigma_1};
 \Omega(\CM_{\Sigma_2};\CC^{\otimes r})) \cr}
}
commutes, where the lower horizontal arrow in induced by
the map
$\CM_{\Sigma_1}\times \CM_{\Sigma_2}\to \CM_{\Sigma_2 \circ
\Sigma_1}$ which expresses concatenation of the conformal
structures.

Many variants of this formulation are possible. For example,
we might
prefer to give a cochain map

\eqn\chncomp{
U_{\Sigma}:C.(\CM_{\Sigma})\to(\CC^{\otimes p})^*\otimes \CC^{\otimes q},
}
where $C.(\CM_{\Sigma})$ is, say, the complex of
smooth singular
chains of $\CM_{\Sigma}$. We may also prefer to use the moduli
spaces of Riemannian structures instead of conformal structures.

There is no difficulty in passing from the closed-string picture
just presented to an open and closed theory. We shall not discuss
these cochain-level theories in any depth in this work, but it is
important to realize that they are the real objective. We shall now
point out a few basic things about them. A much fuller discussion
can be found in Costello \CostelloEI.

For each pair $a,b$ of boundary conditions we shall still have a
vector space --- indeed a cochain complex --- $\CO_{ab}$, but it is
no longer the space of morphisms from $b$ to $a$ in a category.
Rather, what we have is, in the terminology of Fukaya, Kontsevich,
and others, an {\it $A_{\infty}$-category}. This means that instead
of a composition law $\CO_{ab}\times\CO_{bc}\to\CO_{ac}$ we have a
{\it family} of ways of composing, parametrized by the contractible
space of conformal structures on the surface  of \FIGONE. In
particular, any two choices of a composition law from the family are
cochain-homotopic. Composition  is associative in the sense that we
have a contractible family of triple compositions
$\CO_{ab}\times\CO_{bc}\times\CO_{cd}\to\CO_{ad}$, which contains
all the maps obtained by choosing a binary composition law from the
given family and bracketing the triple in either of the two possible
ways.

\bigskip

\noindent{\it Note} \ This is not the usual way of defining an
$A_{\infty}$-structure. According to Stasheff's original definition,
an $A_{\infty}$-structure on a space $X$ consists of a sequence of
choices: first, a composition law $m_2: X\times X \to X$;\  then, a
choice of a map
$$m_3:[0,1] \times X\times X\times X \to X$$
which is a homotopy between $(x,y,z)\mapsto m_2(m_2(x,y),z)$
and
$(x,y,z) \mapsto m_2(x,m_2(y,z))$; then, a choice of a map
$$m_4:C_2 \times X^4 \to X, $$
where $C_2$ is a convex plane polygon whose vertices are
indexed by
the five ways of bracketing a 4-fold product, and $m_4\vert
((\partial C_2)\times X^4)$ is determined by $m_3$; and so on.

There is an analogous definition ---
in fact slightly simpler ---
applying to cochain complexes rather than spaces. These definitions,
however, are essentially equivalent to the one above coming from
2-dimensional field theory: the only important point is to have a
{\it contractible} family of $k$-fold compositions for each $k$. (A
discussion of the relation between the definitions can be found in
\catcoh.)

\bigskip

Apart from the composition law, the essential algebraic
properties we have found in our theories are the non-degenerate
 inner product, and the commutativity of the closed algebra $\CC$.
 Concerning the latter, when we pass to cochain theories the
  multiplication in $\CC$ will of course be commutative up
  to cochain homotopy, but, unlike what happened with the
  open-string composition, the moduli space $\CM_{\Sigma}$
  of closed-string multiplications,
i.e. the moduli space of conformal structures on a pair of pants
$\Sigma$, modulo diffeomorphisms of $\Sigma$ which are the identity
on the boundary circles, is not contractible: it contains a natural
{\it circle} of multiplications, and there are two {\it different}
natural homotopies between the multiplication and the reversed
multiplication. This might  be a clue to an important difference
between stringy and classical space-times. The closed string cochain
complex $\CC$ is the string-theory substitute for the de Rham
complex of space-time, an algebra whose multiplication is
associative and (graded-)commutative on the nose. Over the rationals
or the real or complex numbers, such cochain algebras are known by
the work of Sullivan \sullivan\ and Quillen \quillen\ to
model\foot{In this and the following sentence we are overlooking
subtleties related to the fundamental group.} the category of
topological spaces up to homotopy, in the sense that to each
such
algebra $\CC$ we can associate a space $X_{\CC}$ and a homomorphism
of cochain algebras from $\CC$ to the de Rham complex of $X_{\CC}$
which is a cochain homotopy equivalence. If we do not want to ignore
torsion in the homology of spaces we can no longer encode the
homotopy type in a strictly commutative cochain algebra. Instead, we
must replace commutative algebras with so-called
$E_{\infty}$-algebras, i.e., roughly, cochain complexes $\CC$ over
the integers equipped with
 a multiplication which is associative and commutative up to given
  arbitrarily high-order homotopies.
  An arbitrary space $X$ has an $E_{\infty}$-algebra $\CC_X$ of
  cochains, and conversely one can associate a space $X_{\CC}$
  to each $E_{\infty}$-algebra $\CC$. Thus we have a pair of adjoint functors,
  just as in rational homotopy theory.
  A long evolution in algebraic topology has culminated in
  recent theorems of Mandell \mandell\ which show that the actual homotopy
category of topological spaces is more or less equivalent to the
category of $E_{\infty}$-algebras. The cochain algebras of
closed
string theory have less higher
  commutativity than do $E_{\infty}$-algebras, and this may be an
  indication that we are dealing with non-commutative spaces
  in Connes's sense: that fits in well  with the interpretation
  of the $B$-field of a string background as corresponding to a
  bundle of matrix algebras on space-time. At the same time, the
   nondegenerate inner product on $\CC$ --- corresponding to
   Poincar\'{e} duality --- seems to show we are concerned
   with {\it manifolds}, rather than more singular spaces.


For readers not accustomed to working with cochain complexes it may
be worth saying a few words about what one gains by doing so. To
take the simplest example, let us consider the category $\CK$ of
cochain complexes of finitely generated {\it free} abelian groups
and cochain-homotopy classes of cochain maps. This is called the
{\it derived category} of the category of finitely generated abelian
groups. Passing to cohomology gives us a functor from $\CK$ to the
category of $\IZ$-graded finitely generated abelian groups. In fact
the subcategory $\CK_0$ of $\CK$ consisting of complexes whose
cohomology vanishes except in degree 0 is actually equivalent to the
category of finitely generated abelian groups.\foot{To an abelian
group $A$ one can associate the cochain complex
$$C_A=( \ \cdots \ \to 0 \to R_A \to F_A \to 0 \to \ \cdots \ ),$$
where $F_A$ is a free abelian group (in degree 0) with a surjective
map $F_A \to A$, and $R_A$ is the kernel of $F_A \to A$. The choice
of $F_A$ is far from unique, but nevertheless the different choices
of $C_A$ are {\it canonically } isomorphic objects of $\CK$.} But
the category $\CK$ inherits from the category of finitely generated
free abelian groups  a duality functor with properties as ideal as
one could wish: each object is isomorphic to its double dual, and
dualizing preserves exact sequences. (The dual $C^*$ of a complex
$C$ is defined by $(C^*)^i={\rm Hom}(C^{-i};\IZ)$.) There is no such
nice duality in the category of finitely generated abelian groups.
Indeed, the subcategory $\CK_0$ is not closed under duality, for the
dual of the complex $C_A$ corresponding to a group $A$ has in
general {\it two} non-vanishing cohomology groups: Hom$(A;\IZ)$ in
degree $0$, and in degree $+1$ the finite group Ext$(A;\IZ)$
Pontrjagin-dual to the torsion subgroup of $A$. This follows from
the exact sequence (not to be confused with the cochain complex):
\eqn\extseq{ 0 \rightarrow {\rm Hom}(A,\IZ) \rightarrow {\rm
Hom}(F_A,\IZ) \rightarrow {\rm Hom}(R_A,\IZ) \rightarrow {\rm
Ext}(A,\IZ) \rightarrow 0 }
%


The category $\CK$ also has a tensor product with
better properties
than the tensor  product of abelian groups (which does not preserve
exact sequences), and, better still, there is a canonical cochain
functor from (locally well-behaved) compact spaces to $\CK$ which
takes Cartesian products to tensor products. (The simplicial, \v
Cech, and other candidates for the cochain complex of a space are
{\it canonically} isomorphic in $\CK$.)

\bigskip

We shall return to this discussion in \S6.

\newsec{Solutions of the algebraic conditions: the semisimple case}

\subsec{Classification theorem}

 We now turn to the question : given
a closed string theory $\CC$, what is the corresponding category of
boundary conditions? In our formulation this becomes the question:
given a commutative Frobenius algebra $\CC$, what are the possible
$\CO_{ab}$'s?

We can answer this question in the case when $\CC$ is
semisimple. We will take $\CC$ to be an algebra over
the complex numbers, and in this case
 the most useful characterization
of semisimplicity is that the ``fusion rules''
\eqn\frls{
\phi_\mu \phi_\nu= N_{\mu\nu}^{~\lambda} \phi_{\lambda}
}
are diagonalizable.\foot{The structure constants $N_{\mu\nu}^{~\lambda} $
need not be integral, though in many interesting examples there is a
basis for the algebra in which they are integral.}
  That is, the matrices $L(\phi_{\mu} )$ of the left-regular
representation,  with
matrix elements $N_{\mu\nu}^{~\lambda} $, are simultaneously diagonalizable.

Equivalently, there is a set of basic idempotents $\varepsilon_x$ such that
\eqn\ssi{ \eqalign{ \CC & = \oplus_x  \IC \varepsilon_x \cr
\varepsilon_x \varepsilon_y & = \delta_{xy} \varepsilon_y \cr} }
Equivalently, yet again, $\CC$ is the algebra of complex-valued
functions on the finite set $X={\rm Spec}(\CC)$ of characters of
$\CC$.

The trace $\theta _\CC :\CC \to \IC$, which
should be thought of as a
``dilaton field" on the finite space-time ${\rm Spec}(\CC)$, is
completely described by the unordered set of non-zero complex
numbers

\eqn\normli{
\theta_x := \theta_\CC(\varepsilon_x)
}
which is the only invariant of a
finite dimensional commutative semisimple Frobenius algebra.

It should be mentioned that the most general finite dimensional
commutative algebra over the complex numbers is of the form $\CC =
\oplus \CC_x$, where $x$ runs through the set ${\rm Spec}(\CC)$, and
$\CC_x$ is a local ring, i.e. $\CC_x = \IC \varepsilon _x \oplus
m_x$, with $\varepsilon _x$ as in \ssi\ , and $m_x$ a nilpotent
ideal. If $\CC$ is a Frobenius algebra, then so is each $\CC_x$, and
there is some $\nu_x$ for which $\theta _{\CC}:m_x^{\nu_x}\to \IC$
is an isomorphism, while $m_x^{\nu_x+1}=0$. Let us write $\omega_x
\in m_x^{\nu_x}$ for the element such that $\theta_\CC
(\omega_x)=1$. The element $\omega$ of $\CC$ with components
$\omega_x$ can be regarded as a ``volume form" on space-time. (A
typical example of such a local Frobenius algebra $\CC_x$ is the
cohomology ring --- with complex coefficients --- of complex
projective space $\IP^n$ of dimension $n$. The cohomology ring is
generated by a single 2-dimensional class $t$ which satisfies
$t^{n+1}=0$. The trace is given by integration over $\IP^n$, and
takes $t^k$ to  1 if $k=n$, and to 0 otherwise. Thus $\omega_x =
t^n$ here.)

A useful technical fact about Frobenius algebras --- not necessarily
commutative --- is that, in the notation of \dblttw\ , the ``Euler"
element $\chi = \sum \psi_{\mu}\psi^\mu$ is invertible if and only
if the algebra is semisimple \foot{To see this, one observes that
for any element $\psi$ of the algebra we have $\theta (\psi \chi
)={\rm tr}(\psi)$, where ${\rm tr}(\psi)$ denotes the trace of
$\psi$ in the regular representation. As the pairing $(\psi_1 ,
\psi_2)\mapsto \theta (\psi_1 \psi_2)$ is nondegenerate, it follows
that the trace-form $(\psi_1 , \psi_2)\mapsto {\rm tr} (\psi_1
\psi_2)$ is nondegenerate if and only if $\chi$ is invertible, and
non-degeneracy of the trace-form is well-known to be a criterion for
a finite dimensional algebra to be semisimple. There are several
definitions of semisimplicity, and their equivalence amounts to the
classical theorem of Wedderburn. For our purposes, a semisimple
algebra is just a sum of full matrix algebras.}, which in the
general case means that the algebra is isomorphic to a sum of full
matrix algebras.  The element $\chi$ always belongs to the centre of
the algebra; in the commutative case it has components $\dim
(\CC_x)\omega_x$.

\bigskip

In the semisimple case we have the following complete
characterization of the possible open algebras $\CO_{aa}$ compatible
with a fixed closed algebra $\CC$. Unfortunately,
though, the
arguments we use do not work for graded Frobenius algebras.

\bigskip
\noindent
{\bf Theorem 2}: If $\CC$ is semisimple then $\CO =\CO_{aa}$ is semisimple
for each $a$
and necessarily of the form $\CO = {\rm End}_\CC(W)$ for some
finite dimensional representation $W$ of $\CC$.
\bigskip
{\it Proof}: The images   $\iota_a(\varepsilon_x) = P_x$ are central
simple idempotents. Therefore $\CO_x = P_x \CO= P_x \CO P_x $ is an
algebra over the  Frobenius algebra $\CC_x =\varepsilon_x\CC \cong
\IC$, and so it suffices to work over a single space-time point.
Then $\iota^a(1_{\CO_x})= \alpha 1_{\CC_x}$ for  some element
$\alpha \in \IC$. By the Cardy condition
\eqn\ccdon{ \alpha 1_{\CO_x} = \chi_{\CO_x} = \sum \psi_\mu \psi^\mu
}
Applying $\theta$ we find $\alpha = \dim \CO_x$, and hence
 $\chi_{\CO_x} $ is invertible if $\CO_x \neq 0$. It follows that $\CO_x$ is
semisimple at each point $x$,  i.e. a sum of matrix algebras
$\oplus_i {\rm End}(W_i)$. In fact, the Cardy condition shows that
there can be at most one summand $W_i$ at each point, i.e. the
algebra is simple. For the map $\pi:\CO_x \to \CO_x$ must take
each
summand End$(W_i)$ into itself, and cannot factor through the
1-dimensional $\CC_x$ if more than one $W_i$ is non-zero.
$\spadesuit$

\bigskip

According to Theorem 2 the most general $\CO_{aa}$ is obtained by
choosing a vector space $W_{x,a}$ for each basic idempotent
$\varepsilon_x$, i.e. a {\it vector bundle} on the finite space-time
$X={\rm Spec}(\CC)$, and forming:
\eqn\ssii{ \eqalign{ \CO_{aa} & = \oplus_x  {\rm End}(W_{x,a}). \cr}
} But let us notice that when we have an algebra of the
form End$(W)$
the vector space $W$ is determined by the algebra only up to
tensoring with an arbitrary complex line: any irreducible
representation of the algebra will do for $W$.

Elements $\psi\in \CO_{aa}$ will be denoted $\psi = \oplus \psi_x$.
Let $P_x$ be the projection operator onto the  $x^{th}$ summand.
From the equation
\eqn\project{ \iota_a(\varepsilon_x)  = P_x }
the adjoint relation and the Cardy condition determine the
relations: \eqn\fixes{ \eqalign{ \theta_a(\psi)  & = \sum_x
\sqrt{\theta_x} {\rm Tr}(\psi_x) \cr \iota^a(\psi) & = \oplus_x {\rm
Tr}(\psi_x) {\varepsilon_x \over \sqrt{\theta_x}}\cr
\pi_{b}^{~~a}(\psi_{aa}) & = \oplus_x {1\over \sqrt{\theta_x}} {\rm
Tr}_{W_{x,a}}(\psi_{x,aa})P_{x,b}\cr} }
(one must use  the same square-root in the formula for  $\theta_\CO$
and $\iota^a$.) Note that $\theta_\CC({\varepsilon_x \over
\sqrt{\theta_x}} {\varepsilon_y \over \sqrt{\theta_y}}) =
\delta_{x,y}$, i.e. the elements ${\varepsilon_x \over
\sqrt{\theta_x}}$ form a natural orthonormal basis for $\CC$. Thus,
a boundary condition $a$ gives us a tuple of {\it positive integers}
$w_x = \dim W_x$, one for each basic idempotent, as well as a choice
of the square-root $\sqrt{\theta_x}$. The relation \cyclic, however,
shows that these square-roots are an intrinsic property of the
Frobenius category $\CB$, and do not depend on which particular
object in it we are considering.

Let us now determine the
 $\CO_{aa}\times \CO_{bb}$ bimodules $\CO_{ab}$ associated
to a pair of boundary conditions $a,b$.
These are again fixed by the Cardy condition.

\bigskip
\noindent
{\bf Lemma}: When $\CC$ is semisimple we have
\eqn\bimod{ \CO_{ab} \cong \oplus_x {\rm Hom}(W_{x,a};W_{x,b}) }

\noindent {\it Proof}: Restricting to each $\CO_{aa}$ we can invoke
Theorem 2.  Then the $\iota_a(\varepsilon_x)\CO_{ab} =
\CO_{ab}\iota_{b}(\varepsilon_x)$ are bimodules for the simple
algebras $\CO_{x,aa}$ and $\CO_{x,bb}$. We restrict to a single
idempotent and drop the $x$, that is, we take $\CC=\IC$.   The only
irreducible representation of $\CO_{aa}={\rm End}(W_{a})$ is $W_{a}$
itself, and the only $\CO_{aa}\times\CO_{bb}$-bimodule is
$W_a^*\otimes W_b$. Therefore, $\CO_{ab} \cong n_{ab} W_a^* \otimes
W_b$, where $n_{ab}$ is a nonnegative integer.  Let us work out the
Cardy condition. If $v_m$ is a basis for $W_a$ and $w_n$ is a basis
for $W_b$ then a basis for $\CO_{ab}$ is $v_{m,\alpha}^*\otimes
w_{n,\alpha}$ where $\alpha=1, \dots, n_{ab}$. Then $\pi(\psi) =
n_{ab} {\rm tr}_{W_a}(\psi) P_b$. Comparing to
$\iota_b\iota^a(\psi)$ we get $n_{ab}=1$.  $\spadesuit$

\bigskip

We can now describe the maximal category $\CB$ of boundary
conditions. We first observe that if $p\in \CO_{aa}$ is a projection
---i.e.
$p^2=p$ ---we can assume that $a = b\oplus c$ in $\CB$, where
$b$ is the image of $p$. For we can adjoin images of projections to
any additive category in much the same way as we adjoined direct
sums. If the closed algebra $\CC$ is semisimple we can therefore
choose an object $a_x$ of $\CB$ for each space-time point $x$ so
that $a_x$ is supported at $x$ --- i.e.
$\iota_{a_x}(\varepsilon_x)\CO_{a_xa_x} = \CO_{a_xa_x}$ --- and is
simple, i.e. $\CO_{a_x a_x}=\IC$. For any object $b$ of $\CB$ we
then have a canonical morphism
\eqn\canondecomp{
\oplus _x \CO_{ba_x} \otimes a_x \to b,
}
where on the left we have used the possibility of tensoring
any
object of a linear category by a finite dimensional vector space.
Furthermore, it follows from the lemma that the morphism
\canondecomp \ is an isomorphism, for both sides have the same space
of morphisms into any other object $c$. Finally, notice that $a_x$
is unique up to tensoring with a line $L_x$, for if $a_x '$ is
another choice then $a_x' \cong a_x \otimes L_x$, where $L_x =
\CO_{a_xa_x'}$.

\bigskip
\noindent {\bf Theorem 3}

\item{(i)} If $\CC$ is semisimple, corresponding
to a space-time $X$, then the category $\CB$ of boundary conditions
is equivalent to the category Vect$(X)$ of vector bundles on $X$, by
the inverse functors
\eqn\funct{
\{W_x\} \mapsto \oplus W_x \otimes a_x,
}

\eqn\functinv{
a \mapsto \{ \CO_{a_xa}\}.
}
\item{(ii)}  The equivalence of $\CB$ with Vect$(X)$ is
unique up to transformations Vect$(X) \to {\rm Vect}(X)$ given
by
tensoring with a line bundle $L=\{L_x\}$ on $X$.
\item{(iii)}  The Frobenius structure on $\CB$ is
determined by choosing a square-root $\{\sqrt\theta_x\}$ of
the
dilaton field. It is therefore unique up to multiplication by an
element $\sigma \in \CC$ such that $\sigma ^2 =1$.

\bigskip
{\bf Remarks}

\item{1.} A boundary condition $a$ has a {\it support}
\eqn\supportbc{ {\rm supp}(a) = \{x\in X:W_x \neq 0 \} }
 contained
in $X= {\rm spec}(\CC)$. If two boundary conditions $a$ and $b$ have
the same support then $\CO_{ab}$ is a Morita equivalence bimodule
between $\CO_{aa}$ and $\CO_{bb}$. The reader might wish to compare
this discussion to section 6.4 of \seibergwitten. Note that it is
necessary to invoke the Cardy condition to draw this conclusion.

\item{2.}   Examples of semisimple Frobenius algebras in physics include:

a) The fusion rule algebra (Verlinde algebra) of a RCFT.

b) The chiral ring of an $\CN=2$ Landau-Ginzburg theory for generic
superpotential $W$ (that is, as long as all the critical points of
$W$ are Morse critical points). This is the case when the IR theory
is massive.

c) Generic quantum cohomology of manifolds.

\bigskip

\subsec{Comment on $B$-fields}

We can see from this discussion just where the idea of a $B$-field
would appear, though in fact on a 0-dimensional space-time any
$B$-field must be trivial. We showed that there is a category of
boundary conditions associated to each point of space time, and that
it is isomorphic to the category of finite dimensional vector
spaces, though not canonically. More precisely, it contains minimal
--- i.e. irreducible --- objects from which any other object can be
obtained by tensoring with a finite dimensional vector space.

Now a B-field is in essence a bundle of {\it categories} on
space-time in which the fibre-categories are all isomorphic but not
canonically. We can suppose that each fibre is isomorphic to the
category of finite dimensional vector spaces. The crucial feature is
that the ambiguity in identifying each fibre with the standard fibre
is a ``group" --- in this case actually a category --- of
equivalences whose elements are complex lines and in which
composition is given by the tensor product. Our category of boundary
conditions is precisely the category of ``sections" of a bundle of
categories with this structural group.

It may be helpful to think of this in the
following way. An
electromagnetic field is a line bundle with connection on
space-time. It is something we can think of as part of the structure
of space-time, and makes sense in the absence of fermions. But in a
theory with fermions there is a spinor-space at each point of
space-time, and the electromagnetic field is ``really" the
information about how the spinor spaces are connected together from
point to point of space-time. In this sense the electromagmetic
field ``is" the spinor-bundle with its connection. A B-field
similarly ``is"  the bundle of boundary conditions.

 On a general topological space  $X$ the classes
 of $B$-fields are classified by the elements of
 the cohomology group $H^3(X;\IZ)$, which can be
 understood as $H^1(X;\CG)$, where $\CG$ is the ``group" of line
 bundles under tensor product, which in algebraic topology
 is an Eilenberg-Maclane object of type $K(\IZ,2)$.
 We shall return to this topic in \S7.

\bigskip

\subsec{ Reconstructing the closed algebra}

\bigskip

When we have an open and closed TFT each element $\xi$ of the closed
algebra $\CC$ defines an endomorphism $\xi_a = i_a(\xi)\in\CO_{aa}$
of each object $a$ of $\CB$, and $\eta \circ \xi_a = \xi_b \circ
\eta$ for each morphism $\eta \in \CO_{ba}$ from $a$ to $b$. The
family $\{\xi_a\}$ thus constitutes a natural transformation from
the identity functor $1_\CB : \CB \to \CB$ to itself.

For any $\IC$-linear category $\CB$ we can
consider the ring $\CE$ of
natural transformations of $1_\CB$. It is automatically commutative,
for if $\{\xi_a\},\{\eta_a\}\in \CE$ then $\xi_a\circ\eta_a
=\eta_a\circ\xi_a$ by the definition of naturality. If $\CB$ is a
Frobenius category then there is a map
$\pi_a^{~~b}:\CO_{bb}\to\CO_{aa}$ for each pair of objects $a,b$,
and we can define $j^b:\CO _{bb}\to\CE$ by
$j^b(\eta)_a=\pi_a^{~~b}(\eta)$ for $\eta \in \CO_{bb}$. In other
words, $j^b$ is defined so that the Cardy condition $\iota_a\circ
j^b =\pi_a^{~~b}$ holds. But the question arises whether we can
define a trace $\theta :\CE\to\IC$ to make $\CE$ into a Frobenius
algebra, and with the property that
\eqn\newadjoint{
\theta_a(\iota_a(\xi ).\eta)=\theta(\xi.j^a(\eta))
}
for all $\xi\in\CE$ and $\eta\in\CO_{aa}$. This is certainly true if
$\CB$ is a semisimple Frobenius category with finitely many simple
objects, for then $\CE$ is just the ring of complex-valued functions
on the set of classes of these simple elements, and we can readily
  define $\theta: \CE \to \IC$ by $\theta(\varepsilon_a) =
\theta_a(1_a)^2$, where $a$ is an irreducible object, and
$\varepsilon_a \in \CE$ is the characteristic function of the point
$a$ in the spectrum of $\CE$.  Nevertheless, a Frobenius category
need not be semisimple, and we cannot, unfortunately, take $\CE$ as
the closed string algebra in the general case. If, for example,
$\CB$ has just one object $a$, and $\CO_{aa}$ is a commutative local
ring of dimension greater than 1, then $\CE =\CO_{aa}$, and so
$\iota_a :\CE \to \CO_{aa}$ is an isomorphism, and its adjoint map
$j^a$ ought to be an isomorphism too. But that contradicts the Cardy
condition, as $\pi_a^{~~a}$ is multiplication by $\sum \psi_i
\psi^i$, which must be nilpotent. In \S6 we shall give an example of
two {\it distinct} closed string Frobenius algebras which admit the
same open string algebra $\CO_{aa}$.

The commutative algebra $\CE$ of natural endomorphisms of the
identity functor of a linear category $\CB$ is called the {\it
Hochschild cohomology} $HH^0(\CB)$ of $\CB$ in degree 0. The groups
$HH^p(\CB)$ for $p> 0$, whose definition will be given in a moment,
vanish if $\CB$ is semisimple, but in the general case they appear
to be relevant to the construction of a closed string algebra from
$\CB$. Let us notice meanwhile that for any Frobenius category $\CB$
there is a natural homomorphism $K(\CB)\to HH^0(\CB)$ from the
Grothendieck group\foot{I.e. the group formed from the semigroup of
isomorphism classes of objects of $\CB$ under $\oplus$.} of $\CB$,
which assigns to an object $a$ the transformation whose value on $b$
is $\pi_b^{~~a}(1_a)\in \CO_{bb}$. In the semisimple case this
homomorphism induces an isomorphism $K(\CB)\otimes \IC \to
HH^0(\CB)$.

For any additive category $\CB$ the Hochschild cohomology is defined
as the cohomology of the cochain complex in which a $k$-cochain $F$
is a rule that to each composable $k$-tuple of morphisms
\eqn\prehochsch{
Y_0\buildrel\phi_1\over\rightarrow Y_1\buildrel\phi_2
\over \to \cdots \buildrel\phi_k \over \to Y_k
}
assigns
$F(\phi_1,\ldots ,\phi_k) \in {\rm Hom}(Y_0;Y_k).$ The differential
in the complex is defined by
\eqn\hochdiff{ \eqalign{ (dF)(\phi_1,\ldots ,\phi_{k+1}) =&
F(\phi_2,\ldots,\phi_{k+1}) \circ\phi_1 + \cr    & +
\sum_{i=1}^{k}(-1)^iF(\phi_1,\ldots,\phi_{i+1}\circ \phi_i ,\ldots
,\phi_{k+1})\cr & +(-1)^{k+1}\phi_{k+1}\circ F(\phi_1,\ldots ,\phi_k
).\cr}}
 (Notice, in particular, that a 0-cochain assigns an
endomorphism $F_Y$ to each object $Y$, and is a cocycle if the
endomorphisms form a natural transformation. Similarly, a 2-cochain
$F$ gives a possible infinitesimal deformation $F(\phi_1,\phi_2)$ of
the composition-law $(\phi_1,\phi_2)\mapsto \phi_2 \circ \phi_1$ of
the category, and the deformation preserves the associativity of
composition if and only if $F$ is a cocycle.)

In the case of a category $\CB$ with a single object whose algebra
of endomorphisms is $\CO$ the cohomology just described is usually
called the Hochschild cohomology of the algebra $\CO$ with
coefficients in $\CO$ regarded as a $\CO$-bimodule. This must be
carefully distinguished from the Hochschild cohomology with
coefficients in the dual $\CO$-bimodule $\CO^*$. But if $\CO$ is a
Frobenius algebra it is isomorphic as a bimodule to  $\CO^*$, and
the two notions of Hochschild cohomology need not be distinguished.
The same applies to a Frobenius category $\CB$: because
Hom$(Y_k;Y_0)$ is the dual space of Hom$(Y_0;Y_k)$ we can think of a
$k$-cochain as a rule which associates to each composable $k$-tuple
\prehochsch  \ of morphisms a linear function of an element $\phi_0
\in {\rm Hom}(Y_k;Y_0)$. In other words,  a $k$-cochain is a rule
which to each ``circle" of $k+1$ morphisms
\eqn\hochsch{
\cdots \buildrel\phi_0\over \to Y_0 \buildrel\phi_1\over
\to Y_1 \buildrel\phi_2\over\to \cdots \buildrel\phi_k\over\to Y_k
\buildrel\phi_0\over\to \cdots
}
assigns a complex number $F(\phi_0,\phi_1,\ldots,\phi_k)$.

\ifig\FIGTWELVESUBFIVE{A cyclic pairing of a closed string state
$\phi$ with $k+1$ open string states.  }
{\epsfxsize2.0in\epsfbox{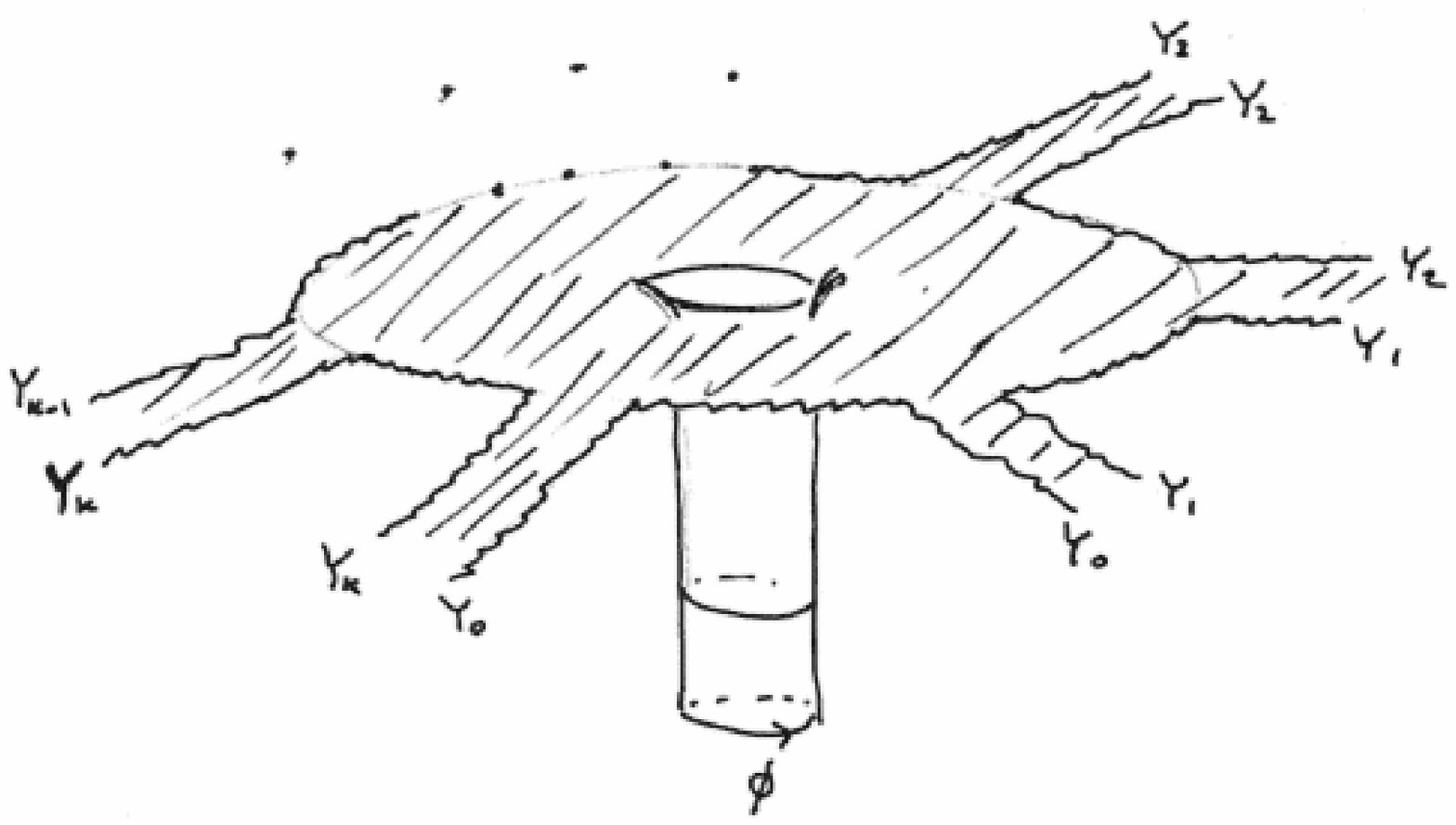}}

If in this description we restrict ourselves to cochains which
are
cyclically invariant under rotating the circle of morphisms
$(\phi_0,\phi_1,\ldots,\phi_k)$ then we obtain a sub-cochain-complex
of the Hochschild complex whose cohomology is called the {\it cyclic
cohomology} $HC^*(\CB)$ of the category $\CB$. The cyclic cohomology
--- which evidently maps to the Hochschild cohomology --- is a more
natural candidate for the closed string algebra associated to $\CB$
than is the Hochschild cohomology (for a state represented by the
vector \hochsch\ pairs in a cyclically invariant way with a closed
string state to give a number, in virtue of \FIGTWELVESUBFIVE). In
our baby examples the cyclic and Hochschild cohomology are
indistinguishable, but it is worth pointing out\foot{As we learnt
from Kontsevich} that while $HH^2(\CB)$ is, as indicated above, the
space of infinitesimal deformations of $\CB$ as a category, the
group $HC^2(\CB)$ is its space of infinitesimal deformations as a
Frobenius category.


A very natural Frobenius category on which to test these
ideas is the
category of holomorphic vector bundles on a compact Calabi-Yau
manifold: that example will be discussed in \S 6.

\subsec{ Spin theories and mod 2 graded categories}

Let us give a brief outline, without proofs, of the modifications of
the preceding discussion which are needed to describe the category
of boundary conditions for a topological-spin theory as defined in
\S2.6.

There is just one spin structure on an interval, and its
automorphism group is $(\pm 1)$, so for each pair of boundary
conditions $a,b$ the vector space $\CO_{ab}$ will have an
involution, i.e. a mod 2 grading. The bilinear composition $\CO_{ab}
\otimes \CO_{bc} \to \CO_{ac}$ will preserve the grading. There is a
non-degenerate trace $\theta_a:\CO_{aa} \to \IC$ which satisfies the
commutativity condition \cyclic\  (without signs).

If the closed theory is described by a Frobenius algebra $\CC =
\CC_{ns} \oplus \CC_r$, as in \S2.6, there will be adjoint maps
\eqn\adhom{ \eqalign{ \iota_a^{ns}:\CC_{ns} & \to \CO_{aa}\cr
\iota^a_{ns}:\CO_{aa} & \to \CC_{ns}\cr
 \iota_a^{r}:\CC_{r} & \to \CO_{aa}\cr
 \iota^a_{r}:\CC_{aa} & \to \CC_r \cr}
 }
which preserve the grading. Moreover and $\iota^{ns}_a$ and
$\iota^r_a$ fit together to define a homomorphism of algebras $\CC
\to \CO_{aa}$. The centrality condition becomes
\eqn\spincentral{ \eqalign{ \iota_a^{ns}(\phi)\psi & = \psi
\iota_a^{ns}(\phi)\cr \iota_a^{r}(\phi)\psi & =(-1)^{\deg\phi
\deg\psi + \deg\psi}  \psi \iota_a^{r}(\phi)\cr} }
Thus, $\iota^{ns}$ maps into the naive center of the algebra
$\CO_{aa}$. The reason we get the naive centre here, rather than the
graded-algebra centre, and also the reason that the trace is naively
commutative, is the same as that  given in \S2.6 for the naive
commutativity of the algebra $\CC$. The sign for $\iota^{r}$ is
obtained by carefully following the choices of sections of the spin
bundle one chooses under the diffeomorphism in figure 8.

There are two Cardy conditions
\eqn\spincardy{ \eqalign{ \iota^{ns}_a \iota^b_{ns}(\psi) & =
\pi_b^a(\psi) := \sum (-1)^{\deg\psi_\mu\deg \psi} \psi_\mu \psi
\psi^\mu \cr \iota^{r}_a \iota^b_{r}(\psi) & = \tilde \pi_b^a(\psi)
:= \sum (-1)^{\deg\psi_\mu(\deg \psi+1)} \psi_\mu \psi \psi^\mu.
\cr} }

\bigskip

If we assume the closed algebra is semisimple then, just as before,
we can assume that $\CC_{ns}$ is the algebra of functions on  a
finite set $X$, and we can determine the category of boundary
conditions point-by-point. In other words, we can assume that $\CC =
\IC[\eta]$, where the generator $\eta$ of $\CC_r$ satisfies
$\eta^2=1$, but may have either even or odd degree. In either case,
the argument we have already used shows (by means of the first Cardy
formula) that the algebra $\CO_{aa} $ is the full matrix algebra of
a  vector space $W$. If the degree of $\eta$ is even then
$\iota^r(\eta)=P$ with $P$ even, $P^2=1$, and $P\psi P =
(-1)^{\deg\psi} \psi$.  In this case the category of boundary
conditions at the point is equivalent to the category of mod $2$
graded vector spaces. If, on the other hand, the degree of $\eta$ is
odd, then $P$ is odd, $P^2=1$ and $P$ is (naive) central.   The
involution of the algebra $\CO_{aa}$ corresponds to an involution of
the module $W$, and the action of $P$ is an isomorphism between the
two halves of the grading. The even subalgebra of $\CO_{aa}$ is a
full matrix algebra.  Thus the category of boundary conditions is
the equivalent to the category of graded representations of the
superalgebra $\IC[\eta]$, which in turn is equivalent simply to the
category of ungraded vector spaces. The Frobenius structure of the
open algebra determines that of the closed algebra by taking the
square, as in the ungraded case. The two cases $\deg \eta=0$ and
$\deg \eta=1$ are roughly analogous to the distinction between the
even and odd degree Clifford algebras over the complex numbers.

\bigskip

Suppose, conversely, that we have an arbitrary semisimple mod $2$
graded category $\CB$, i.e. a linear category equipped with an
involutory functor $S$ which one thinks of as the flip of the
grading. Such a category has two kinds of simple object $P$: those
such that $S(P) \cong P$, and those for which this is not true. The
first kind of object generates a subcategory of $\CB$ isomorphic to
the category of vector spaces, and the second kind generates a
subcategory isomorphic to the category of graded vector spaces. Thus
any semisimple graded category $\CB$ is the category of boundary
conditions for a unique topological-spin theory.

\newsec{Vector bundles, K-theory, and ``boundary states''}

In the semisimple case there is a nice
geometrical interpretation of the category $\CB$ of boundary conditions: the possible objects correspond to the vector bundles over the ``space-time" $X={\rm Spec}(\CC)$
associated to $\CC$, which is just a finite set of points. The fiber above a point $x$ is just the vector
space $W_x$.
\bigskip

\ifig\FIGTHIRTEEN{Correlations on the upper half plane with boundary
condition $a$ are the same as the closed string amplitude for an
insertion of a boundary state $B_a$. }
{\epsfxsize2.0in\epsfbox{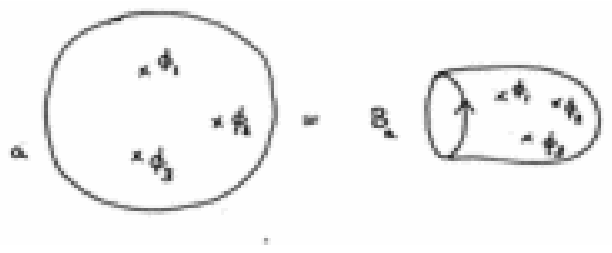}}

Let us now make some comments on ``boundary states''. In the
conformal field theory literature one associates to a boundary
condition $a$  a corresponding ``state'' $B_a$ in the closed string
state space. (Strictly, $B_a$ is an element of the algebraic dual.)
Translated to the present context $B_a \in \CC$. The defining
property of the boundary state is that the correlation functions of
operators on a disk with the boundary condition $a$ are equal to the
correlation functions of the closed theory on the sphere obtained by
capping off the disk with another disk and inserting the state $B_a$
at the centre of the cap. This is illustrated in \FIGTHIRTEEN.

In equations,
\eqn\boundryst{
\theta_{a}(\iota_a\bigl(\phi_1) \cdots \iota_a(\phi_n) \bigr)
= \theta_{\CC}\bigl(B_a \phi_1 \cdots \phi_n \bigr)
}
for all $\phi_1, \dots, \phi_n$.
Using the adjoint relation and
 nondegeneracy of the trace we find that
\eqn\bpid{
B_a = \iota^a(1_{\CO_{aa}})
}

The map  $a\mapsto B_a$ is a natural homomorphism
\eqn\char{
K(\CB)\to \CC.
}
%


The operator which adds $g$ handles and $h=\sum h_a$ holes, where
$h_a$ of the holes have the boundary condition $a$, is just $\chi^g
\prod(B_a)^{h_a}$, where $\chi$ is the Euler element of $\CC$.

Let us record one simple property of these boundary states.
First, using the Cardy condition we have
\eqn\cardydim{
\eqalign{
\theta_\CC(B_a B_b)
& = \theta_a(\iota_a(B_b) ) \cr
& =\theta_a(\iota_{a} \iota^b(1_b)) \cr
& = \theta_a(\pi_a^{~~b}(1_b) ) \cr
& = \dim \CO_{ab}\cr}
}

In the semisimple case we can give an
explicit formula for the
the ``boundary state'' in terms of the basic idempotents:
\eqn\bsi{
B_\CO = \out(1_\CO) = \sum_x (\dim W_x) {\varepsilon_x \over \sqrt{\theta_x}}
}

The formula shows that the boundary states form a positive cone in
the unimodular lattice $\CL_B$ spanned by the orthonormal basis
${\varepsilon_x \over \sqrt{\theta_x}}$ in the closed algebra $\CC$.
In particular it follows from \bsi\ that boundary states can only be
added with positive integral coefficients. They are therefore {\it
not} like quantum mechanical states of branes. The fundamental
integral structure is a result of the Cardy condition.

\bigskip

It is natural to speculate whether there
should be an operation of
``multiplication" of boundary conditions. There are arguments both
for and against. The original perspective on D-branes, according to
which they are viewed as ``cycles" in space-time on which open
strings can begin and end, suggests that there should be a
multiplication, corresponding to the intersection of cycles. As no
multiplication seems to emerge from the toy structure we have
developed in this paper one may wonder whether an important
ingredient has been omitted. Against this there are the following
considerations. Our boundary conditions seem to correspond more
closely to vector bundles --- i.e. to $K$-theory classes --- on
space-time than to homology cycles: that will be plainer when we
consider the equivariant situation in \S 7. Now the $K$-theory
classes of a ring have a product, coming from the tensor product of
modules, only when the ring is commutative; and we have already
remarked that the B-fields which are part of the closed string model
of space-time seem to encode a degree of noncommutativity. More
precisely,  D-branes seem to define classes in the {\it twisted}
$K$-theory of space-time, twisted by the B-field, and the twisted
$K$-theory of a space does not form a ring: the product of two
twisted classes is a twisted class corresponding to the sum of the
twistings of the factors. But in string theory there is no concept
of ``turning off" the B-field to find an underlying untwisted
space-time. For example, the conformal field theory corresponding to
a torus with a non-zero B-field can be isomorphic by ``T-duality" to
a theory coming from another torus with no B-field.

Another reason for not expecting a multiplication operation on
D-branes also comes from T-duality in conformal field theory. There
the closed string theories defined by a Riemannian torus $T$ and its
dual $T^*$ are isomorphic, and we do indeed have a $K$-theory
isomorphism $K(T)\cong K(T^*)$, but it is not compatible with the
multiplication in $K$-theory. Furthermore, in some examples of TFTs
coming from $N = 2$ supersymmetric sigma models the category of
boundary conditions does seem to be a tensor category.

The formula \bsi\ for the boundary state shows that the lattice
$\CL_B$, which is picked out inside $\CC$ by the dilaton field
$\theta$, is not closed under multiplication in $\CC$ unless
$\theta_x =1$ for all points $x$; but the lattices corresponding to
different dilaton fields multiply into each other just as happens
with twisted $K$-classes. Nevertheless, in the semisimple case, if
we define an element $S:=\sum_x \sqrt{\theta_x} \varepsilon_x$, then
the operation
\eqn\possibmult{ (B_1,B_2) \mapsto SB_1B_2 }
 does define a
multiplication on boundary states, though its significance is
unclear.

\subsec{``Cardy states'' vs. ``Ishibashi states''}

The formula \bsi\ for the boundary state is reminiscent of what  is
what is known as  a ``Cardy state'' in the construction of conformal
field theories with boundary. That leaves the question: what are the
``Ishibashi'' or ``character states''?  This question can be nicely
addressed in the topological framework of this exposition.   In the
basis of primary fields in closed (or chiral) conformal field theory
{\it the fusion rules are in general not diagonal}. Thus the usual
basis $\phi_\mu$, $\mu = 1, \dots, N$
 for $\CC$   is different from the
basis $\varepsilon_i$, $i=1,\dots, N$ used above. The fusion rules are diagonal
in the basis $\varepsilon_i$.

The analogy with CFT is the following. In boundary conformal
field theory one associates an ``Ishibashi'' or character
state to every primary field of the chiral algebra. Formally
these states are solutions to $(T-\bar T) \vert B \rangle =0$,
or the generalization of this to the case of other chiral
algebras. They are best thought of as intertwiners between
left and right chiral representations. In the present context
there is no chiral algebra, and we should think of {\it every}
element of $\CC$ as a solution to $(T-\bar T) =0$, and its
generalizations.

The basis of   ``Ishibashi states'' associated to definite representations
of the chiral algebra is naturally associated to  the basis
$\phi_\mu$ analogous to primary fields. In this basis the algebra is
given by
\eqn\fusionrules{
\phi_\mu \phi_\nu = N_{\mu\nu}^{~~\lambda } \phi_\lambda
}
with positive integral $N_{\mu\nu}^{~~\lambda}$.
Using these formulae we recover, essentially, Cardy's formula for Cardy boundary
states in terms of character boundary states. Note that there is no need
to use any relation to the modular group.

We close with one further brief remark. It is nice to see the
standard relation that the closed string coupling is  the square of
the open string coupling in the present context. If we scale
$\theta_\CC \rightarrow \lambda^{-2} \theta_\CC$ then $\chi_\CC =
\sum_\mu \phi_\mu \phi^\mu \rightarrow \lambda^{+2} \chi_\CC$. We
may therefore interpret $\lambda^2$ as the closed string couplling.
On the other hand, the squareroot of $\theta_i$ in $B_\CO$ shows
that $B_\CO \rightarrow \lambda B_\CO$, and therefore $\lambda$ is
the open string coupling. Indeed,
 the partition function for a surface with $g$ handles and $h$ holes
is $Z(\Sigma) = \theta_{\CC}((\chi_\CC)^g (B_\CO)^h)$, and therefore
scales as $Z(\Sigma)  \rightarrow \lambda^{-\chi(\Sigma)}
Z(\Sigma)$, as expected, where $\chi(\Sigma)=2-2g-h$ is the Euler
number of $\Sigma$..

\newsec{Landau-Ginzburg theories}

\lref\KapustinBI{
  A.~Kapustin and Y.~Li,
  ``D-branes in Landau-Ginzburg models and algebraic geometry,''
  JHEP {\bf 0312}, 005 (2003)
  [arXiv:hep-th/0210296].
}
\lref\DOrlov{D. Orlov, ``Triangulated categories of singularities and
D-branes in Landau-Ginzburg models,'' math.AG/0302304}
%
\lref\BrunnerDC{
  I.~Brunner, M.~Herbst, W.~Lerche and B.~Scheuner,
  ``Landau-Ginzburg realization of open string TFT,''
  arXiv:hep-th/0305133.
}
\lref\KapustinGA{
  A.~Kapustin and Y.~Li,
  ``Topological correlators in Landau-Ginzburg models with boundaries,''
  Adv.\ Theor.\ Math.\ Phys.\  {\bf 7}, 727 (2004)
  [arXiv:hep-th/0305136].
}
\lref\LazaroiuZI{
  C.~I.~Lazaroiu,
  ``On the boundary coupling of topological Landau-Ginzburg models,''
  JHEP {\bf 0505}, 037 (2005)
  [arXiv:hep-th/0312286].
}
\lref\HerbstAX{
  M.~Herbst and C.~I.~Lazaroiu,
  ``Localization and traces in open-closed topological Landau-Ginzburg
  models,''
  JHEP {\bf 0505}, 044 (2005)
  [arXiv:hep-th/0404184].
}
\lref\KapustinDF{
  A.~Kapustin and L.~Rozansky,
  ``On the relation between open and closed topological strings,''
  Commun.\ Math.\ Phys.\  {\bf 252}, 393 (2004)
  [arXiv:hep-th/0405232].
}
%
\lref\DijkgraafQH{
  R.~Dijkgraaf,
  ``Intersection theory, integrable hierarchies and topological field
theory,''
  arXiv:hep-th/9201003.
}
%
\lref\SeibergAT{
  N.~Seiberg and D.~Shih,
  ``Minimal string theory,''
  Comptes Rendus Physique {\bf 6}, 165 (2005)
  [arXiv:hep-th/0409306].
}

D-branes can be defined in general two-dimensional
 N=2 Landau-Ginzburg theories \HIV.  Such theories can be topologically
twisted, producing topological Landau-Ginzburg theories.
It is interesting to compare with the D-branes obtained
from our results applied to the resulting closed topological
theory. Here we confine ourselves to a few very elementary
remarks. In the past few years, following an initial suggestion by
Kontsevich,  an elaborate theory of
categories of topological Landau-Ginzburg branes has been
developed. We refer to  \refs{\KapustinBI,\DOrlov,\BrunnerDC,\KapustinGA,
\LazaroiuZI,\HerbstAX,\KapustinDF}  for details. These categories
are thought to capture more physical information about the D-branes.
In the case when all the critical points of the superpotential are Morse
there is a functor to the category of branes we construct.

Let us recall the definition of a topological LG theory.
One begins with a superpotential $W(X_i)$ which is a
holomorphic function of chiral superfields $X_i$.
When $W$ is a polynomial the Frobenius algebra is simply
the Jacobian ideal
\eqn\jacideal{
\CC = \IC[X_i]/(\p_i W)
}
The Frobenius structure is defined by a residue formula.
For example, in  the one-variable case we define
\eqn\lgtrace{
\theta(\phi) := {\rm Res}_{X=\infty} {\phi(X)\over W'(X)}
}

If the critical points of $W$ are all Morse critical
points then the algebra \jacideal\ is semisimple.
Physically Morse critical points correspond to massive
theories, while non-Morse critical points renormalize
to nontrivial 2d CFT's in the infrared.

If all the critical points are Morse then
 the trace is easily written in terms of the
critical points $p_a$ as
\eqn\lgtraceii{
\theta(\phi) = + \sum_{dW(p_a)=0} {\phi(p_a)\over \det(\p_i\p_j
W\vert_{p_a}) }
}

In the semisimple one-variable case we can construct the
basic idempotents as follows.
Let
\eqn\idemi{
dW = \prod_{\alpha=1}^n(X-r_\alpha)
}
where we assume all the roots are distinct. Then it is
easy to check that
\eqn\idemii{
\varepsilon_{\beta} := \prod_{\alpha: \alpha \not= \beta} {(X-r_\alpha)
\over (r_\beta  - r_\alpha)}
}
are basic idempotents. (To prove this, write $(X-r_\alpha) =
(X-r_{\beta}) + (r_{\beta}
- r_{\alpha})) $.

\bigskip{\bf Example:}  $W= {1\over 3} t^3 - q t$.
For $n=2$ we can explicitly write
\eqn\expld{
\eqalign{
\varepsilon_1 & = {\sqrt{q} + t \over 2 \sqrt{q}} \cr
\varepsilon_2 & = {\sqrt{q} - t \over 2 \sqrt{q}} \cr}
}
Note that $\theta_1 = 1/(2\sqrt{q})$ and $\theta_2 = -1/(2\sqrt{q})$.

Then from the general result above one finds
$\CO = {\rm End}(W_1) \oplus {\rm End}(W_2)$ and
\eqn\opnstf{
\eqalign{
\theta_\CO(\Psi) & = {1\over \sqrt{\theta_1 }}   {\Tr}(\Psi_1)
+  {1\over \sqrt{\theta_2 }} {\Tr}(\Psi_2)  \cr
\iota^*(\Psi)& = \sqrt{\theta_1} {\Tr}(\Psi_1) \varepsilon_1
+
\sqrt{\theta_2} {\Tr}(\Psi_2) \varepsilon_2\cr}
}
Thus, the general boundary state is
\eqn\genbs{
B = w_1 {\varepsilon_1\over \sqrt{\theta_1}} +
 w_2  {\varepsilon_2\over \sqrt{\theta_2}}
}
where $w_1, w_2$ are integers.
Note the interesting monodromy as $q \rightarrow e^{2\pi i} q$.
Branes of type $(w_1,w_2)$ map to branes of type
$(w_2, -w_1)$.

Clearly, there will be similar phenomena for general
Landau-Ginzburg theories. The space of superpotentials
$W$ has a codimension one ``discriminant locus'' where
it  has non-Morse critical points. Analytic continuation
around this locus will permute the $\varepsilon_i$,
but will only permute the $\sqrt{\theta_i}$ up to sign.
One may understand in this elementary way some of the
brane permutation/creation phenomena discussed in
numerous places in the literature.

The ``vector bundles on spacetime'' that we have found
can be taken quite literally in the context of the
theory of strings moving in less than one dimension
which was worked out in 1988-1991. (For a reviews
\ginspargmoore\dvvtst\DijkgraafQH. )  Strings moving in a spacetime
of $n$ disjoint points can be modelled by
matrix chains or by topological field theory.
The latter point of view  is described
in, for example,  \dvvtst\DijkgraafQH.
In the latter point of view, one considers
topological gravity coupled to topological matter.
For $n$ spacetime points the topological matter
can be taken to be the $N=2$ Landau-Ginzburg
theories associated to $W$ given by
  the unfolding of $A_n$ singularities:
\eqn\aenn{
W = {x^{n+1} \over n+1} +  a_n x^n + \cdots + a_0
}
For generic $W$ we find vector bundles on $n$
spacetime points. This is of course what we
expect for the branes in such spacetimes!

It is worth mentioning that in these simplest of
string theories (the ``minimal string theories'')
considerable progress has been made in recent
years in understanding the full spectrum of D-branes,
going beyond the topological field theory truncation.
See \SeibergAT\ for a review.

\newsec{Going beyond semisimple Frobenius algebras}

The examples of topological field theories coming from $N=2$
conformal field theories --- Landau-Ginzburg models  and
the quantum
cohomology  rings of Calabi-Yau manifolds --- suggest that it is of
interest to understand the possible solutions of the algebraic
conditions in the case when $\CC$ is not semisimple. In this section
we shall make some partial progress with this problem, and shall
also explain how it should perhaps be viewed in a wider context.

\subsec{Examples related to the cohomology of manifolds}

A natural example of a graded commutative Frobenius
algebra is the  cohomology with complex coefficients
 of a compact oriented manifold $X$.
 Thus, for $\CC$ we can take the algebra $\CC= H^*(X;\IC)$ with
trace $\theta(\phi) = \int_X \phi$. What are the corresponding $\CO$'s?

A natural guess, which turns out to be wrong, but for interesting
reasons,  is that
we should take $\CO=\CC\otimes {\rm Mat}_N(\IC)= {\rm Mat}_N(\CC)$
for some $N>0$, together with
\eqn\opensturff{
\eqalign{
\theta_\CO(\psi) & = \int_X \Tr(\psi) \cr
 }
}
While $\CO$ is indeed a Frobenius algebra, the only natural candidate
for the map $\iota_*$ is
$\iota_*(\phi)  = \phi\otimes 1_N$.
However, this fails to satisfy the Cardy condition:
one  computes $\iota^*(\psi)  = \Tr(\psi) $ from the adjoint
relation, and hence $\iota_* \iota^*(\psi) = \Tr\psi\otimes 1_N$.
On the other hand, one also
  finds
\eqn\pimap{
\eqalign{
\pi(\psi) & = \sum (-1)^{\deg \omega_i(\deg \psi+\deg \omega^i )}
\omega_i\otimes e_{lm}  \wedge \psi \wedge \omega^i\otimes e_{ml} \cr
& =  \chi(TX) \wedge \Tr\, \psi  \cr}
}
Here $\omega_i$ is a basis for $H^*(X;\IC)$, $e_{ml}$ are matrix
units, and  $\chi(TX)\in H^{\rm top}(X;\IC)$ is the Euler class of $TX$.
The map $\pi$ annihilates
forms of positive degree, and cannot agree with
$\iota_* \iota^*$.

This example can be modified to give an open and closed theory by
taking $\CO$ to be associated with a {\it submanifold} of $X$. This
is, after all, the standard picture of D-branes! Let us work in the
algebraic category of $\IZ$-graded vector spaces, and continue to
take $\CC= H^*(X;\IC)$, with $X$ a compact connected oriented
$n$-dimensional manifold, and the trace $\theta_{\CC}(\phi) = \int_X
\phi$ of degree ${-n}$ as above. Let us look for an open algebra of
the form $\CO = {\rm Mat}_N(\CO_0)$, with $\CO_0$  commutative. Then
$\CO_0$ is a Frobenius algebra, and we may as well assume that it is
$H^*(Y;\IC)$ for some compact oriented manifold\foot{In fact we need
to allow $Y$ to have orbifold singularities to ensure this.} $Y$ of
dimension $m$, and that $\iota :\CC \to\CO_0$ is $f^*$ for some map
$f:Y\to X$.

Thus $\CO = H^*(Y;\IC) \otimes Mat_N(\IC)$ with
open string trace
\eqn\ytrac{
\theta_{\CO}(\Psi) = \theta_o \int_Y {\Tr}(\Psi)
}
of degree ${-m}$, where $\theta_o$ is a constant. This is a
non-commutative Frobenius algebra.

By the adjoint relation $\iota^*$ is then determined
to be
\eqn\pshfrwd{
\iota^*(\Psi) =\theta_o f_*({\Tr}(\Psi))
}
where $f_*$ is the adjoint of the ring
homomorphism $f^*:H^*(X)\to
H^*(Y)$ with respect to Poincar\'{e} duality. Thus $\iota^*$ has
degree $n-m$. On the other hand , one sees at once that $\pi :\CO
\to \CO$ has degree $m$, so if the Cardy condition is to hold we
must have $n=2m$. If that is true, then we can assume, by making a
small generic perturbation of $f$, that $f$ is an immersion of $Y$
in $X$. We can now make the the adjoint map $f_*$ more explicit:
\eqn\Gysin{
f_*(\psi)=\pi^*(\psi)\wedge\Phi_{\CN},
}
where $\pi: \CN \to Y$ is the projection of the normal bundle
(identified with a tubular neighborhood of $Y$ in $X$) and
$\Phi_{\CN}$ is the Thom class of the bundle, compactly supported in
the tubular neighborhood,
which represents the cohomology class of
$Y$ in $X$. One easily finds that

\eqn\pipsi{
\iota_* \iota^*(\Psi) = \theta_o \chi(\CN Y)\wedge {\Tr}(\Psi) \otimes 1.
}
where $\chi(\CN Y)$ is the Euler class of the normal bundle of
$Y\hookrightarrow X$, i.e. the homological self-intersection of $Y$ in $X$.

On the other hand,
\eqn\psipsi{
\eqalign{
\pi(\Psi) & = {1\over \theta_o} {\Tr}(\Psi) \chi(TY)   \cr}
}
where $\chi(TY) \in H^{\rm top}(Y;\IC)$ is the Euler class of the
tangent bundle $TY$, whose
integral is the Euler number of $Y$.

Evidently the Cardy conditions are satisfied
if we choose $\theta_o$
so that $\chi(TY) = \theta_o^2\chi(\CN Y)$. This is always possible
if $\chi(\CN Y)$, which is the self-intersection number of $Y$ in
$X$, is non-zero, and also possible if $Y$ is a Lagrangian
submanifold of a symplectic manifold $X$, for then $\CN \cong TY$.
The boundary state is $B=\theta_o N \Phi_{\CN}$.

One immediate consequence of this discussion is that if we
start,
say, with $\CO = H^*(\IC P^2)$ as our open algebra we can
easily find two different closed algebras compatible with it, by
regarding $Y$ as a submanifold  either  of $X=\IC P^4$ or of $X'=\IH
P^2$.

Unfortunately we do not know how to describe the {\it category} of
boundary conditions for $\CC = H^*(X)$. But it seems likely, in any
case, that to get a significant result one would have to consider
the theory on the cochain level. We next turn our attention to that
case.

\subsec{ The Chas-Sullivan theory}

There is an interesting example ---  due to Chas and Sullivan
\chassullivan --- on the cochain level of a structure a little
weaker than that of our open and closed theories which may
illuminate the use of cochain theories. Let us start with a compact
oriented manifold $X$, which we shall take to be connected and
simply connected.  We can define a category $\CB$ whose objects are
the oriented submanifolds of $X$, and whose vector space of
morphisms from $Y$ to $Z$ is $\CO_{YZ}={\rm
Ext}^*_{H^*(X)}(H^*(Y);H^*(Z))$ --- the cohomology, as usual,  has
complex coefficients, and $H^*(Y)$ and $H^*(Z)$ are regarded as
$H^*(X)$-modules by restriction. The composition of morphisms is
given by the Yoneda composition of Ext groups. With this definition,
however, it will not be true that $\CO_{YZ}$ is dual to $\CO_{ZY}$.
(To see this it is enough to consider Ext$^0=$ Hom, when, say, $Y=X$
and $Z$ is a point.)

We can do better by defining a cochain complex $\hat{\CO}_{YZ}$ of ``morphisms" by
\eqn\barmorph{
\hat{\CO}_{YZ}=\CB_{\Omega(X)}(\Omega(Y);\Omega(Z)),
}
where $\Omega(X)$ denotes the usual de Rham complex of a manifold
$X$, and $\CB_A(B;C)$, for a differential graded algebra $A$ and
differential graded modules $B$ and $C$, is the usual cobar
resolution
\eqn\barres{
\Hom (B;C)\to \Hom(A\otimes B;C)\to\Hom(A\otimes A \otimes B;C)\to \ldots,
}
(in which the differential is given by
\eqn\bardiffl{ \eqalign{ df(a_1 \otimes \ldots \otimes a_k \otimes
b) & = a_1f(a_2 \otimes \ldots \otimes a_k \otimes b) +\cr + &  \sum
(-1)^if(a_1 \otimes \ldots \otimes a_ia_{i+1}\otimes \ldots \otimes
a_k\otimes b) + \cr + & (-1)^k f(a_1 \otimes \ldots \otimes a_{k-1}
\otimes a_kb) \  \ )\cr} }
 whose cohomology is Ext$_A(B;C)$. This is
different from $\CO_{YZ}={\rm Ext}^*_{H^*(X)}(H^*(Y);H^*(Z))$, but
related to it by a spectral sequence whose $E_2$-term is $\CO_{YZ}$
and which converges to $H^*(\hat{\CO}_{YZ})= {\rm
Ext}_{\Omega(X)}(\Omega(Y);\Omega(Z))$. But more important is that
$H^*(\hat{\CO}_{YZ})$ is the homology of the space $\CP_{YZ}$ of
paths in $X$ which begin in $Y$ and end in $Z$. To be precise,
$H^p(\hat{\CO}_{YZ})\cong H_{p+d_Z}(\CP_{YZ})$, where $d_Z$ is the
dimension of $Z$. On the cochain complexes the Yoneda composition is
associative up to cochain homotopy, and defines a structure of an
$A_\infty$-category $\hat{\CB}$. The corresponding composition of
homology groups
\eqn\concat{
H_i(\CP_{YZ})\times H_j(\CP_{ZW}) \to H_{i+j-d_Z}(\CP_{YW})
}
is the composition of the Gysin map associated to
the inclusion of
the codimension $d_Z$ submanifold $\CM$ of pairs of composable paths
in the product $\CP_{YZ}\times \CP_{ZW}$ with the concatenation map
$\CM \to \CP_{YW}$.


Let us try to fit a ``closed string" cochain
algebra $\CC$ to this
$A_\infty$ category. The algebra of endomorphisms of the identity
functor of $\CB$, denoted $\CE$ in \S 3, is easily seen to be just
the cohomology algebra $H^*(X)$. We have mentioned in Section 2 that
this is the Hochschild cohomology $HH^0(\CB)$.

The definition of Hochschild cohomology for a
linear category $\CB$
was given at the end of \S 3. In fact the definition of the
Hochschild complex makes sense for an $A_{\infty}$ category such as
$\hat{\CB}$, and  it is one candidate for the closed algebra $\CC$.

In the present situation $\CC$ is
equivalent to the usual Hochschild
complex of the differential graded algebra $\Omega(X)$, whose
cohomology is the homology of the free loop space $\CL X$ with its
degrees shifted downwards\foot{Thus the identity element of the
algebra, in $H^0(\CC)$, is the fundamental class of $X$, regarded as
an element of $H_n(\CL X)$ by thinking of the points of $X$ as point
loops in $\CL X$.}  by the dimension $d_X$ of $X$, so that the
cohomology $H^i(\CC)$ is potentially non-zero for $-d_X \leq i <
\infty$. This algebra was introduced by Chas and Sullivan in
precisely the present context --- they were trying to reproduce the
structures of string theory in the setting of classical algebraic
topology. There is a map $H^i(X)\to H^{-i}(\CC)$ which embeds the
ordinary cohomology ring of $X$ in the Chas-Sullivan ring, and there
is also a ring homorphism $H^i(\CC) \to H_i(\CL_0X)$ to the
Pontrjagin ring of the based loop space $\CL_0 X$, based at any
chosen point in $X$.

The other candidate for $\CC$ mentioned in
Section 2 was the cyclic
cohomology of the algebra $\Omega(X)$, which is well-known \petrack\ to be the
{\it equivariant} homology of the free loop space $\CL X$ with
respect to its natural circle-action. This may be an improvement on
the non-equivariant homology.


The structure we have arrived at is, however, {\it not} a
cochain-level open and closed theory, as we have no trace maps
inducing inner products on $H^*(\hat{\CO}_{YZ})$. When one tries to
define operators corresponding to cobordisms it turns out to be
possible only when each connected component of the cobordism has
non-empty outgoing boundary. (A theory defined on this smaller
category is often called a {\it non-compact} theory.) The nearest
theory in our sense to the Chas-Sullivan one is the so-called
``$A$-model" defined for a symplectic manifold $X$. There the
$A_{\infty}$ category is the {\it Fukaya category}, whose objects
are the Lagrangian submanifolds of $X$ equipped with bundles with
connection, and the cochain complex of morphisms from $Y$ to $Z$ is
the Floer complex which calculates the ``semi-infinite" cohomology
of the path space $\CP_{YZ}$. In good cases the cohomology of this
Floer complex has a vector space basis indexed by the points of
intersection of $Y$ and $Z$, and the cohomology of the corresponding
closed complex is just the ordinary cohomology of $X$. From our
perspective the essential feature of the Floer theory is that it
satisfies Poincar\'{e} duality for the infinite dimensional manifold
$\CL X$.
\bigskip

\subsec{Remarks on the $B$-model}

Let $X$ be a complex variety of complex
dimension $d$ with a trivialization of
its canonical bundle. That is, we assume there is a
nowhere-vanishing holomorphic $d$-form $\Omega$.
The {\it B-model} \wittenab\ is a $\IZ$-graded topological field theory arising from the N=2 supersymmetric $\sigma$-model of $X$. The natural boundary conditions for the theory are provided by holomorphic vector bundles on $X$.

The category of holomorphic vector bundles is
not a Frobenius
category. There is, however, a very natural $\IZ$-graded Frobenius
category associated to $X$: the category $\CV_X$ whose objects are
the vector bundles on $X$, but whose space of morphisms from $E$ to
$F$ is
\eqn\bmorph{
\CO_{EF}={\rm Ext}^*_X(E;F)=H^{0,*}(X;E^*\otimes F).
}
%
%
The trace $\theta_E:\CO_{EE}\to\IC$, of degree $-d$, is defined by
\eqn\btrace{
\theta_E(\Psi)=\int_X \Tr (\Psi)\wedge\Omega .
}
This is nondegenerate by Serre duality, but the
category is still not
semisimple --- in fact the non-vanishing of groups Ext$^i$ for $i>0$
precisely expresses the non-semisimplicity of the category. (For a
non-zero element of ${\rm Ext}^1_X(E;F)$ corresponds to an exact
sequence $0 \to F \to G \to E \to 0$ which does not split, i.e. to a
vector bundle $G$ with a subbundle $F$ with no complementary
bundle.)

What are the endomorphisms of the identity functor of
$\CV_X$?
Multiplication by any element of $H^{0,*}(X)$ clearly
defines such an endomorphism. A holomorphic vector field $\xi$ on
$X$ also defines an endomorphism of degree 1, for any bundle $E$ has
an ``Atiyah class"\foot{Corresponding to the extension of bundles
$E\otimes T^*_X \to J^1E \to E$, where $J^1E$ is the bundle of
1-jets of holomorphic sections of $E$.} $a_E\in {\rm
Ext}^1(E;E\otimes T^*_X)$ --- its curvature --- which we can
contract with $\xi$ to give $e_{\xi} = \iota_{\xi}a_E \in {\rm
Ext}^1(E;E)$. More generally, a class
$$\eta \in H^{0,q}(X;\bigwedge^p T_X)={\rm Ext}^q(\bigwedge^p T^*_X;\IC)$$
 can be contracted with
 $(a_E)^p\in {\rm Ext}^p(E;E\otimes (T^*_X)^{\otimes p})$ to give
$$e_{\eta} = \iota_{\eta}(a_E)^p \in {\rm Ext}^{p+q}(E;E).$$
 Now Witten   has shown in \wittenab\ that $H^{0,*}(X;\bigwedge ^* T_X)$
 is indeed the closed string algebra of the B-model.
 To understand this in our context we must once again
 pass to the cochain-level theory of which the Ext groups
 are the cohomology. A good way to do this is to replace a
 holomorphic vector bundle $E$ by its $\overline{\partial}$-complex
  $\hat{E}=\Omega^{0,*}(X;E)$, which is a differential graded module
  for the differential graded algebra $A=\Omega^{0,*}(X)$.
  Then we define $\hat{\CO}_{EF}$ as the cochain complex Hom$_A(\hat{E};\hat{F})$,
  whose cohomology groups are Ext$^*_X(E;F)$. If we are going to do this,
  it is natural to allow a larger class of objects, namely all finitely
  generated projective differential graded $A$-modules.
  Any coherent sheaf $E$ on $X$ defines such a module:
  one first resolves $E$ by a complex $E^*$ of vector bundles,
  and then takes the total complex of the double complex $\hat{E}^*$.
  The resulting enlarged category is essentially the {\it bounded derived category}
  of the category of coherent sheaves on $X$. In this setting,
  we find without difficulty that the endomorphisms
  of the identity morphism are given, just as in the
  topological example above, by the Hochschild complex
$$\hat{\CC} = \{A\to A\otimes A \to A\otimes A \otimes A \to \ldots \ \},$$
whose cohomology is $H^*(X;\bigwedge ^* T_X)$. There is still, however,
 work to do to understand the trace maps on $\hat{\CC}$, and the
 adjoint maps $\iota_E$ and $\iota^E$. We feel that this has not yet
 been properly elucidated in the literature.
 For some progress on this question see \markarian\caldararu.

\newsec{Equivariant 2-dimensional topological open and closed theory}

An important construction in string theory is the ``orbifold''
construction. Abstractly, this can be carried out whenever the
closed string background has a group $G$ of automorphisms.
%
%
There are two steps in defining an orbifold theory.
First, one must extend the theory by introducing ``external'' gauge
fields, which are $G$-bundles (with connection) on the world-sheets.
Next,  one must construct a new theory  by summing
over all possible $G$-bundles (and connections).

We begin by describing carefully the first step in forming the
orbifold theory. The second step --- summing over the $G$-bundles
--- is then very easy in the case of a finite group $G$.

\subsec{Equivariant closed theories  }

Let us begin with some general remarks. In $d$-dimensional
topological field theory one begins with a category $\CS$ whose
objects are oriented $(d-1)$-manifolds and whose morphisms are
oriented cobordisms. Physicists say that a theory admits a group $G$
as a {\it global symmetry group} if $G$ acts on the vector space
associated to each $(d-1)$-manifold, and the linear operator
associated to each cobordism is a $G$-equivariant map. When we have
such a ``global" symmetry group $G$ we can ask whether the symmetry
can be ``gauged", i.e. whether elements of $G$ can be applied
``independently" --- in some sense --- at each point of space-time.
Mathematically the process of ``gauging" has a very elegant
description: it amounts to extending the field theory functor  from
the category $\CS$ to the category $\CS_G$ whose objects are
$(d-1)$-manifolds equipped with a principal $G$ bundle, and whose
morphisms are cobordisms with a $G$-bundle.\foot{We are assuming
here that the group $G$ is {\it discrete}: if $G$ is a Lie group we
should define $\CS_G$ as the category of manifolds equipped with a
principal $G$-bundle with a connection.} We regard $\CS$ as a
subcategory of $\CS_G$ by equipping each $(d-1)$-manifold $S$ with
the trivial $G$-bundle $S \times G$. In $\CS_G$ the group of
automorphisms of the trivial bundle $S \times G$ contains $G$, and
so in a gauged theory $G$ acts on the state space $\CH (S)$: this
should be the original ``global" action of $G$. But the gauged
theory has a state space $\CH (S,P)$ for each $G$-bundle $P$ on $S$:
if $P$ is non-trivial one calls $\CH (S,P)$ a ``twisted sector'' of
the theory. In the case $d=2$, when $S=S^1$ we have the bundle $P_g
\to S^1$ obtained by attaching the ends of $[0,2\pi] \times G$ via
multiplication by $g$. Any bundle is isomorphic to one of these, and
$P_g$ is isomorphic to $P_{g'}$ if and only if $g'$ is conjugate to
$g$. But note that the state space depends on the bundle and {\it
not} just its isomorphism class, so we have a twisted sector state
space $\CC_g = \CH (S,P_g)$ labelled by a group element $g$ rather
than by a conjugacy class.

We shall call a theory defined on the category $\CS_G$ a {\it
$G$-equivariant} TFT. It is important to distinguish the equivariant
theory  from the corresponding ``gauged theory,'' described below.
In physics, the equivariant theory is obtained by coupling to
nondynamical background gauge fields, while the gauged theory is
obtained by ``summing'' over those gauge fields in the path
integral.

An alternative and equivalent viewpoint which is especially
useful in the two-dimensional case is that $\CS_G$ is
the category whose objects are oriented $(d-1)$-manifolds
 $S$ equipped with a
 map $p: S\to BG$, where
$BG$ is the classifying space of $G$. In this viewpoint we have a
bundle over  the space Map$(S,BG)$ whose fiber at $p$ is $\CH_p$. To
say that $\CH_p$ depends only on the $G$-bundle $p^*BG$ on $S$
pulled back from the universal $G$-bundle $EG$ on $BG$ by $p$ is the
same as to say that the bundle on Map$(S,BG)$ is equipped with a
flat connection allowing us to identify the fibres at points in the
same connected component by parallel transport; for the set of
bundle isomorphisms $p_0^*EG \to p_1^*EG$ is the same as the set of
homotopy classes of paths from $p_0$ to $p_1$.  When $S=S^1$ the
connected components of the space of maps correspond to the
conjugacy classes in $G$: each bundle $P_g$ corresponds to a
specific point $p_g$ in the mapping space, and a group-element $h$
defines a specific path from $p_g$ to $p_{hgh^{-1}}$.

The second viewpoint makes clear that $G$-equivariant topological
field theories are examples of ``homotopy topological field
theories'' in the sense of Turaev \turaev. We shall use his two main
results: first, an attractive generalization of the theorem that a
two-dimensional TFT ``is" a commutative Frobenius algebra, and,
secondly, a classification of the ways of gauging a given global
$G$-symmetry of a semisimple TFT.
 We shall now briefly review
his work .

\bigskip

\ifig\FIGFOURTEEN{Definition of the product in the $G$-equivariant
closed theory. The heavy dot is the basepoint on $S^1$. To specify
the morphism unambiguously we must indicate  consistent holonomies
 along a set of curves whose complement consists of simply
connected pieces. This means that the product is not commutative. We
need to fix a convention for holonomies of a composition of curves,
i.e. whether we are using left or right path-ordering. We will take
$h(\gamma_1\circ \gamma_2) = h(\gamma_1) \cdot h(\gamma_2)$. }
{\epsfxsize2.0in\epsfbox{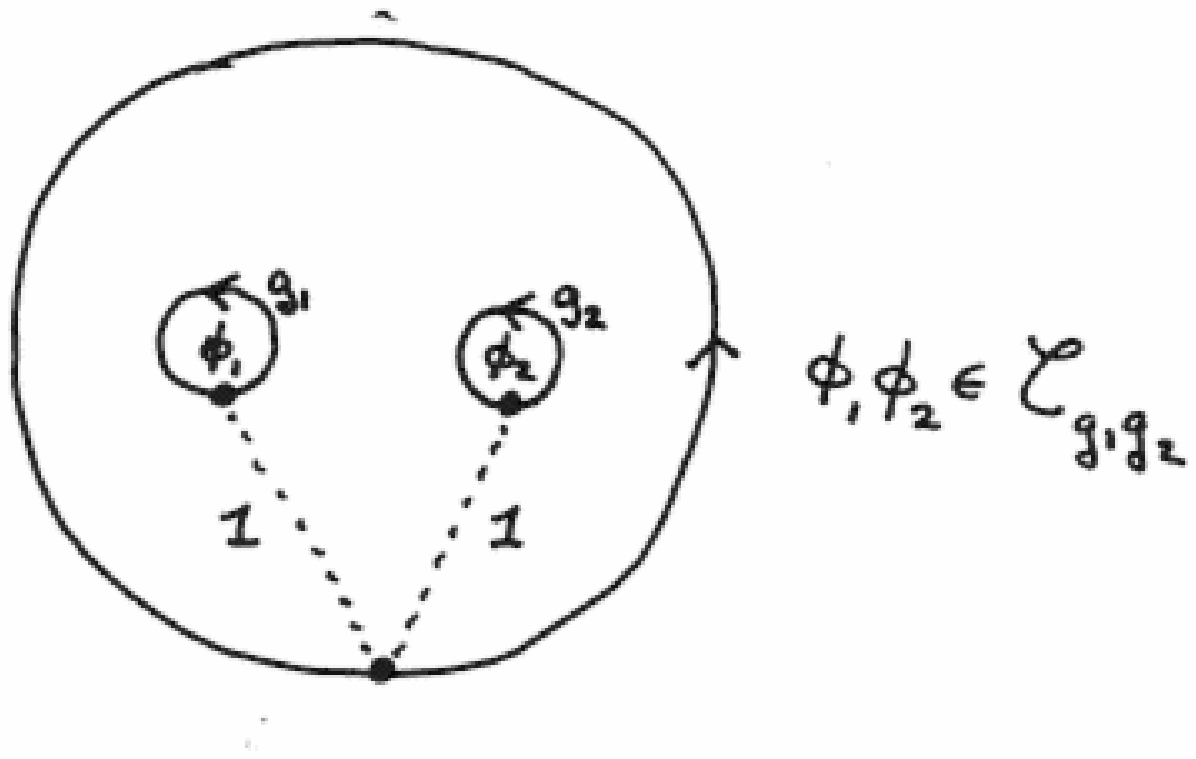}}

\ifig\FIGFIFTEEN{(a) The action of $\alpha_h$ on a state $\phi\in
\CC_g$. This can also be represented by the cylinder as in (b). }
{\epsfxsize2.0in\epsfbox{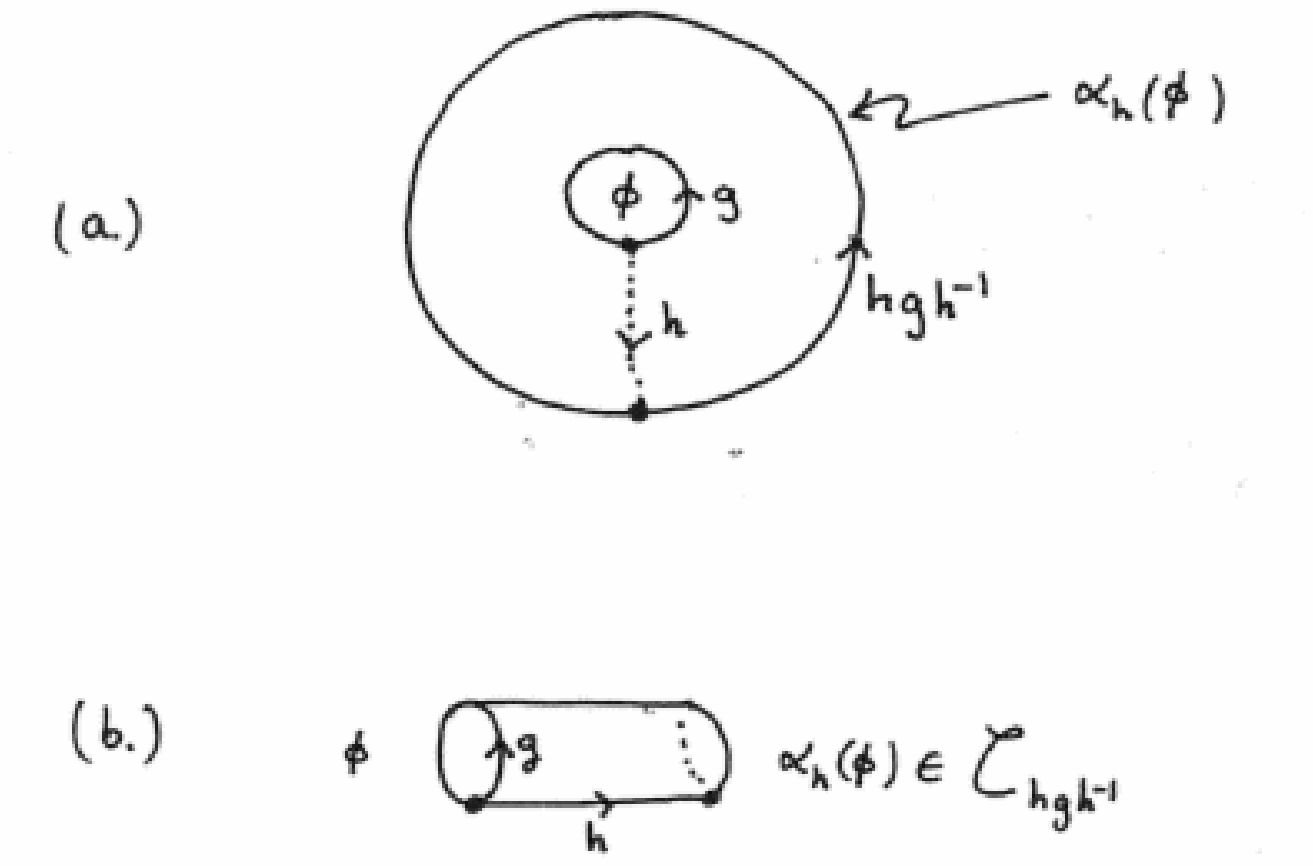}}

A $G$-equivariant TFT gives us for  each element $g\in G$  a vector space $\CC_g$, associated to the circle equipped with the bundle $P_g$ whose holonomy is $g$. The usual pair-of-pants cobordism, equipped with the evident $G$-bundle which restricts to $P_{g_1}$ and $P_{g_2}$ on the two incoming circles, and to $P_{g_1g_2}$ on the outgoing circle, induces a product

\eqn\equivprodct{
\CC_{g_1} \otimes \CC_{g_2} \to \CC_{g_1 g_2}
}
making $\CC:=\oplus_{g\in G} \CC_g$ into a
$G$-graded algebra.

As in the usual case there is a trace $\theta: \CC_1 \to \IC$
defined by the disk diagram with one ingoing circle. Note that
the holonomy around the boundary of the disk must be $1$. Making the standard assumption
that the cylinder corresponds to the unit operator we obtain a
nondegenerate pairing $\CC_g \otimes \CC_{g^{-1}} \to \IC$.

A new element in the equivariant theory is that $G$ acts as
an automorphism group on $\CC$. That is, there is a
a homomorphism   $\alpha: G \to {\rm Aut}(\CC)$ such that
\eqn\automorphi{
\alpha_h: \CC_{g} \to \CC_{hgh^{-1} }.
}
Diagramatically, $\alpha_h$ is defined by the surface in
\FIGFIFTEEN.

\ifig\FIGSIXTEEN{If the holonomy along path $\CP_2$ is $h$ then the
holonomy along path $\CP_1$ is $1$. However, a Dehn twist around the
inner circle maps $\CP_1$ into $\CP_2$. Therefore, $\alpha_h(\phi) =
\alpha_1(\phi)= \phi$, if $\phi\in \CC_h$. }
{\epsfxsize2.0in\epsfbox{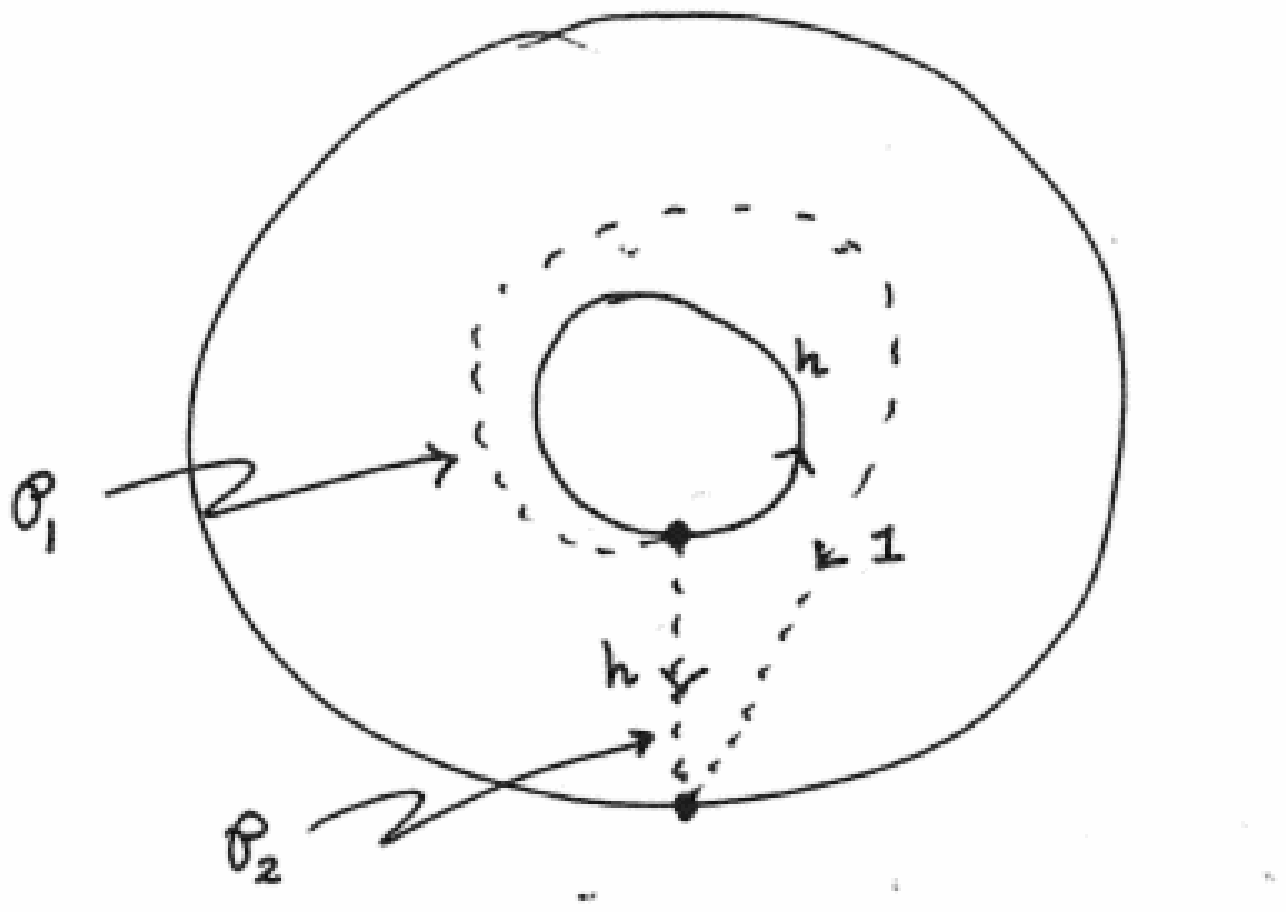}}

\ifig\FIGSEVENTEEN{Demonstrating twisted centrality.  }
{\epsfxsize2.0in\epsfbox{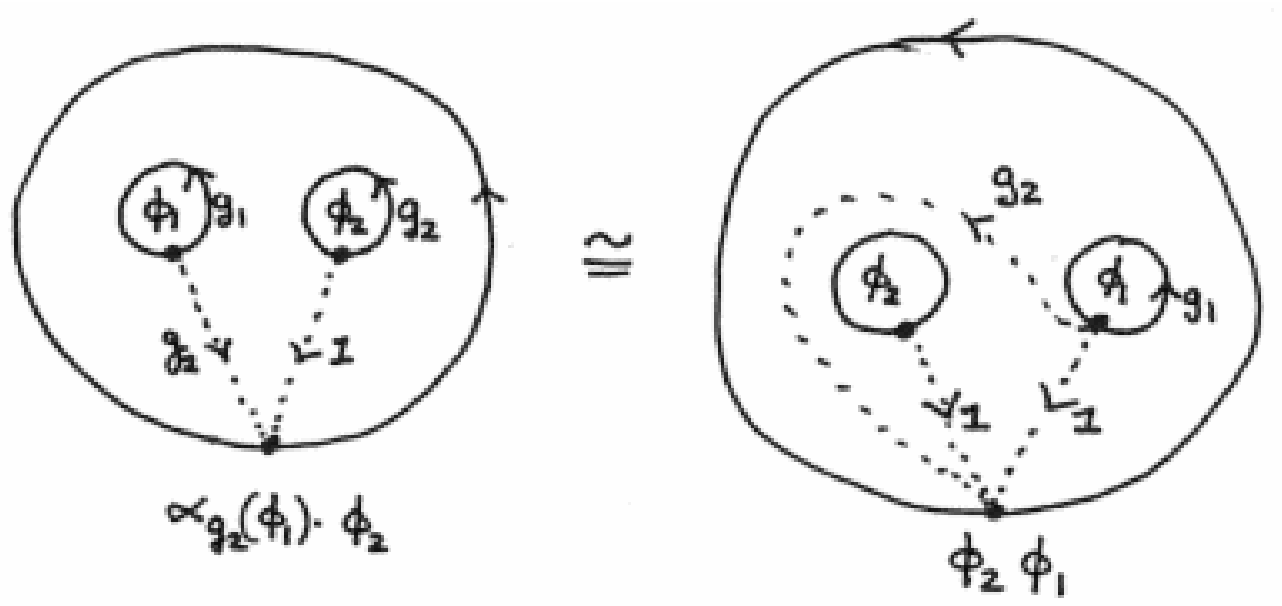}}

\ifig\FIGEIGHTEEN{Deforming the LHS of (a) into a spacetime
evolution diagram yields (b), whose value is ${\Tr}_{\CC_h}( L_\psi
\alpha_g) $. Similarly deforming the RHS of (a) gives a diagram
whose value is $ {\Tr}_{\CC_{g^{-1}}}  (R_\psi \alpha_h) $.   }
{\epsfxsize2.0in\epsfbox{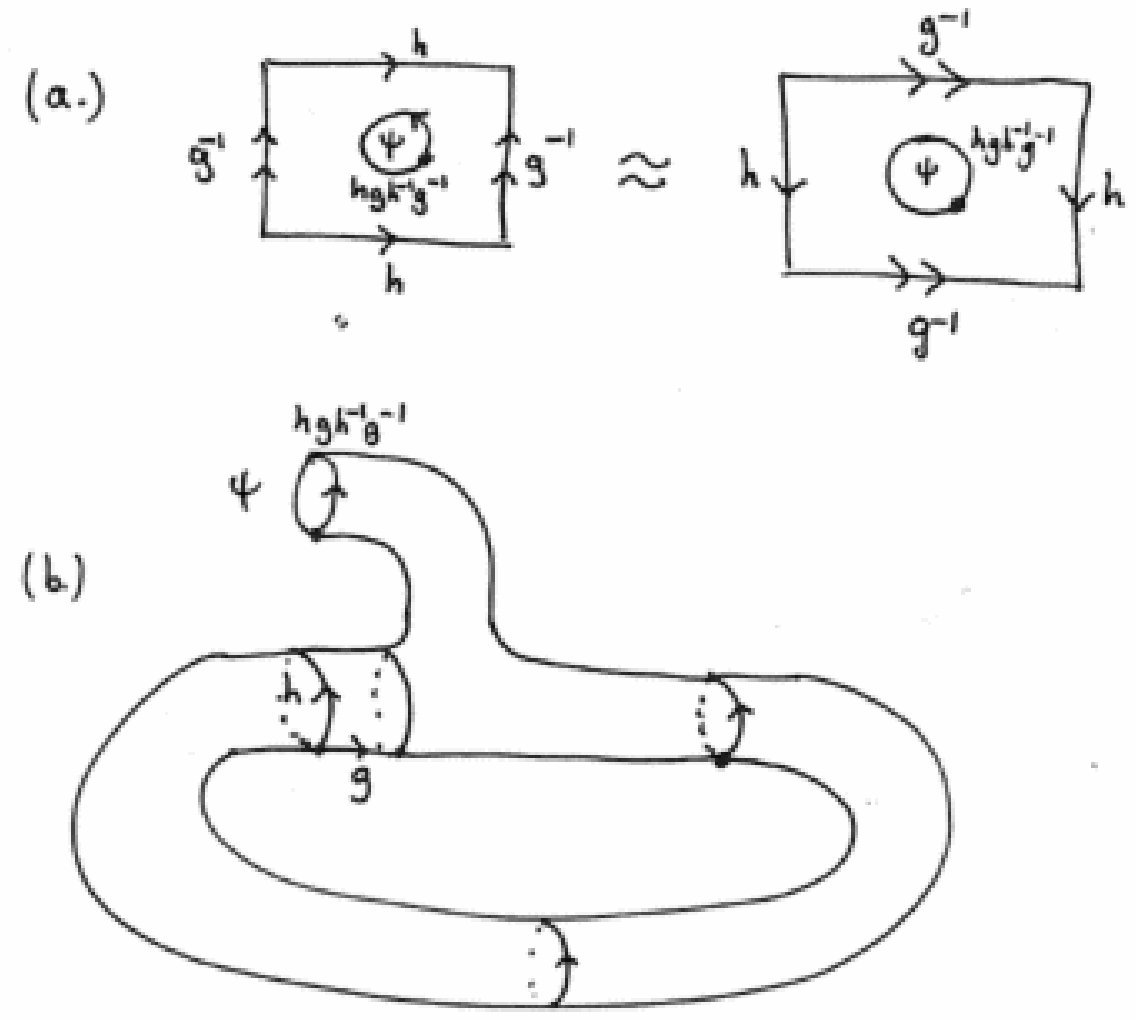}}

 Now let us note some
properties of $\alpha$. First, if $\phi\in \CC_h$ then
$\alpha_h(\phi)=\phi$. The reason for this is explained in
\FIGSIXTEEN.

Next, while $\CC$ is not commutative, it is ``twisted-commutative''
in the following sense.  If $\phi_1\in \CC_{g_1}$ and $\phi_2\in \CC_{g_2}$ then
\eqn\twistedcomm{
\alpha_{g_2}(\phi_1)\phi_2 = \phi_2 \phi_1.
}
The necessity of this condition is illustrated in \FIGSEVENTEEN.

The last property we need is a little more complicated. The trace of the identity map of $\CC_g$ is the partition function of the theory on a torus with the bundle with holonomy $(g,1)$. Cutting the torus the other way, we see that this is the trace of $\alpha_g$ on $\CC_1$. Similarly, by considering the torus with a bundle with holonomy $(g,h)$, where $g$ and $h$ are two commuting elements of $G$, we see that the trace of $\alpha_g$ on $\CC_h$ is the trace of $\alpha_h$ on $\CC_{g^{-1}}$. But we need a strengthening of this property. Even when $g$ and $h$ do not commute we can form a bundle with holonomy $(g,h)$ on a torus with one hole, around which the holonomy will be $c=hgh^{-1}g^{-1}$. We can cut this torus along either of its generating circles to get a cobordism operator from $\CC_c \otimes \CC_h$ to $\CC_h$ or from $\CC_{g^{-1}} \otimes \CC_c$ to $\CC_{g^{-1}}$. If $\psi\in \CC_{hgh^{-1}g^{-1}}$ let us
introduce two linear transformations $L_{\psi},R_{\psi}$ associated to left- and
right-multiplication by $\psi.$ On the one hand,
$L_\psi \alpha_g: \phi \mapsto \psi \alpha_g(\phi)$
is a map $ \CC_h \to \CC_h$. On the other hand  $R_\psi \alpha_h : \phi \mapsto \alpha_h(\phi)\psi$
is a  map $\CC_{g^{-1}} \to \CC_{g^{-1}}$.
The last sewing condition states that
these two endomorphisms must have equal traces:
\eqn\equaltrc{
{\Tr}_{\CC_h}\biggl( L_\psi \alpha_g \biggr)= {\Tr}_{\CC_{g^{-1}}}
\biggl( R_\psi \alpha_h  \biggr).
}
The reason for this can be deduced by pondering the diagram in
\FIGEIGHTEEN.

\ifig\FIGNINETEEN{A simpler axiom than Turaev's torus axiom.   }
{\epsfxsize2.0in\epsfbox{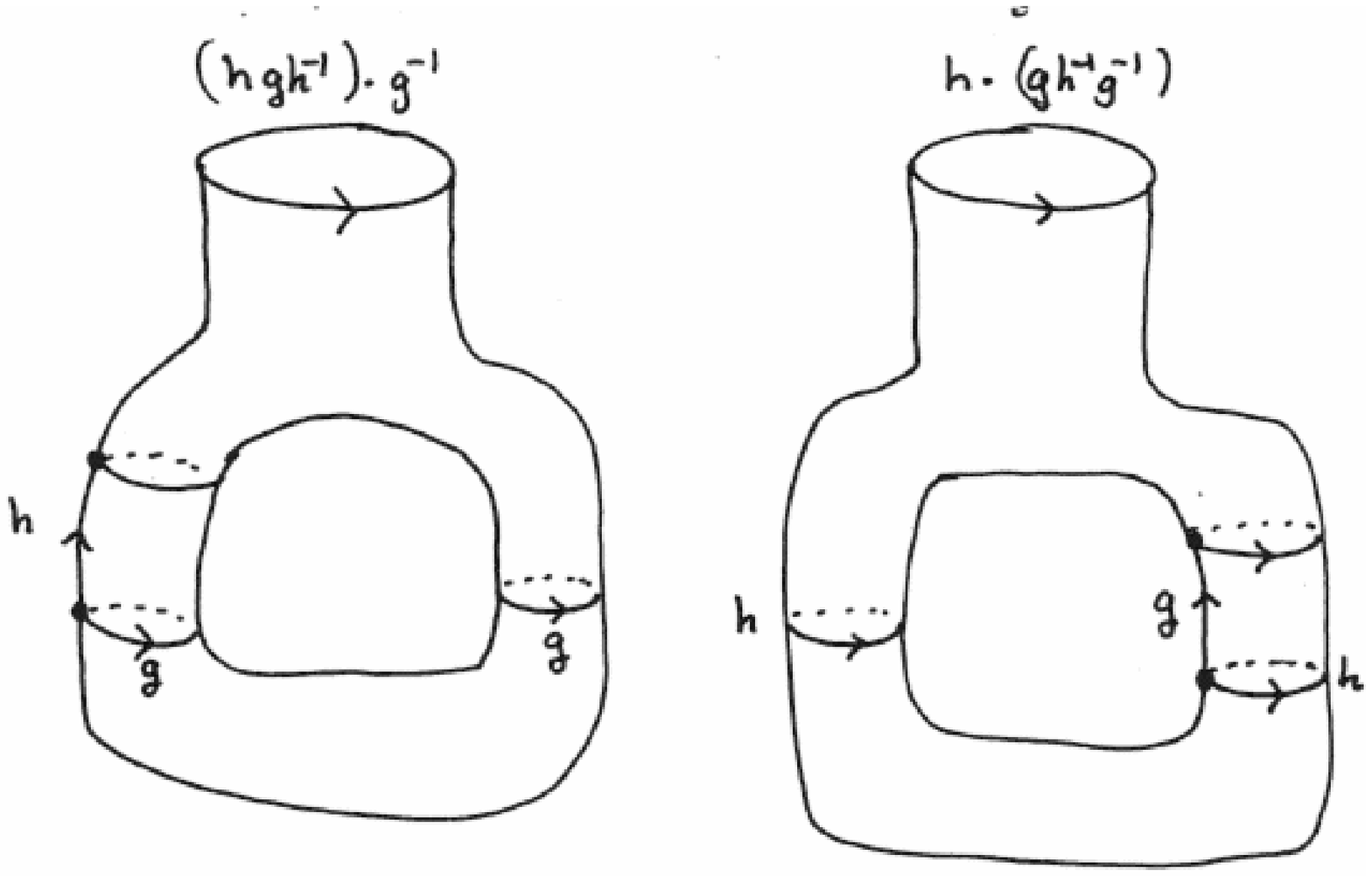}}

The equation (\equaltrc) was taken by Turaev as one of his axioms.
It can, however, be reexpressed in a way that we shall find more
convenient. Let $\Delta_g \in \CC_g \otimes \CC_{g^{-1}}$ be the
``duality" element corresponding to the identity cobordism of
$(S^1,P_g)$ with both ends regarded as outgoing. We have $\Delta_g =
\sum \xi_i \otimes \xi^i$, where $\xi_i$ and $\xi^i$ run through
dual bases of $\CC_g$ and $\CC_{g^{-1}}$. Let us also write
$$\Delta_h = \sum \eta_i \otimes \eta^i \in \CC_h \otimes \CC_{h^{-1}}.$$
Then (\equaltrc) is easily seen to be equivalent to
\eqn\newax{ \sum\alpha_h(\xi_i)\xi^i = \sum \eta_i \alpha_g(\eta^i)
,}
in which both sides are elements of $\CC_{hgh^{-1}g^{-1}}$. This
equation is illustrated by the isomorphic cobordisms of
\FIGNINETEEN.

In summary, the sewing theorem for $G$-equivariant 2d topological
field theories is given by the following theorem:

\bigskip
\noindent{\bf Theorem 4} (\turaev) \  To give a 2d $G$-equivariant
topological field theory is to give a  $G$-graded algebra $\CC =
\oplus_g \CC_g$ together with a group homomorphism $\alpha: G \to
{\rm Aut}(\CC)$ such that

\item{1.} There is a $G$-invariant  trace  $\theta:\CC_1 \to \IC$ which
induces a nondegenerate pairing $\CC_g \otimes \CC_{g^{-1}}\to \IC$.

\item{2.} The restriction of $\alpha_h $ to $\CC_h$ is the identity.

\item{3.} For all $\phi\in \CC_g, \phi'\in \CC_h$, $\alpha_h(\phi) \phi'
= \phi' \phi$.

4. For all  $g,h\in G$ we have
\eqn\newaxp{ \sum\alpha_h(\xi_i)\xi^i = \sum \eta_i
\alpha_g(\eta^i)\in \CC_{hgh^{-1}g^{-1}} , }
 where $\Delta_g = \sum
\xi_i \otimes \xi^i \in \CC_g \otimes \CC_{g^{-1}}$ and $\Delta_h =
\sum \eta_i \otimes \eta^i \in \CC_h \otimes \CC_{h^{-1}}$ as above.

\bigskip
\noindent
{\bf Remarks}:

\item{1.} We will give a proof of the sewing theorem in the
appendix.

\item{2.}
Warning: Turaev calls the above a
 {\it crossed $G$ Frobenius algebra}, but it is not a
crossed-product algebra in the sense of $C^*$ algebras
(see below).
We will refer to an algebra satisfying the conditions of
the theorem as a {\it Turaev algebra}.

\item{3.} Axioms 1 and 3 have counterparts in the non-equivariant
theory, but axioms 2 and 4 are new elements.

\bigskip

\subsec{\it The orbifold theory}

Before going any further, let us describe
how we obtain the orbifold
theory from the Turaev algebra.

Let us return to the general discussion at the beginning of
$\S{7.1}$, where we outlined the definition of an equivariant
theory. Roughly speaking, the gauged theory is obtained from the
equivariant theory by summing over the gauge fields. More precisely,
the state-space which a gauged theory associates to a
$(d-1)$-manifold $S$ consists of ``wave-functions" $\psi$ which
associate to each $G$-bundle $P$ on $S$ an element $\psi_P$ of the
state-space $\CH_{S,P}$ of the equivariant theory. The map $\psi$
must be ``natural" in the sense that when $\theta :P\to P'$ is a
bundle isomorphism the induced isomorphism $\CH_{S,P}\to \CH_{S,P'}$
takes $\psi_P$ to $\psi_{P'}$. This is often referred to as the
``Gauss law.'' In the two-dimensional case, the Gauss law amounts to
saying that the state-space $\CC_{\rm orb}$ for the circle is the
$G$-invariant part of the Turaev algebra $\CC = \oplus \CC_g$. In
other words,

\eqn\orbstates{
\CC_{\rm orb} = \oplus \{\CC_g\}^{Z_g},
}
where now $g$ runs through a set of representatives for the
conjugacy classes in $G$, and we take the invariant part of $\CC_g$
under the centralizer $Z_g$ of $g$ in $G$. The algebra $\CC_{\rm
orb}$ is not a graded algebra if $G$ is non-abelian. One must check
that the product in $\CC_{\rm orb}$ is simply the restriction of the
product in $\CC$. The trace $\CC_{\rm orb} \to \IC$ is the
restriction of the trace $\CC \to \IC$ which is the given trace on
$\CC_1$ and is zero on $\CC_g$ when $g\neq 1$. Then $\CC_{\rm orb}$
is a commutative Frobenius algebra which encodes the orbifold
theory.

\subsec{Solutions of the closed string $G$-equivariant
sewing conditions}

Having found the sewing conditions in the $G$-equivariant case we
can ask what  examples there are of the structure. The Frobenius
algebra $\CC_1$ with its $G$-action corresponds to a topological
field theory with a global $G$-symmetry. In the case when $\CC_1$ is
a semisimple Frobenius algebra --- and therefore the algebra of
functions on a finite $G$-set $X$ --- Turaev finds a nice answer:
ways of gauging the symmetry, i.e. of extending $\CC_1$ to a Turaev
algebra, correspond to {\it equivariant $B$-fields} on $X$, i.e. to
equivariant 2-cocycles of $X$ with values in $\IC^{\times}$, and two
such $B$-fields define isomorphic Turaev algebras if and only if
they represent the same class in $H^2_G(X;\IC ^{\times}) \cong
H^3_G(X;\IZ)$. We now review this result and take the opportunity to
introduce a more geometric picture of Turaev's algebra $\CC$ (in the
semisimple case). We shall first recall some very general
constructions.

\subsubsec{{\it General constructions}}

Whenever a
group $G$ acts on a set $X$  we can form a category
$X//G$, whose objects are the points  $x$ of $X$, and whose
morphisms $x_0 \to x_1$ are
\eqn\homct{
{\rm Hom}(x_0,x_1):=\{g\in G: gx_0 = x_1\}.
}

\ifig\FIGTWENTY{An oriented two-simplex $\Delta_{x,g_1,g_2}$ in the
space $\vert X//G\vert$.   } {\epsfxsize2.0in\epsfbox{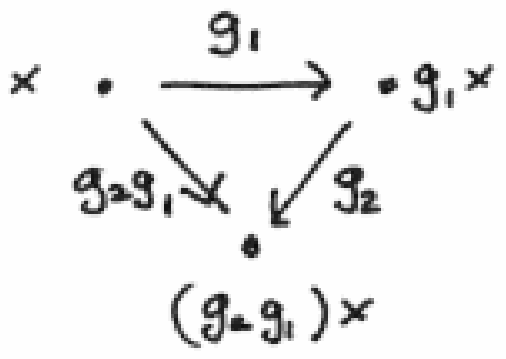}}

\ifig\FIGTWENTYONE{An oriented 3-simplex in $\vert X//G\vert$.  }
{\epsfxsize2.0in\epsfbox{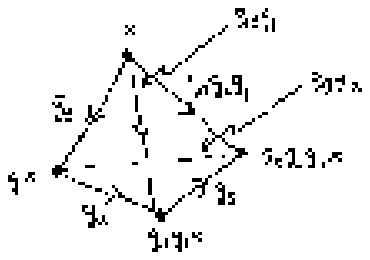}}

Next, for any category ${\bf \CC}$, one can form the {\it space of
the category}, denoted $\vert {\bf \CC}\vert $. This is an oriented
simplicial complex whose $p$-simplexes are in 1-1 correspondence
with the composable $p$-tuples of morphisms in the category. To be
specific, the vertices are the objects of the category. The  edges
are the morphisms. Triples of morphisms $(f_1, f_2,f_3)$ with $f_3 =
f_2\circ f_1$ correspond to  2-simplices, and so forth. In the
present case, when we form the simplicial complex $\vert X//G\vert$
the 2-simplices are the triples $(g_1,g_2,x)$ illustrated in
\FIGTWENTY. Three-simplices are shown in \FIGTWENTYONE, etc.

The space $\vert X//G\vert$ is a model for $(X \times EG)/G$.
Hence the (cellular) cohomology of this space $H^*(\vert X//G\vert;\IC^{\times})$ is
the equivariant cohomology $H_G^*(X;\IC^{\times})$.

Another object which we can associate to any category $C$ is its {\it algebra}
$A(C)$ over the field $\IC$. This has a vector space basis $\{\varepsilon_f\}$ indexed by the morphisms of $C$, and the product is given by $\varepsilon_{f_1}\varepsilon_{f_2} = \varepsilon_{f_1\circ f_2}$ when $f_1$ and $f_2$ are composable, and $\varepsilon_{f_1}\varepsilon_{f_2} = 0$ otherwise. For the category $X//G$ the algebra $A(X//G)$ is the usual {\it crossed-product} algebra $A(X)\sdtimes G$ in the sense of operator algebra theory, where $A(X)$ is the algebra of complex-valued functions on the set $X$. \foot{For any commutative algebra $A$ with $G$-action, $A\sdtimes G$ is spanned by elements $a\otimes g$ with $a\in A$ and $g\in G$, and the product is given by
\eqn\crprod{
(a_1\otimes g_1)(a_2\otimes g_2) = a_1g_1(a_2)\otimes g_1g_2.
}
The isomorphism $A(X//G)\to A(X)\sdtimes G$ takes $\varepsilon_{g,x}$ to $\chi_{gx}\otimes g$, where $\chi_x$ is the characteristic function supported at $x$.}

The construction of the category-algebra $A(C)$ can be generalized.
A {\it $B$-field} on a category $C$ is a rule which associates  a
complex line $L_f $  to each morphism $f$ of $C$, and associative
isomorphisms
$$
L_{f_1} \otimes L_{f_2} \to L_{f_1\circ f_2}
$$
to each pair $(f_1 ,f_2)$ of composable morphisms .
In concrete terms, to give such a product is to give a 2-cocycle
on the space $\vert C \vert$. Indeed, choosing
basis elements $\ell_f \in L_f$, we must have
\eqn\algebr{
\ell_{f_1} \cdot \ell_{f_2} = b(f_1, f_2,f_3) \ell_{f_3}
}
where $b(f_1, f_2, f_3) \in \IC^{\times}$ defines a 2-cochain on
$\vert C \vert$. (We choose values in $\IC^{\times}$ rather than
$\IC$ so the product is not degenerate.) Associativity of \algebr\
holds iff $b$ is a 2-cocycle. A change of basis of the $L_f$
modifies $b$ by a coboundary. Hence the isomorphism classes of
$B$-fields  on $C$ are in 1-1 correspondence with cohomology classes
$[b]\in H^2(\vert C \vert ;\IC^{\times})$. When we have a $B$-field
$b$ on $C$ we can form a {\it twisted} category-algebra $A_b (C)$,
which as a vector space is $\oplus L_{f}$, and where the
multiplication is defined by means of the associative maps $L_{f_1}
\otimes L_{f_2} \to L_{f_1\circ f_2}$.

Applying the above construction
to   the category $X//G$, an associative
product on the lines $L_{g,x}$ is the same thing as
a 2-cocycle in  $H^2_G(X;\IC^{\times})$.
In terms of the basis elements $\ell_{g,x}$ for
the lines $L_{g,x}$ we shall write the multiplication
\eqn\multrule{
\eqalign{
\ell_{g_2,x_2} \ell_{g_1,x_1}
& = b_{x_1}(g_2,g_1) \ell_{g_2g_1,x_1}
\qquad\qquad {\rm if}\qquad x_2 = g_1x_1 \cr
&= 0 \qquad\qquad\qquad\qquad\qquad\qquad {\rm otherwise} \cr}
}
Here $b_{x_1}(g_2,g_1) = b(\Delta_{x,g_1,g_2})$ is the value of the
cocycle on the oriented 2-simplex of \FIGTWENTY.

Notice that if $G_x$ is the isotropy group of some point $x\in X$
then restricting \multrule\ to the elements $\ell_{g,x}$ with $g\in
G_x$ shows that $b_x$  defines an element of the group cohomology
$H^2(G_x;\IC^{\times})$, corresponding to the central extension of
$G_x$ by $\IC^{\times}$ whose elements are pairs $(g,\lambda)$ with
$g\in G_x$ and $\lambda \in L_x-\{0\}$. This central extension of
the isotropy group $G_x$ does not in general extend to any central
extension of the whole group $G$. It does so, however, in the
particular case when the $B$-field $b$ is pulled back from a
2-cocycle of $G$ by the map $X\to ({\rm point})$, i.e. when
$b_x(g_2,g_1)$ is independent of $x$. In general the cocycle
$b:G\times G\times X \to \IC^{\times}$ can be regarded as a cocycle
of the group $G$ with values in the abelian group $A(X)^{\times} =
{\rm Map}(X;\IC^{\times})$ with its natural $G$-action. Thus it
defines a (non-central) extension
$$1 \to A(X)^{\times} \to \tilde G \to G \to 1.$$

One technical point to notice is that for any $B$-field we have $L_f
=\IC$ canonically when $f$ is an identity morphism. Thus $L_{g,x}
=\IC$ when $g=1$. We shall always choose $\ell_{g,x}=1$ when $g=1$,
thereby normalizing the cocycle so that $b_x(g_1,g_2)=1$ if either
$g_1$ or $g_2$ is 1.

The algebra $A_b(X//G)=\oplus_{g\in G, x\in X} L_{g,x}$ with the
multiplication rule defined by \multrule\ can be identified with
the twisted crossed-product algebra $A(X)\sdtimes_b G$ via
$$
\ell_{g,x} \mapsto  \chi_{gx} \otimes g,
$$
where $\chi_x$ is the characteristic function supported at $x$.
The twisted cross-product is defined by

\eqn\twisprod{
(f_1 \otimes g_1) (f_2 \otimes g_2) = \alpha_{g_1g_2}(b(g_1,g_2))   f_1\,\alpha_{g_1}(f_2) \otimes g_1 g_2,
}
where $b(g_1,g_2)$ denotes the function $x \mapsto b_x(g_1,g_2)$ in $A(X)^{\times}$, and the group $G$ acts on $A(X)$ in the natural way
$$\alpha_g(f)(x) = f(g^{-1} x),$$
so that $g\cdot \chi_x = \chi_{gx}$.

If we wish to apply these considerations to the spin case described
in sections 2.6 and 3.4 then we must consider the lines $L_f$ to be
$\IZ/2$ graded. In this case the theory will admit a further
twisting by $H^1(\vert C\vert;\IZ/2)$. However, we will not discuss
this generalization further.

\ifig\FIGTWENTYTWO{The algebra of little loops for $X=S_3/S_2$,
where $S_n$ is the permutation group on $n$ letters.  }
{\epsfxsize2.0in\epsfbox{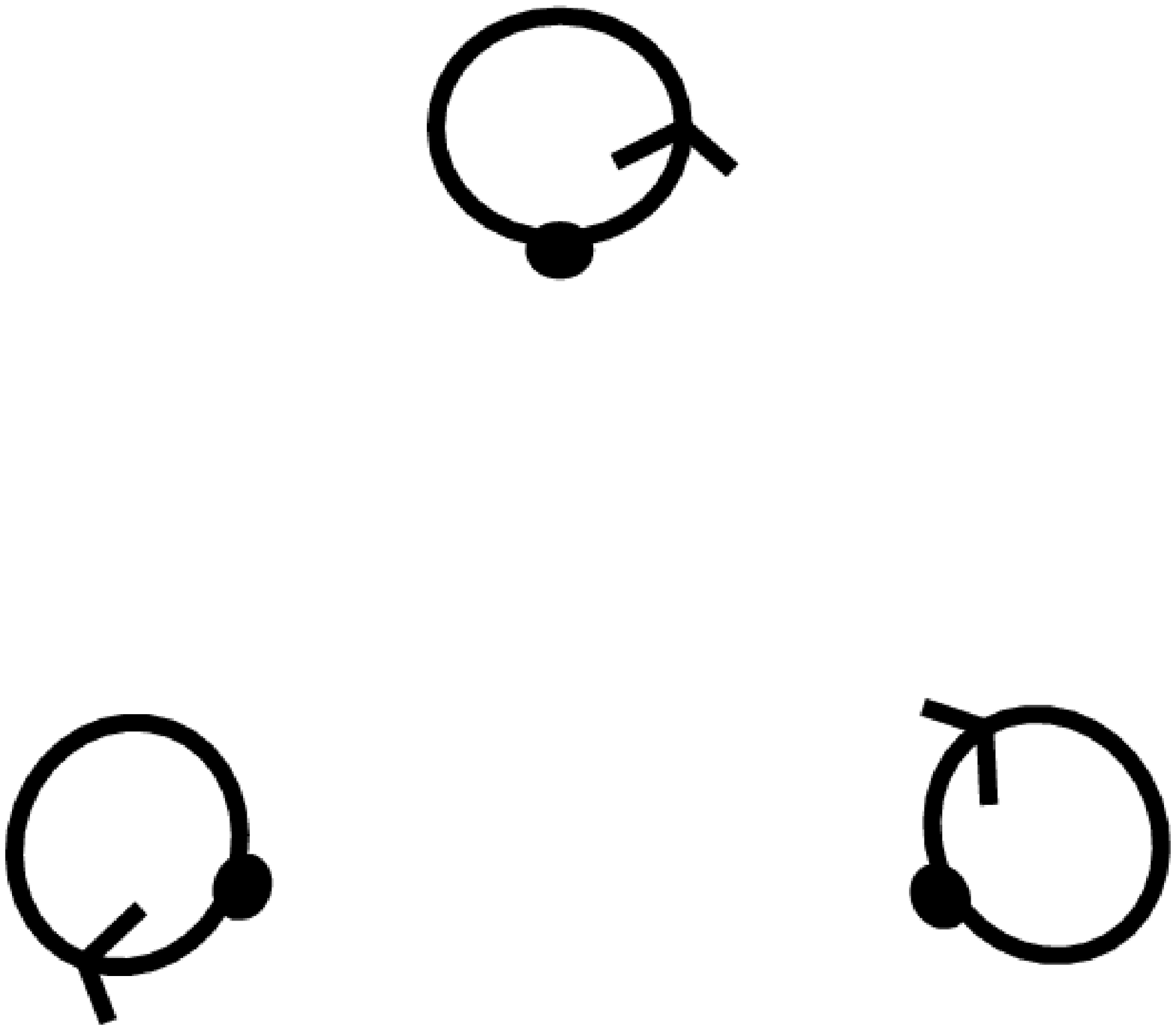}}

\subsubsec{\it The Turaev algebra associated to a $G$-space}

The algebra $A_b(X//G)$ does {\it not} satisfy the sewing conditions
and is not a Turaev algebra. In particular \twistedcomm\ is usually
not satisfied for a crossed-product algebra. However, the
subcategory defined by the morphisms with the same initial and
terminal object does lead to a Turaev algebra for any $B$-field $b$
on $X//G$. We call this the
 ``algebra of little loops".  Thus we
define $\CC = \oplus_g \CC_g\subset A_b(X//G) $ by
\eqn\turaevi{
\CC_g := \oplus_{x: gx = x} L_{g,x}
}
and define the trace by
\eqn\tracequiv{ \theta(\ell_{g,x})  = \delta_{g,1}
\theta(\varepsilon_x) } where on the right $\theta$ is the given
$G$-invariant trace on $\CC_1$, and the $\varepsilon_x$ are the
usual idempotents in the semisimple Frobenius algebra $\CC_1 =
A(X)$, i.e. $\varepsilon_x = 1 \in L_{1,x} = \IC$. The algebra of
little loops can be visualized as in \FIGTWENTYTWO.

An equivalent way to describe $\CC$ is
as the commutant of $\CC_1 =A(X)$ in $A_b(X//G)=A(X)\sdtimes _b G$.
As $A(X)$ is in the centre of $\CC$, it is natural to think of $\CC$
as the sections of a bundle of algebras on $X$; the fibre of
this bundle at $x\in X$ is the twisted group algebra $\IC_{b_x}[G_x]$,
where $G_x$ is the isotropy group of $x$. Furthermore, the bundle of
algebras has a natural $G$-action, covering the $G$-action on $X$.
To see this, notice that the extension
 $\tilde{G} = \{f,g): f \in A(X)^{\times}, g \in G\}$ of
 the group $G$ by the multiplicative group $A(X)^{\times}$
 defined by the $B$-field sits inside the multiplicative group
 of $A(X)\sdtimes _b G$, normalizing the subalgebra $A(X)$.
 As $A(X)$ is in the centre of $\CC$, this means that $G$ acts
 by conjugation on the algebra $\CC$. Notice, however, that
 only $\tilde{G}$ and not $G$ acts on the larger algebra $A_b(X//G).$

 In terms of explicit formulae, the action of $G$ on the algebra $\CC$ is given by
\eqn\actgroup{
\alpha_{g_1}(\ell_{g_2,x})=\ell_{g_1,x}\ell_{g_2,x}\ell_{g_1,x}^{-1}
 = z_x(g_2,g_1)\ell_{g_1g_2g_1^{-1},g_1x},
}
where

\eqn\actgrpii{
z_x(g_2,g_1) = {b_x(g_1,g_2)b_x(g_1g_2,g_1^{-1})\over b_x(g_1,g_1^{-1})}.
}

In this way we obtain a Turaev algebra, which we shall denote by
$\CC = T(X,b,\theta)$. The only
non-trivial point is to verify the
``torus" axiom \equaltrc\ . But in fact it is easy to see that both
sides of the equation are equal to
$$\sum_x\ell_{h,x}\ell_{g,x}\ell_{h,x}^{-1}\ell_{g,x}^{-1},$$
where $x$ runs through the set $\{x\in X:hx = gx=x\}$.

\bigskip

 Turaev has shown that the above construction
 is the most general one possible in the semisimple case.

\bigskip
\noindent {\bf Theorem 5} (\turaev, Theorem 3.6)\ \ \ Let $\CC$ be a
Turaev algebra. If  $\CC_1$ is semisimple then $\CC$ is the twisted
algebra $T(X,b,\theta)$ of little loops on $X= {\rm Spec}(\CC_1)$
for some cocycle $b \in Z^2_G(X;\IC^{\times})$.

\bigskip
\noindent {\it Proof}: If $\CC_1$ is semisimple we may decompose it
in terms of the basic idempotents $\varepsilon_x$. Then $\CC_g$ is a
module over $\CC_1$, and hence it should be identified with the
cross sections of the vector bundle over the finite set $X$ whose
fibre at $x$ is $\CC_{g,x} = \varepsilon_x\CC_g$. (This is a trivial
case of what is called the Serre-Swan theorem.) Now we consider the
torus axiom \equaltrc\ in the case $h=1$. We have $\Delta_1 = \sum
\theta(\varepsilon_x)^{-1}\varepsilon_x\otimes \varepsilon_x$, and
hence
$$\sum \theta(\varepsilon_x)^{-1}\alpha_g(\varepsilon_x)
\varepsilon_x=\sum \theta(\varepsilon_x)^{-1}\varepsilon_x,$$ where
the second sum is over $x$ such that $gx=x$. On the other hand we
readily calculate that if $\{a_{x,i}\}$ is a basis of $\CC_{g,x}$
and $\{a^*_{x,i}\}$ is the dual basis of $\CC_{g^{-1},x}$ then
$a_{g,i}a^*_{g,i}=\theta(\varepsilon_x)^{-1}\varepsilon_x$, so that
the other side of the torus axiom is
$$\sum \dim (\CC_{g,x})\varepsilon_x.$$
Thus the axiom tells us that $\CC_{g,x}$ is a one-dimensional space
$L_{g,x}$ when $gx =x$, and is zero otherwise. The multiplication in
$\CC$ makes these lines into a $G$-equivariant $B$-field on the
category of small loops in $X//G$. Finally, it is not hard to show
that the category of $B$-fields on $X//G$ is equivalent to the
category of $G$-equivariant $B$-fields on the category of small
loops; but we shall omit the details. $\spadesuit$.

\bigskip

Let us now consider the orbifold theory
coming from the gauged theory
defined by the Turaev algebra $\CC =T(X,b,\theta)$. We saw in
Section 7.2 that it is defined by the commutative Frobenius algebra
$\CC_{\rm orb}$ which is the $G$-invariant subalgebra of $\CC$. In
the case of the Turaev algebra of a $G$-space $X$ we have
\bigskip

\noindent{\bf Theorem 6} \ \ The orbifold algebra $\CC_{\rm orb}$ is
the centre of the crossed-product algebra $A(X)\sdtimes_b G$. It is
the algebra of functions on a finite set $(X/G)_{\rm string}$ which
is a ``thickening" of the orbit space $X/G$ with one point for each
pair $\xi,\rho$ consisting of an orbit $\xi$ and an irreducible
projective representation $\rho$ of the isotropy group $G_x$ of a
point $x\in\xi$, with the projective cocycle $b_x$ defined by the
$B$-field.

\bigskip
\noindent{\it Proof} \ The Turaev algebra $\CC$
consists of the
elements of $A(X)\sdtimes_b G$ which commute with $A(X)$. But an
element of $A(X)\sdtimes_b G$ belongs to its centre if an only if it
commutes with $A(X)$ and also commutes with the elements of $G$,
i.e. is $G$-invariant.

Now we saw that $\CC$ is the product over the points $x\in X$ of the twisted group-algebras $\IC_{b_x}[G_x]$. The invariant part is therefore the product over the orbits $\xi$ of the $G_x$-invariant part of $\IC_{b_x}[G_x]$, i.e. of the centre of $\IC_{b_x}[G_x]$, which consists of one copy of $\IC$ for each irreducible representation $\rho$ with the cocycle $b_x$. \ $\spadesuit$

\bigskip

The Turaev algebra $\CC = T(X,b,\theta)$ sits between
$\CC_{\rm orb}$
and $A(X)\sdtimes _b G$. We shall see in Section 7.6 that
$A(X)\sdtimes _b G$ is semisimple, and hence Morita
equivalent\foot{This means that the category of representations of
$A(X)\sdtimes_b G$ is equivalent to the category of representations
of $\CC_{\rm orb}$, uniquely up to tensoring with a ``line bundle"
--- a representation $L$ of $\CC_{\rm orb}$ such that $L\otimes
_{\CC_{\rm orb}}L' \cong \CC_{\rm orb}$ for some $L'$.} to its
centre $\CC_{\rm orb}$.  But the Turaev algebra retains more
information than the orbifold theory: it encodes $X$ and its
$G$-action. The difference is plainest when $G$ --- of order $n$ ---
acts freely on $X$; then $A(X)\sdtimes G$ is the product of a copy
of the algebra of $n \times n$ matrices for each $G$-orbit in $X$,
and provides us with no way of distinguishing the individual points
of $X$. We shall see in section 7.5 that the   category of boundary
conditions for the gauged theory $\CC$ is a natural enrichment of
the category for the orbifold theory, at least in the semisimple
case.

 It might come as a surprise that the cross-product
algebra of spacetime $A(X)\sdtimes G$
 is not the appropriate Frobenius algebra for $G$-equivariant
topological field theory,  in view of the occurrence of the
crossed-product algebra as a central concept in the theory of
D-branes on orbifolds developed in \dm\koscrev\martmoore. In fact,
this fits in very nicely with the philosophy of this paper. The
Turaev algebra remembers the points of
$X$, and so allows only the
``little loops" above. In this way the sewing conditions - which are
meant to formalize worldsheet locality - also encode a crude form of
{\it spacetime} locality.

\bigskip

We shall conclude this section by making  contact with the
usual path integral expression for the orbifold partition
function on a torus.  To do this we compute $\dim \CC_{{\rm orb}}$ by
 computing the projection onto $G$-invariant states in $\CC$.
 Note that $\alpha_g(\ell_{h,x})$  is only proportional to $\ell_{h,x}$
 when $[g,h]=1$ and $gx=x$, and then
 \eqn\prpcse{
\alpha_g(\ell_{h,x}) = {b_x(g,h)\over b_x(h,g)} \ell_{h,x} }
 where we have combined \actgrpii\ with the cocycle identity. Thus we find
 \eqn\dimtwsted{ \dim \CC_{{\rm orb}} =
 {1\over \vert G\vert} \sum_{gh=hg} \sum_{ x=gx=hx} {b_x(g,h)\over b_x(h,g)} }

\bigskip

\ifig\FIGTWENTYTHREE{The wavy line is a constrained boundary. If
there is holonomy $g$ along the dotted path $\CP$ then this morphism
gives the $G$-action on $\CO$. }
{\epsfxsize2.0in\epsfbox{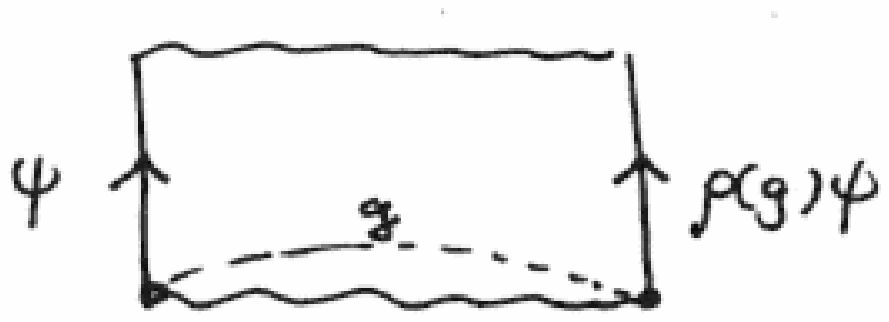}}

\ifig\FIGTWENTYFOUR{The definition of the multiplication in $\CO$.
The holonomy on all dotted paths is $1$. Note the order of
multiplication. } {\epsfxsize2.0in\epsfbox{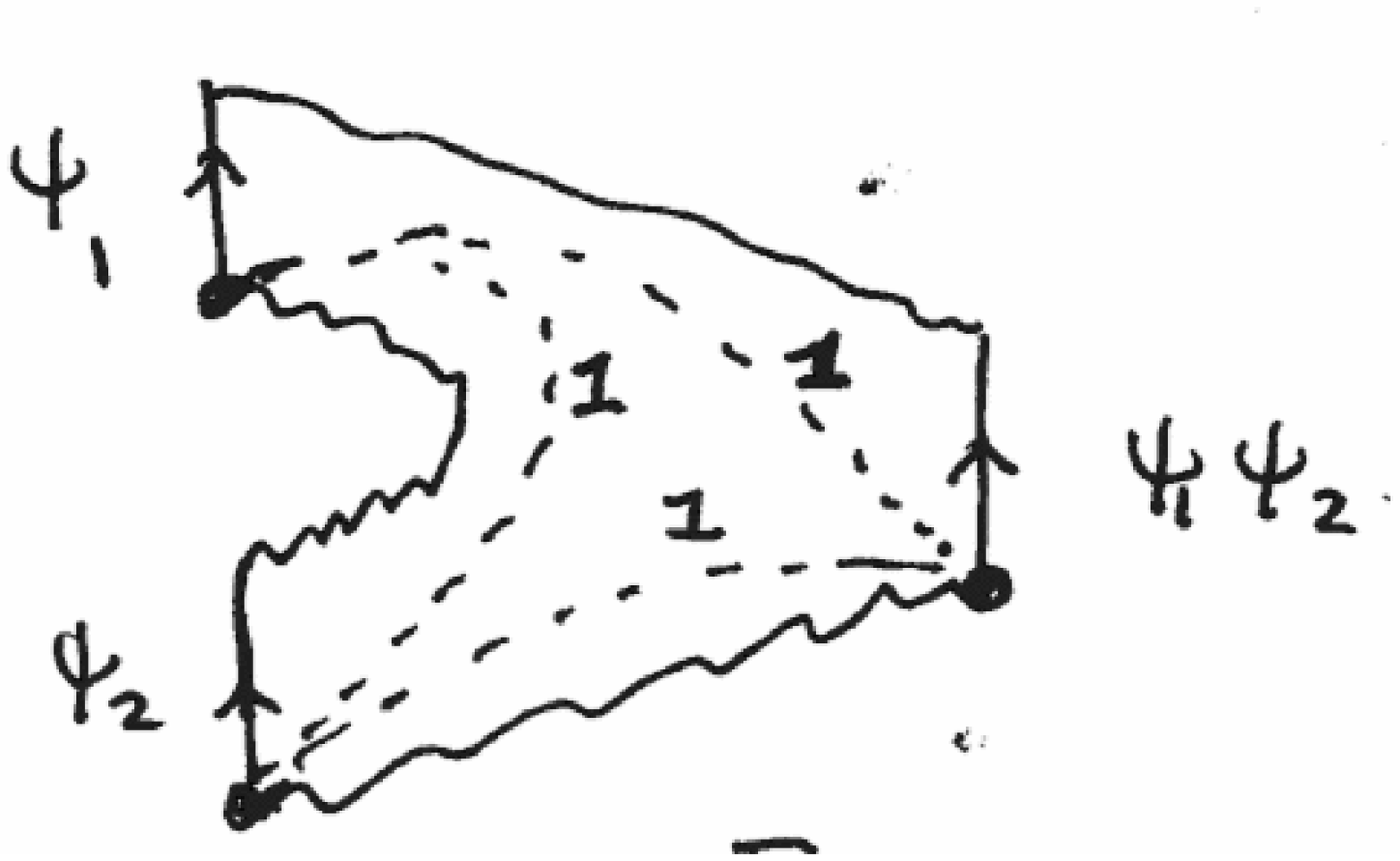}}

\ifig\FIGTWENTYFIVE{Showing that $G$ acts on $\CO$ as a group of
automorphisms. } {\epsfxsize2.0in\epsfbox{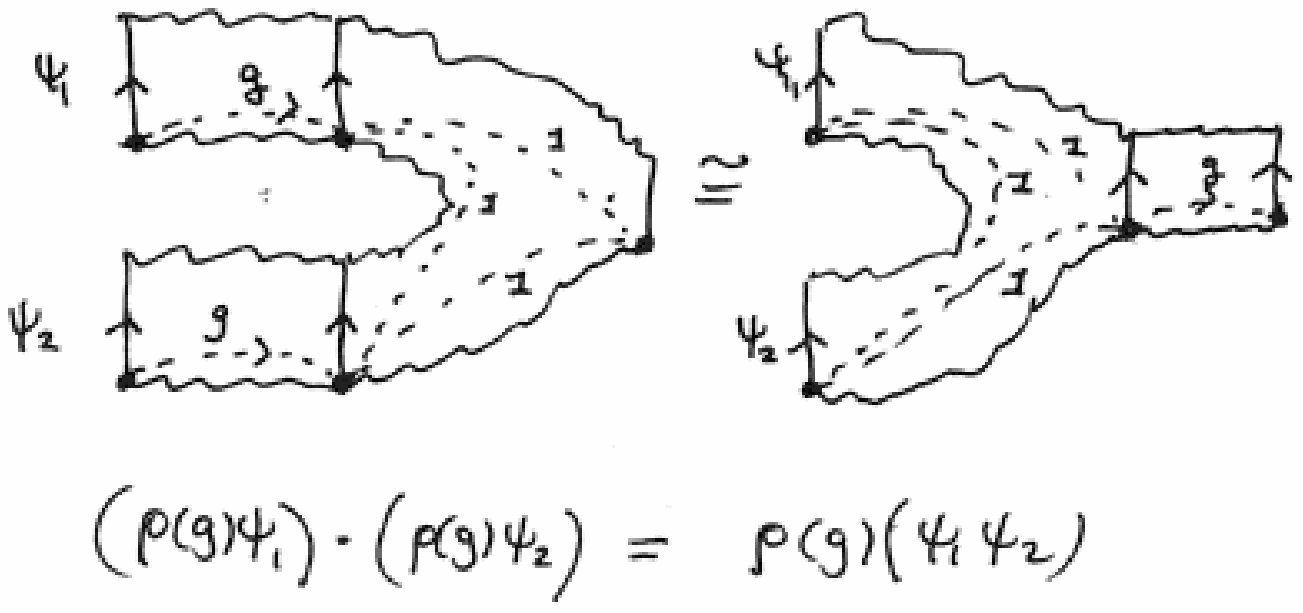}}

\ifig\FIGTWENTYSIX{The open/closed transitions $\iota_g$ and
$\iota^g$. } {\epsfxsize2.0in\epsfbox{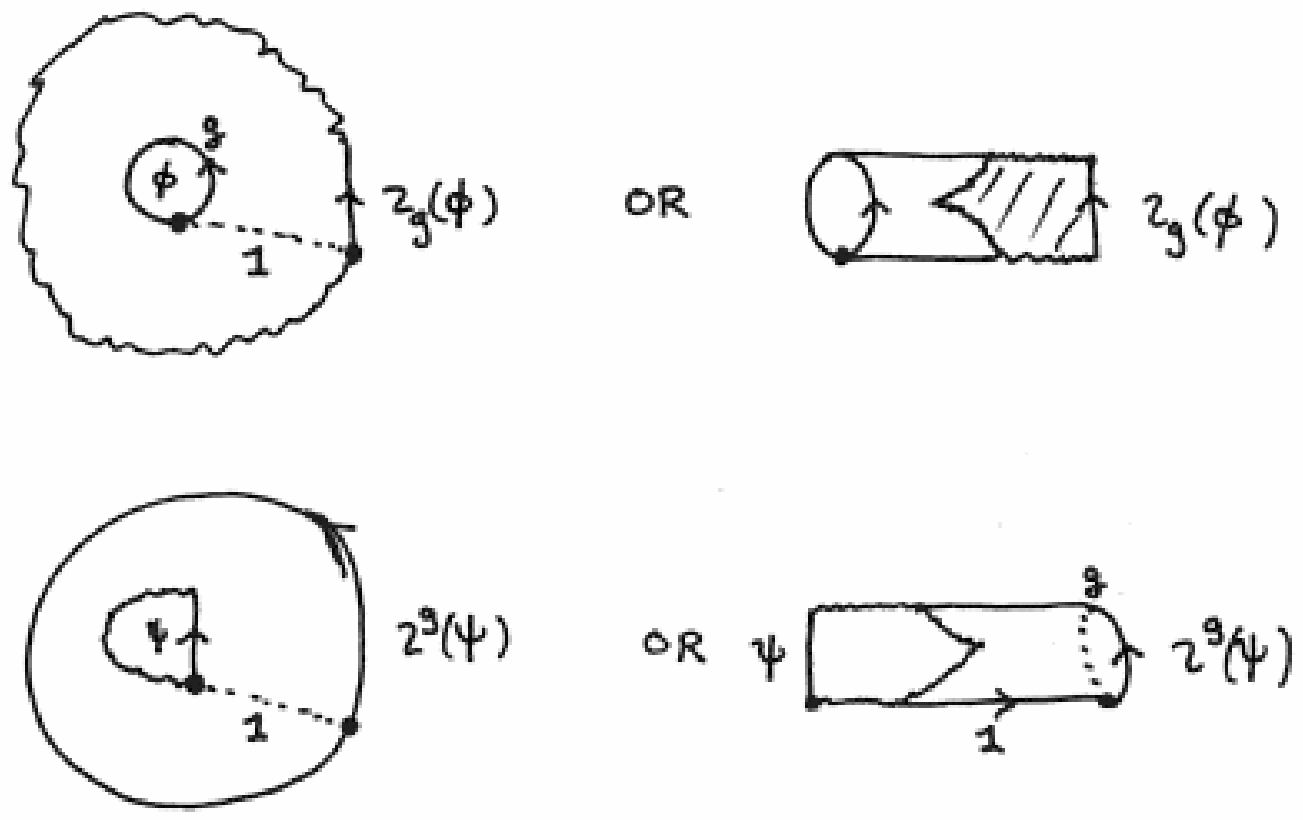}}

\ifig\FIGTWENTYSEVEN{The $G$-twisted centrality axiom. }
{\epsfxsize2.0in\epsfbox{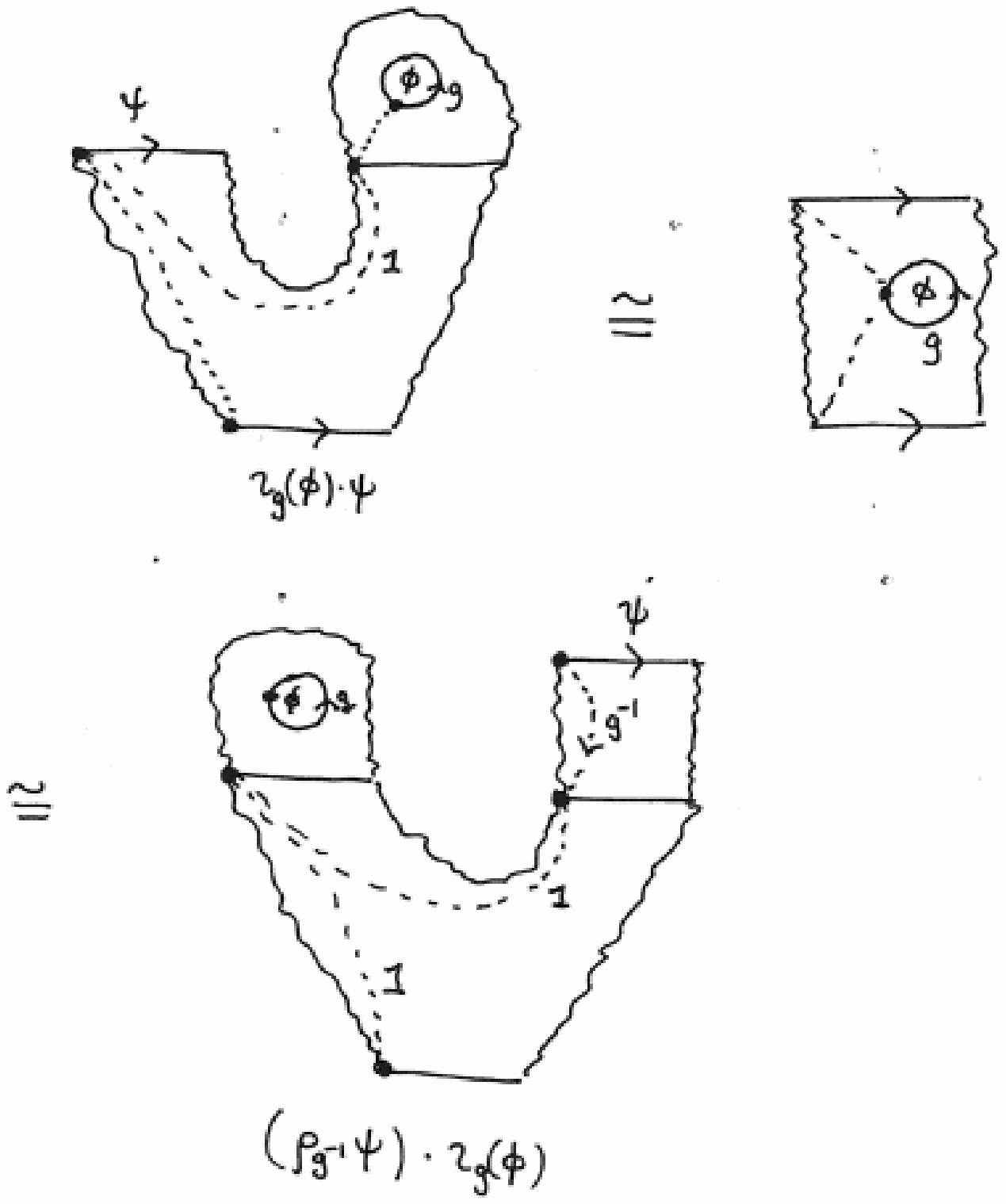}}

\ifig\FIGTWENTYEIGHT{The $G$-twisted adjoint relation. }
{\epsfxsize2.0in\epsfbox{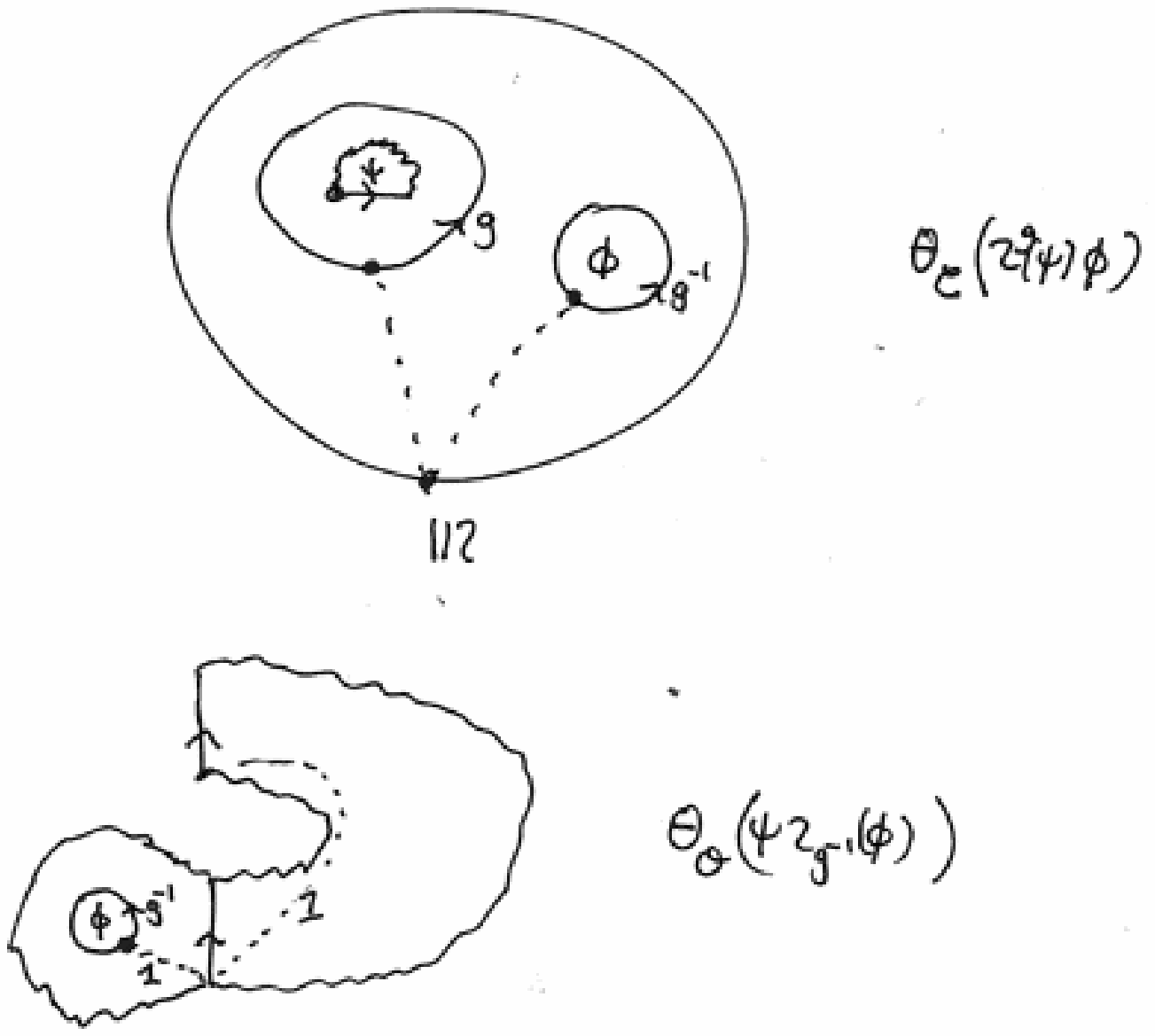}}

\ifig\FIGTWENTYNINE{The $G$-twisted Cardy condition. In the
double-twist diagram the holonomy around $\CP_1$ is $1$ and the
holonomy around $\CP_2$ is $g$. }
{\epsfxsize2.0in\epsfbox{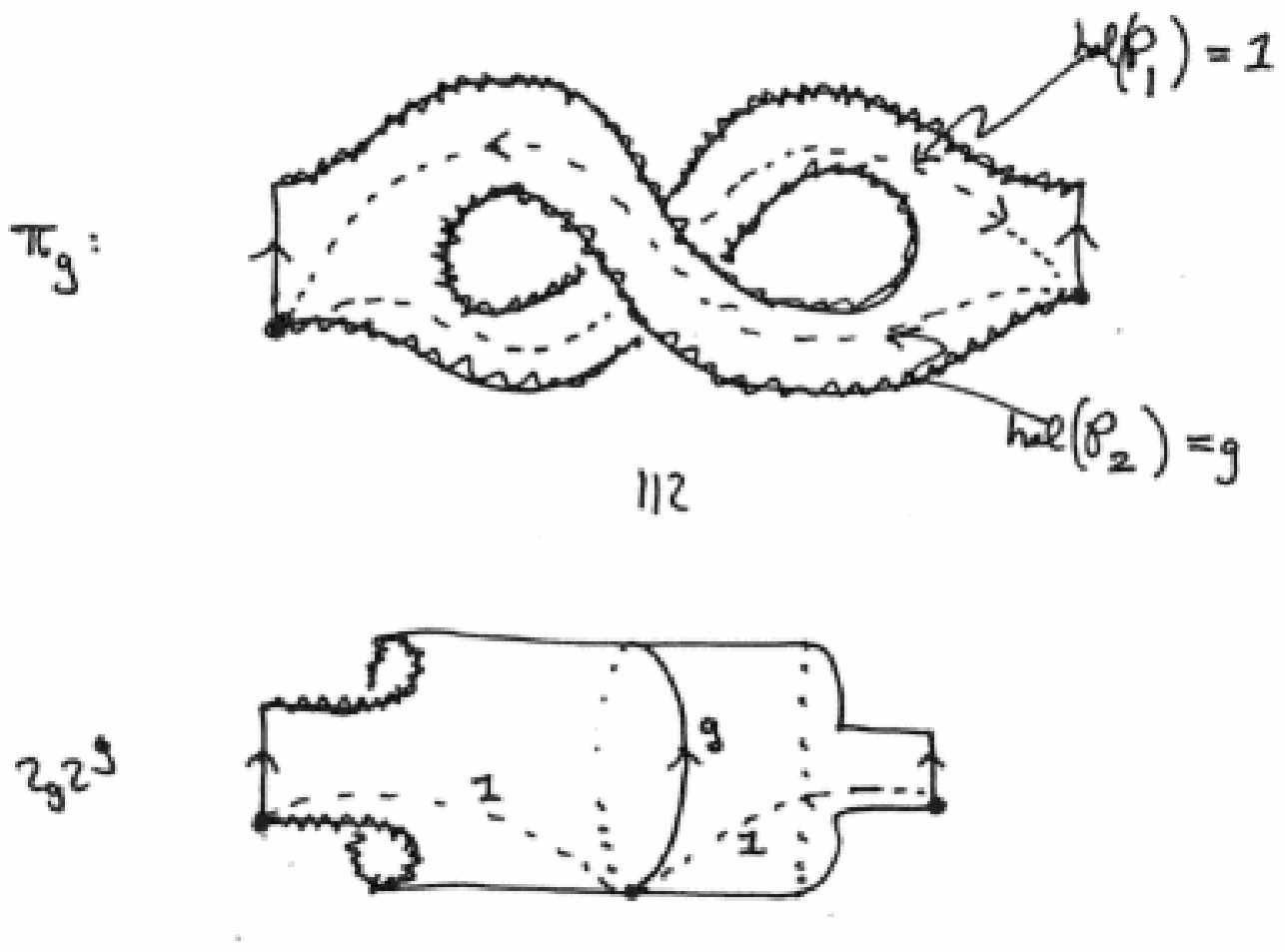}}

\subsec{Sewing conditions for equivariant open and closed theory}

Let us now pass on to consider $G$-equivariant open and closed
theories. We enlarge the category $\CS_G$ so that the objects are
oriented 1-manifolds w ith boundary, with labelled ends, equipped
with principal $G$-bundles. The morphisms are the same cobordisms as
in the non-equivariant case, but equipped with $G$-bundles. Up to
isomorphism there is only one $G$-bundle on the interval: it is
trivial, and admits $G$ as an automorphism group. So an equivariant
theory gives us for each pair $a,b$ of labels a vector space
$\CO_{ab}$ with a $G$-action. The action of $g\in G$ on $\CO_{ab}$
can be regarded as coming from the ``square" cobordism with the
bundle whose holonomy is $g$ along each of its ``constrained" edges.
There is also a composition law $\CO_{ab}\times \CO_{bc} \to
\CO_{ac}$, which is $G$-equivariant. These are illustrated in
\FIGTWENTYTHREE\ and \FIGTWENTYFOUR.

In the open/closed case the conditions analogous to equations
\assoc\ to \cardycon\ are the following.

Focussing first on a single label $a$, we
have a not necessarily commutative Frobenius algebra $(\CO_{aa}= \CO,\theta_{\CO})$
together with a $G$-action $\rho: G \to {\rm Aut}(\CO)$:
\eqn\opeqi{ \rho_g(\psi_1 \psi_2) = \bigl(\rho_g
\psi_1\bigr)\bigl(\rho_g \psi_2\bigr) }
which preserves the trace $\theta_{\CO}(\rho_g \psi) =
\theta_{\CO}(\psi)$. See \FIGTWENTYFIVE.

 There are also $G$-twisted
open/closed transition maps
\eqn\opeqii{
\eqalign{
\iota_{g,a}=\iota_g: \CC_g & \to \CO_{aa}=\CO \cr
\iota^{g,a}=\iota^g: \CO_{aa}=\CO & \to \CC_g \cr}
}
which are equivariant:
\eqn\opeqv{
\matrix{
\CC_{g_1} & ~ {\buildrel \alpha_{g_2} \over \rightarrow} ~ & \CC_{{g_2}g_1{g_2}^{-1}} \cr
 &   &  \cr
\downarrow \iota_{g_1} &  &\qquad  \downarrow \iota_{{g_2}g_1{g_2}^{-1}} \cr
\CO & ~ {\buildrel \rho_{g_2} \over \rightarrow} ~ & \CO \cr}
}
\eqn\opeqvi{
\matrix{
\CC_{{g_2}^{-1} g_1{g_2} } & ~ {\buildrel \alpha_{g_2} \over \rightarrow} ~ & \CC_{g_1} \cr
&  &  \cr
 \iota^{{g_2}^{-1} g_1 {g_2}} \uparrow\qquad   &  & \uparrow \iota^{g_1} \cr
\CO & ~ {\buildrel \rho_{g_2} \over \rightarrow} ~ & \CO \cr}
}
These maps are illustrated in \FIGTWENTYSIX.  The open/closed maps
must satisfy the $G$-twisted versions of conditions 3a-3e of section
2.2. In particular, the map $\iota :\CC \to \CO$ obtained by putting
the $\iota_g$ together is a ring homomorphism, i.e.

\eqn\opeqiii{
\iota_{g_1}(\phi_1) \iota_{g_2}(\phi_2) = \iota_{g_2 g_1}(\phi_2\phi_1) \qquad\qquad\forall \phi_1\in \CC_{g_1},
\phi_2\in \CC_{g_2}. }
Since the identity is in $\CC_1$ the condition \ident\ is unchanged.
The $G$-twisted centrality condition
is
\eqn\opeqiv{
\iota_g(\phi)(\rho_g \psi)) = \psi \iota_g(\phi) \qquad\qquad \forall \phi\in \CC_g, \psi\in \CO.
}
and is illustrated in \FIGTWENTYSEVEN.

The $G$-twisted adjoint condition is
\eqn\opeqvii{ \theta_{\CO}\bigl(\psi \iota_{g^{-1}}(\phi)\bigr) =
\theta_{\CC}\bigl(\iota^g(\psi)\phi)\bigr) \qquad\qquad \forall
\phi\in \CC_{g^{-1}.} }
and is shown in \FIGTWENTYEIGHT.

Finally, the $G$-twisted Cardy conditions
restrict not only each algebra $\CO_{aa}$, but
also the spaces of morphisms $\CO_{ab}$ between labels $b$ and $a$.
For each $g\in G$ we must have
\eqn\opeqviii{
\pi_{g,b}^{~~~a} = \iota_{g,b} \iota^{g,a}
}
Here $\pi_{g,b}^{~~~a}$ is defined by
\eqn\opeqix{
\pi_{g,b}^{~~~a}(\psi) = \sum_\mu \psi^\mu \psi  \bigl(\rho_g \psi_\mu\bigr),
}
where we sum over a basis $\psi_\mu$ for $\CO_{ab}$, and take
$\psi^\mu$ to be the dual basis of $\CO_{ba}$. See \FIGTWENTYNINE.

We may now formulate
\bigskip
\noindent {\bf Theorem 7} \ \ The above conditions form a complete
set of sewing conditions for $G$-equivariant open/closed 2d TFT.

\bigskip

This will be proved in the appendix. Note that the above axioms are
slightly redundant since \opeqv\ and \opeqvii\ together imply
\opeqvi.

\subsec{Solution of the sewing conditions for semisimple $\CC$ }

We now show that, when $\CC$ is semisimple, the
solutions of the above sewing conditions are
provided by $G$-equivariant bundles on $X = {\rm Spec}(\CC_1)$ twisted by
the $B$-field defined by $\CC$.

Let us first say a word about these bundles. To give a finite
dimensional representation of the cross-product algebra
$A(X)\sdtimes G$ is to give a representation of $A(X)$ --- i.e. a
vector bundle $E$ on $X$ --- together with an intertwining action of
$G$. Thus representations of $A(X)\sdtimes G$ are precisely
$G$-vector-bundles on $X$. For a finite group $G$ there are many
equivalent ways of defining the notion of a {\it twisted}
$G$-vector-bundle on $X$, twisted by a $B$-field $b$ representing an
element of $H^2_G(X;\IC^{\times})$: the simplest for our purposes is
to say that a twisted bundle is just a representation of
$A(X)\sdtimes_b G$. (Unfortunately this description does not work
when $G$ is not finite, and so it is not the one used in
\AtiyahSegal\ . We shall explain the relationship with the
description of \AtiyahSegal\  at the end of this subsection.)

The problem is easily reduced to consideration
of   a single $G$-orbit, so we may
assume $X=G/H$ for some
subgroup $H$ of $G$. Accordingly, the closed string
 Frobenius data is
specified by a 2-cocycle $b$
and  a single constant $\theta_c\in \IC^{\times}$ defining
the trace: $\theta(\ell_{g,x}) = \delta_{g,1} \theta_c$.
As usual, the isomorphism class only depends on
  $[b]\in H^2(H;\IC^{\times})$.

\bigskip
\noindent {\bf Theorem 8}\ \ \  Let $\CC = T(X,b,\theta_c)$ be a
Turaev algebra with $\CC_1$ semisimple and $X=G/H$. For a single
label $a$ the most general solution $\CO = \CO_{aa}$ of the sewing
constraints is determined by a  choice of square-root $\theta_o =
\sqrt{\theta_c}$ and a projective representation $V$ of $H$ with the
cocycle $b_o$ which is the restriction of $b$.

The algebra $\CO$ is the algebra of sections of
the $G$-equivariant  bundle of algebras over $X$:
\eqn\defoh{
\CO := \Gamma\bigl(G\times_H ({\rm End}(V))\bigr)
= {\rm Ind}_H^G\bigl({\rm End}(V)\bigr),
}
and the trace is determined by $\theta_o$:
\eqn\ohtrace{
\theta_{\CO}(\Psi) = \theta_o \sum_{x\in G/H} {\Tr}_V(\Psi(x)).
}

\bigskip
\noindent
{\it Proof}\ \

Let  us suppose that we have a
 Turaev algebra $\CC$ with $\CC_1$ semisimple, together with
$\CO,\theta_{\CO},  \iota_g, \iota^g$ satisfying the sewing
conditions. Let $X$ be the $G$-space ${\rm Spec}(\CC_1)$. Then, from
our  results in the non-equivariant case, we know that $\CO = {\rm
End}_{\CC_1}(\Gamma({\rm End}(E)))$ for some vector bundle $E\to X$,
unique up to tensoring wtih a line bundle $L \to X$. Thus $\CO
=\oplus\CO_x$, where $\CO_x ={\rm End}(E_x)$. We also know that the
trace on $\CO$ must be given by \fixes\ . The same square-root
$\theta_o$ of $\theta_x$ must be taken for each $x\in X$ to make
$\theta :\CO \to \IC$ invariant under $G$. Now $G$ acts compatibly
on $\CC_1$ and $\CO$ by algebra isomorphisms, so $g\in G$ maps
$\CO_{x_0}$ to $\CO_x$ by an algebra isomorphism. This proves
\defoh\ , where $V = E_{x_0}$. Finally, the Turaev algebra $\CC$ is
the product $\oplus\CC_x$, where $\CC_x$ is the twisted group-ring
of $G_x$ with the twisting $b_x$. The algebra homomorphism $\CC \to
\CO$ makes $\CC_x$ act on $E_x$, and so $V$ is a projective
representation of $H=G_{x_0}$ with the cocycle $b_{x_0}$.

This proves that $\CO$ is of the form stated.
One must still check
that the definition \defoh\ does provide a solution of the sewing
conditions, but that presents no problems.

\bigskip
\noindent {\it Remark}\ \ \ Although in the hypothesis of
the theorem
we were given a cocycle $b$  representing an element of
$H^2_G(X;\IC^{\times})$, the conclusion uses only its restriction
$b_{x_0}$. This should not surprise us, as cohomologous cocycles $b$
define isomorphic Turaev algebras, and $H^2_G(X;\IC^{\times})$ is
canonically isomorphic to the group cohomology $H^2_H({\rm
point};\IC^{\times})$ when $X=G/H$.

\bigskip

We can now deduce a complete description of the category of boundary
conditions, using exactly the same arguments by which we obtained
Theorem 3 from Theorem 2.

\bigskip

\noindent {\bf Theorem 9}\ \ \ If $\CC$ is a Turaev algebra with
$\CC_1$ semisimple, corresponding to a space-time $X$ with a
$B$-field $b$, then the category of boundary conditions for $\CC$ is
equivalent to the category of $b$-twisted $G$-vector-bundles on $X$,
uniquely up to tensoring with a $G$-line-bundle on $X$. Its
Frobenius structure is determined by a choice of the dilaton field
$\theta$.

\bigskip

The meaning of this theorem needs to be explained. The linear
category of equivariant boundary conditions for a given Turaev
algebra is an example of what is called an ``enriched" category: for
each pair of objects $a,b$ the vector space $\CO_{ab}$ has an action
of the group $G$. Now the category Vect$_G$ of finite dmensional
vector spaces with $G$-action is a symmetric tensor category, with
the neutral object $\IC$. To say that we have a category {\it
enriched} in a tensor category such as Vect$_G$ means that we have

(i) a set of objects,

(ii) for each pair  $a,b$ of objects an object $\CO_{ab}$ of
Vect$_G$, and

(iii) for each triple $a,b,c$ of objects an associative
``composition" morphism
$$\CO_{ab}\otimes \CO_{bc} \to \CO_{ac}$$ of $G$-vector spaces.

The axioms are almost identical to the axioms for a category, but
the space of morphisms has extra structure.  In such a  situation
the category is said to be an enrichment of the ordinary linear
category in which the morphisms from $b$ to $a$ are $F(\CO_{ab})$,
where $F:{\rm Vect}_G \to {\rm Vect}$ is the functor defined by
$F(V)={\rm Hom}_G(\IC;V) = V^G$. There is, however, another ordinary
category associated to the enriched category by simply forgetting
the $G$-action, so that the morphisms from $b$ to $a$ are simply
  $\CO_{ab}$ as a vector space.

An example of a category enriched in Vect$_G$ is the category of
finite dimensional representations of $\tilde G$, where $\tilde G$
is a central extension (with a fixed cocycle) of $G$ by the circle,
where the central circle acts by scalar multiplication. Indeed,
given two such representations $V_1^*\otimes V_2$ is a
representation of $G$.

Theorem 9 should really be expanded as follows. The category of
$b$-twisted $G$-vector-bundles on $X$ has a natural enrichment in
Vect$_G$, in which the $G$-vector space of morphisms consists of the
homomorphisms of $b$-twisted vector bundles which are not
necessarily equivariant with respect to the $G$-action. This
enrichment is equivalent to the category of equivariant boundary
conditions. The underlying ordinary category is the category of
boundary conditions for the orbifold theory.

\bigskip

Theorem 9 has a converse, which is the $G$-equivariant extension of
the discussion of \S3.3

\bigskip

{\bf Theorem 10} \ \ \ {\it If $\CB$ is a linear category enriched
in ${\rm Vect}_G$, with $G$-equivariant traces making it a Frobenius
category, and the linear category obtained from $\CB$ by forgetting
the $G$-action is semisimple with finitely many irreducible objects,
then $\CB$ is equivalent to the category of equivariant boundary
conditions for a canonical equivariant topological field theory. The
Turaev algebra defining the theory is $\oplus_g \CC_g$, where an
element of $\CC_g$ is a family $\phi_a \in \CO_{aa}$, indexed by the
objects $a$ of $\CB$, satisfying
\eqn\reconturaev{ \phi_a \circ f = (g.f) \circ \phi_b }
 for each $f
\in \CO_{ab}$.}

To prove this, one must show that \reconturaev\  really does define
a Turaev algebra. The details are straightforward and we will omit
them.

\subsec{Equivariant boundary states}

 To conclude our discussion, let
us consider the equivariant analogues of the ``boundary states"
discussed in \S 4. Our notion of the category of boundary conditions
for a $G$-gauged theory is intrinsically $G$-invariant, and we have
already pointed out that it gives us exactly the same category as we
would obtain from the orbifold theory in which we have summed over
the gauge fields. To reformulate this in terms of boundary states we
begin with the definition.

In the gauged theory associated to a Turaev algebra $\CC =
T(X,b,\theta)$ the observables at any point of the world-sheet are
precisely the elements of $\CC$. The boundary state $B_a \in \CC$
associated to a boundary condition $a$ is characterized  by the
property that the correlation function of observables $\phi_1,
\ldots , \phi_k$ evaluated at points of a surface $\Sigma$ with
boundary $S^1$ with the boundary condition $a$ (and arbitrary
holonomy around the boundary) is equal to that of the same
observables on the closed surface obtained by capping-off the
boundary, with the additional insertion of $B_a$ at the centre of
the cap. It suffices (because of the factorization properties of a
field theory) to check the case when $\Sigma$ is a disc. The
correlation function on the disc is obtained by propagating $\phi_1
\ldots \phi_k \in \CC_g$ to $\CH(\emptyset)= \IC$ by the annulus
whose non-incoming boundary circle is constrained by the condition
$a$, along which the holonomy is necessarily $g$. Our rules tell us
that the result is
$$\theta_{\CO_{aa}}(\iota _{g,a}(\phi_1\ldots \phi_k)).$$
Equating this to $\theta_{\CC_1}(\phi_1 \ldots \phi_k B_a)$, we see
that
$$B_a = \sum_g \iota_{g,a}(1).$$

The map $a \mapsto B_a$ evidently has its
image in the $G$-invariant
part --- i.e. the centre --- of the Turaev algebra. It extends to a
homomorphism
$$K_{G,h}(X) \to T(X,b,\theta)^G,$$
and we have

\bigskip
{\bf Theorem 11} The $G$-invariant boundary states generate a
lattice in $T(X,b,\theta)^G$ related to the twisted equivariant
$K$-theory via:
\eqn\turkay{
 K_{G,h}(X)\otimes \IC = T(X,b,\theta)^G.
}

\bigskip

{\bf Remark}

\item{1.} Equation \turkay\ is related to an old observation of \dvvv.
If $X=G$ with
$G$ acting on itself by conjugation then $T(X,0)^G$ is the Verlinde algebra
occuring in the conformal field theory of orbifolds for chiral
algebras with one representation \dvvv. The different orbits
are the conjugacy classes of $G$. Focusing on one conjugacy
class $[g]$ we can compare with the above results. One basis of states
is provided by a choice of a character of the centralizer of
$g$. These are just the $G$-invariant boundary states found above.

\item{2.} The translation of the above
results to the language of branes at orbifolds is the following. The
boundary states corresponding to the different $b$-irreps $V_i$ are
the ``fractional branes'' of \dm.  The use of projective
representations was proposed in \dm, and further explored in
\douglas. A different proof of the fact that the cocycle for the
open sector  and that of the closed sector $b$ are cohomologous can
be found in \aspinwall.

\bigskip

To conclude this section, let us return to  explain the relation
between the definition of twisted equivariant $K$-theory by
$A(X)\sdtimes _b G$-modules and the definition given in
\AtiyahSegal\ .

In \AtiyahSegal\  the elements of  the twisted equivariant theory
are described as follows. First, the twisting class $b \in
H^3_G(X;\IZ)$ is represented by a bundle $P$ of projective Hilbert
spaces on $X$ equipped with a $G$-action covering the $G$-action on
$X$. Then elements of $K_{G,P}(X)$ are represented by families
$\{T_x\}_{x\in X}$ of fibrewise Fredholm operators in the bundle
$P$. Let us show how to associate such a pair $(P,\{T_x\})$ to a
finitely generated $A(X)\sdtimes _b G$- module. Such a module is the
same thing as a finitely generated $A(X)$-module equipped with a
compatible action of the extended group $\tilde G$ associated to $b$
which we have already described. Equivalently, it is a finite
dimensional vector bundle $E$ on $X$ with an action of $\tilde G$ on
the total space which covers the action of $G$ on $X$. Let us choose
a fixed infinite dimensional Hilbert space $\CH$. Then $\hat E = E
\otimes \CH$ is a Hilbert bundle on $X$, and the associated bundle
$P = \IP (\hat E)$ of projective spaces has a natural  action of
$G$, and it represents the class of $b$ in $H^3_G(X;\IZ)$. (Cf. the
proof of Proposition 6.3 in \AtiyahSegal\ .) If $T:\CH \to \CH$ is a
fixed surjective Fredholm operator with a one-dimensional kernel,
then (identity)$_E \otimes T:P \to P$ represents an element of
$K_{G,P}(X)$ according to the definition of \AtiyahSegal\ .

If the cocycle $b$ is a coboundary --- or even
if $b_x(g_1,g_2) $ is
independent of $x \in X$ --- it is plain that the two rival
definitions of equivariant $K$ coincide. A Mayer-Vietoris argument
can then be used to show that they coincide for all $b$.

The essential point here is that when $X$ and $G$
are finite the
twisting class $b$ is of finite order, and that makes it possible to
represent the $K$-classes by families of Fredholm operators of
constant rank, and hence by finite dimensional vector bundles.

\appendix{A}{Morse theory proof of the sewing theorems}

In this appendix we shall use Morse theory to give uniform proofs of
four theorems. The first is the very well-known result that a
two-dimensional topological field theory is precisely encoded in a
commutative Frobenius algebra. The second is the corresponding
statement for open and closed theories: this is Theorem 1 of Section
2 above. The third and fourth are the equivariant analogues of the
first two, i.e. Theorems 4 and 7 of Section 7.

\bigskip

\subsec{The classical theorem}

We wish to prove that when we have a commutative Frobenius algebra
$\CC$ we can assign to an oriented cobordism $\Sigma$ from $S_0$ to
$S_1$ a linear map
$$U_{\Sigma}:\CC^{\otimes p}\to \CC^{\otimes q},$$
where the oriented 1-manifolds $S_0$ and $S_1$ have $p$ and $q$
connected components respectively.

We can always choose a smooth function $f:\Sigma \to [0,1] \subset
\IR$ such that $f^{-1}(0)=S_0$ and $f^{-1}(1)=S_1$, and which has
only ``Morse" singularities, i.e. the gradient $df$ vanishes at only
finitely many points $x_1,\ldots,x_n \in \Sigma$, and

(i) \ \ the Hessian $d^2f(x_i)$ is a non-degenerate quadratic form
for each $i$, and

(ii) \ \ the critical values $c_1=f(x_1),\ldots ,c_n=f(x_n)$ are
distinct, and not equal to 0 or 1.

\noindent Each critical point has an {\it index}, equal to 0, 1, or
2, which is the number of negative eigenvalues of the Hessian
$d^2f(x_i)$.

The choice of the function $f$ gives us a decomposition of the
cobordism into ``elementary" cobordisms. If
$$0=t_0<c_1 <t_1 <c_2 <t_2< \ldots <c_n <t_n = 1,$$
and $S_t=f^{-1}(t)$, then each $S_{t_i}$ is a collection of, say,
$m_i$ disjoint circles, with $m_i = m_{i-1}\pm 1$, and $\Sigma_i =
f^{-1}([t_{i-1},t_i])$ is a cobordism from $S_{t_{i-1}}$ to
$S_{t_i}$ which is trivial (i.e. a union of cylinders) except for
one connected component of one of the four forms of \FIGTWO.

\noindent For a given Frobenius algebra $\CC$ we know how to define
an operator
$$U_{\Sigma_i}:\CC^{\otimes m_{i-1}}\to \CC^{\otimes m_i}$$
 in each case. (In the third case the map we assign is

$$\phi  \mapsto \sum \phi \phi_i \otimes \phi^i ,$$
where $\{\phi_i\}$ and $\{\phi^i\}$ are dual bases of $\CC$ such
that $\theta_{\CC}(\phi^i \phi_j)=\delta_{ij}$.) We should notice
two points. First, we need $\CC$ to be commutative, for otherwise we
would need to have an {\it order} on the two incoming circles of a
pair of pants, and no such order is given. Secondly, the assignments
we make have the property that reversing the direction of time in a
cobordism replaces the operator by its adjoint with respect to the
Frobenius inner product on the state-spaces. This property will be a
firm principle in all our constructions, and it reduces the number
of cases we have to check in the tedious arguments below.

\ifig\FIGTHIRTY{  } {\epsfxsize2.0in\epsfbox{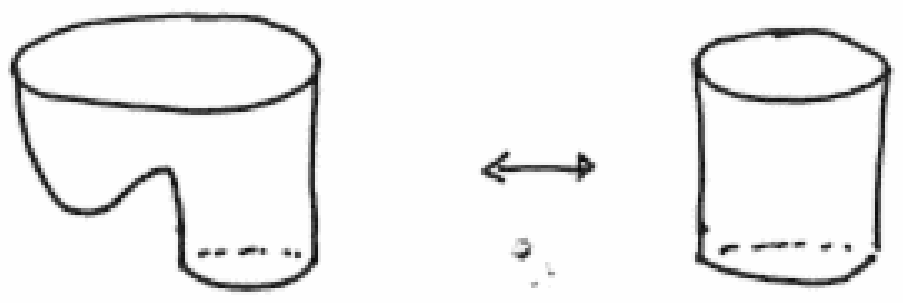}}

 The important task  now is
to show that the composite operator $U_{\Sigma_n}\circ \cdots \circ
U_{\Sigma_1}$ is independent of the chosen Morse function $f$.

Two Morse functions $f_0$ and $f_1$ can always be connected by a
smooth path $\{f_s\}_{0\leq s \leq 1}$ in which $f_s$ is a Morse
function except for a finite set of parameter values $s$ at which
one of the following two things happens:

(i) \ \ $f_s$ has one degenerate critical point where in local
coordinates $(u,v)$ it has the form $f_s(u,v)=\pm u^2 +v^3$, or

(ii) \ \ two distinct critical points $x_i,x_j$ of $f_s$ have the
same critical value $f_s(x_i)=f_s(x_j)=c.$

\noindent In the first case, two critical points of adjacent indices
are created or annihilated as the parameter  passes through the
non-Morse value $s$, and the cobordism changes by \FIGTHIRTY.

\noindent or vice-versa, or by the time-reversal of these pictures.
The well-definedness of $U_{\Sigma}$ under this kind of change is
ensured by the identity $1.a=a$ in the algebra $\CC$.

Case (ii) is more problematical. Because operators of the form $U
\otimes 1$ and $ 1\otimes U'$ commute, we easily see that there is
nothing to prove unless the two critical points $x_i$ and $x_j$ are
connected in the ``bad" critical contour $S_c$, in which case they
must both have index 1.

 Let us consider the resulting two-step cobordism which is factorized in
different ways before and after the critical parameter value $s$. It
will have just one non-trivial connected component, which, because
an elementary cobordism changes the number of circles by 1, must be
a cobordism from $p$ circles to $q$ circles, where
$(p,q)=(1,1),(2,2),(1,3)$ or $(3,1)$. We need to check only one of
(1,3) and (3,1), as they differ only by time-reversal. Because the
Euler number of a cobordism is the number of critical points of its
Morse function (counted with the sign $(-1)^{\rm index}$), the
non-trivial component has Euler number $-2$, so is a 2-holed torus
when $(p,q)=(1,1)$ and a 4-holed sphere in the other cases.

\ifig\FIGTHIRTYONE{  } {\epsfxsize2.0in\epsfbox{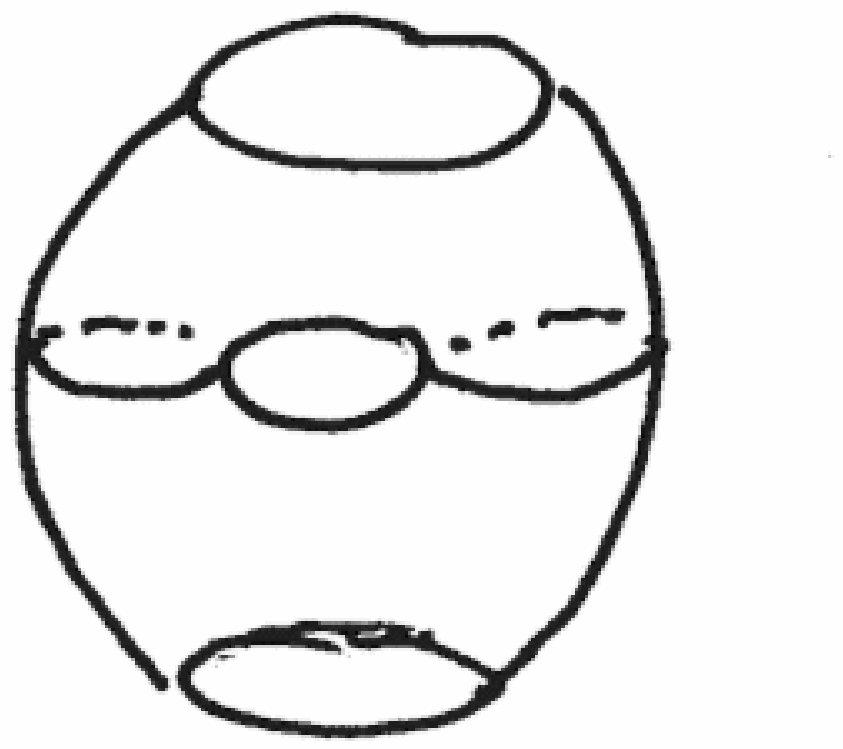}}

\ifig\FIGTHIRTYTWO{  } {\epsfxsize2.0in\epsfbox{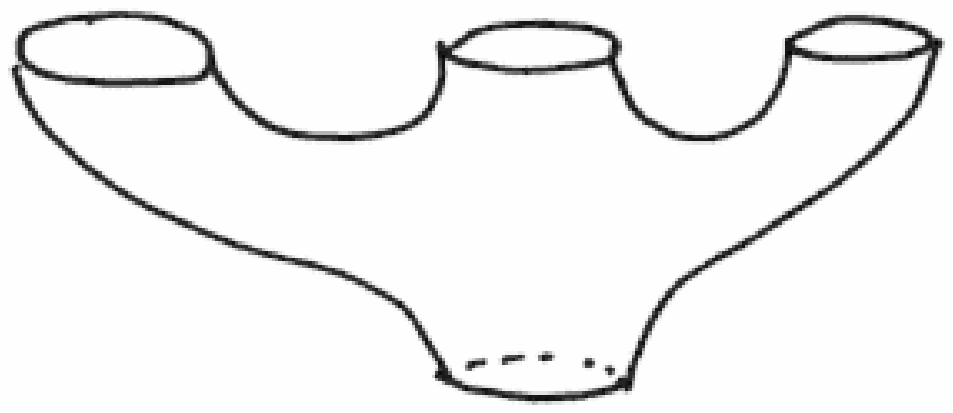}}

\ifig\FIGTHIRTYTHREE{  } {\epsfxsize2.0in\epsfbox{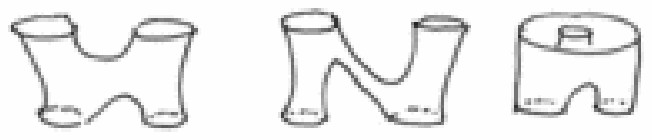}}

\ifig\FIGTHIRTYFOUR{  } {\epsfxsize2.0in\epsfbox{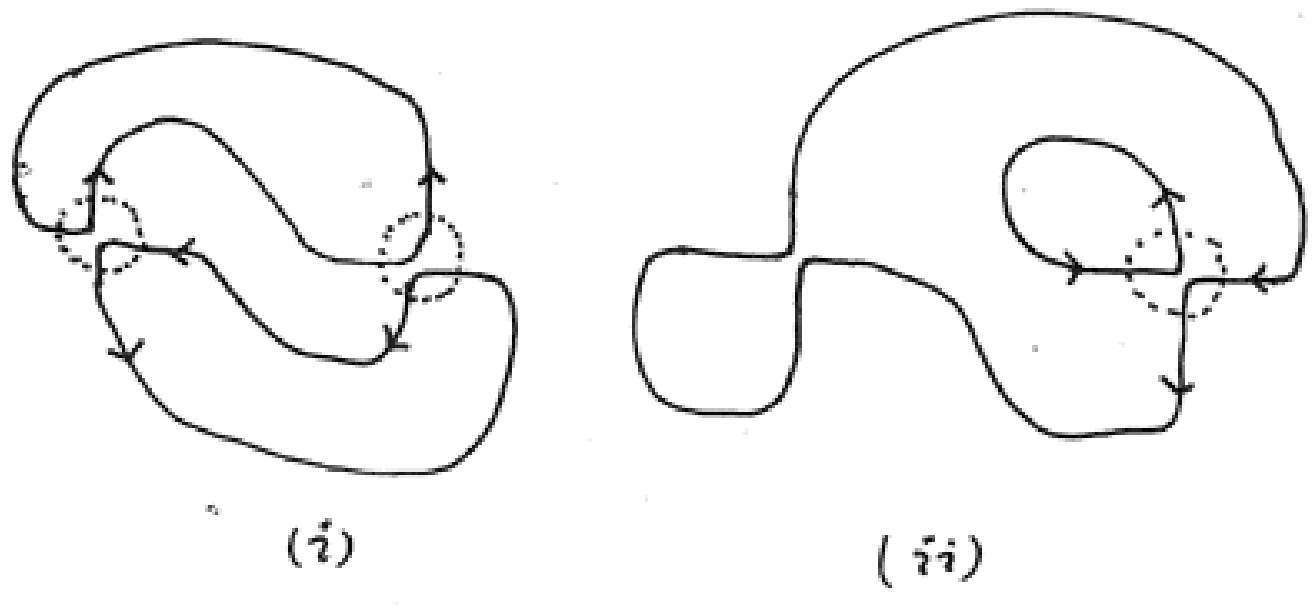}}

 In the
case (1,1), depicted in \FIGTHIRTYONE, a circle splits into two
which then recombine. There is nothing to check, because, though a
torus with two holes can be cut into two pairs of pants by many
different isotopy classes of cuts, there is only one possible
composite cobordism, and we have only one possible composite map
$\CC \to \CC\otimes \CC \to \CC.$

In the case (3,1),  two circles of the three combine, then the
resulting circle combines with the third. The picture is
\FIGTHIRTYTWO. Clearly this case is covered by the associative law
in $\CC$.

In the case (2,2) we are again factorizing a 4-holed sphere into two
elementary cobordisms. This can be done in many ways, as we see from
the pictures \FIGTHIRTYTHREE.   The best way of making sure we are
not overlooking any possibility is to think of the contour just
below the doubly-critical level, which, if it consists of two
circles, must have one of the two forms $(i)$ or $(ii)$ in
\FIGTHIRTYFOUR. (Consider the possible ways of connecting the
``terminals'' inside the dotted circles.) But, whatever happens, the
only algebraic maps the cobordism can lead to are

$$\CC \otimes \CC \to \CC \to \CC \otimes \CC$$
and
$$\CC \otimes \CC \to \CC \otimes \CC \otimes \CC \to \CC \otimes \CC,$$
given by
$$\phi \otimes \phi ' \mapsto \phi\phi ' \mapsto \sum \phi\phi '\phi_i
\otimes \phi^i$$ and
$$\phi \otimes \phi' \mapsto \sum \phi \phi_i \otimes \phi^i \otimes
\phi' \mapsto \sum \phi \phi_i \otimes \phi^i \phi'$$ respectively,
where $\{\phi_i\}$ and $\{\phi^i\}$ are dual bases of $\CC$ such
that $\theta_{\CC}(\phi^i \phi_j)=\delta_{ij}$. These two maps are
equal because of the identity
\eqn\twototwo{ \sum \phi'\phi_i \otimes \phi^i = \sum \phi_i \otimes
\phi^i\phi', }
which holds in any Frobenius algebra because the inner product of
each side with $\phi^j\otimes \phi_k$ is
$\theta_{\CC}(\phi^j\phi'\phi_k).$

That completes the proof of the theorem: notice that we have used
all the axioms of a commutative Frobenius algebra.

\subsec{Open and closed theories}

As in the preceding argument we consider a cobordism $\Sigma$ from
$S_0$ to $S_1$, but now $S_0$ and $S_1$ are collections of circles
and intervals, and the boundary $\partial\Sigma$ has a constrained
part $\partial_{\rm constr}\Sigma$, which we shall abbreviate to
$\partial'\Sigma$, which is a cobordism from $\partial S_0$ to
$\partial S_1$. We choose $f:\Sigma \to [0,1]$ as before, but now
there are two kinds of critical points of $f$: interior points of
$\Sigma$ at which the gradient $df$ vanishes, and  points of
$\partial'\Sigma$ at which the gradient of the restriction of $f$ to
the boundary vanishes. For an internal critical point,
``nondegenerate" has its usual meaning. A critical point $x$ on the
boundary is called nondegenerate if it is a nondegenerate critical
point of the restriction of $f$ to $\partial'\Sigma$, and in
addition the derivative of $f$ normal to the boundary does not
vanish at $x$.

As before, $f$ is a {\it Morse function} if all its critical points
are non-degenerate, and all the  critical values are distinct and
$\neq 0,1$. We can  always choose such a function.

There are now four kind of boundary critical points, which we can
denote $0\pm,1\pm$, recording the index and the sign of the normal
derivative. Six things can happen as we pass through one of them. At
those of type $0+$ or $1-$, an open string is created or
annihilated. At type $0-$ either two open strings join end-to-end,
or else an open string becomes a closed string. Type $1+$ is the
time-reverse of $0-$. If we have a Frobenius catgegory $\CB$, we
know what to do in each of the six cases.

\ifig\FIGTHIRTYFIVE{  } {\epsfxsize2.0in\epsfbox{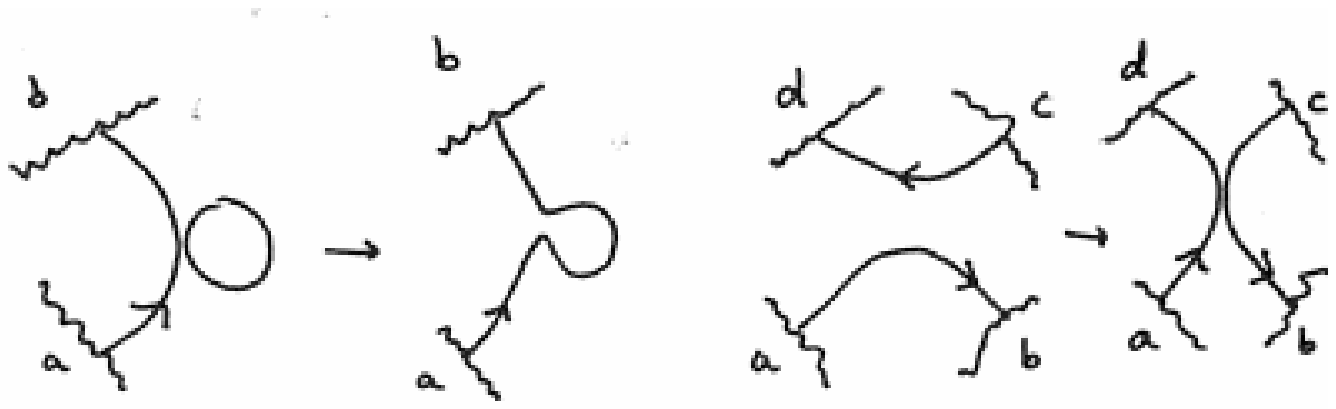}}

An internal critical point has index 0,1, or 2, as before. Only if
the index is 1 can the corresponding cobordism involve an open
string. Up to time reversal, there are three index 1 processes: two
closed strings can become one, an open string can ``absorb" a closed
string, and two open strings can ``reorganize themselves" to form
two new open strings as in \FIGTHIRTYFIVE.

 For a given Frobenius
category $\CB$, we assign to (open)+(closed)$\to$(open) the map
$$\CO_{ab}\otimes \CC \to \CO_{ab}$$
given by $\phi \otimes \psi \mapsto \phi\psi.$ \ (Here, as we
usually do, we are regarding $\CO_{ab}$ as a $\CC$-module, writing
$$\phi\psi =\iota_a(\phi)\psi = \iota_b(\phi)\psi.)$$
To (open)+(open)$\to$(open)+(open) we assign the map
$$\CO_{ab}\otimes \CO_{cd}\to\CO_{ad}\otimes\CO_{cb}$$
given by
$$\psi\otimes\psi'\mapsto \sum \psi\psi_i \otimes \psi' \psi^i,$$
where $\psi_i$ and $\psi^i$ are dual bases of $\CO_{bd}$ and
$\CO_{db}$.

We must now consider what happens when we change the Morse function.
As before, two Morse functions can be connected by a path $\{f_s\}$
in which each $f_s$ is a Morse function except for finitely many
values of $s$ at which either one critical point is degenerate or
else two critical values coincide. We begin with the degenerate
case. There are now three kinds of degeneracy which we must allow,
for besides internal degeneracies which are just as in the closed
string case we can have two kinds of degeneracy on the boundary:
either $f\vert \partial'\Sigma$ has a cubic inflexion, or else the
normal derivative vanishes at a boundary critical point.

\ifig\FIGTHIRTYSIX{  } {\epsfxsize2.0in\epsfbox{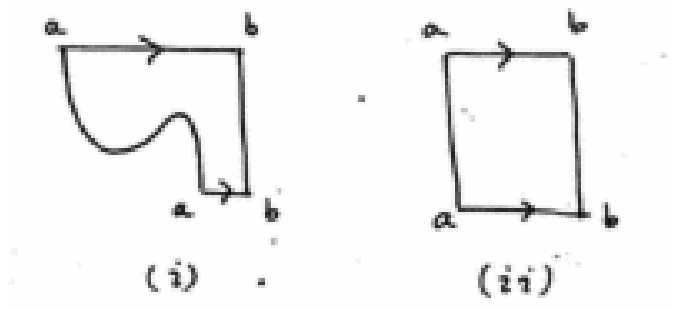}}

When $s$ passes through a boundary inflexion, two nondegenerate
boundary critical points of opposite index but with normal
derivatives of the same sign are created or annihilated. This means
that the cobordism changes between figures $(i)$ and $(ii)$ of
\FIGTHIRTYSIX\ (or the time-reversal). These changes are covered by
the axiom that the category $\CB$ has identity morphisms.

When the normal derivative vanishes at a boundary critical point
what happens is that an internal critical point has moved ``across
the boundary of $\Sigma$, i.e. it moves into coincidence with a
boundary critical point and changes the sign of the normal
derivative there. There are four cases:
$$(0-) \ + \ ({\rm index} \  0) \ \to \ (0+),$$
$$(0+) \ + \ ({\rm index} \  1) \ \to \ (0-),$$
and the time-reversals of these. In the first case, the composite
cobordism in which a small closed string is created and then breaks
open is replaced by the elementary cobordism in which an open string
is created. This corresponds to the axiom that $\CC \to \CO_{aa}$
takes $1_{\CC}$ to $1_a$. In the second case, in the composite
cobordism, an open string is created, and then it either ``absorbs"
an existing closed string or else ``rearranges" itself with an
existing open string; these composites are to be equivalent,
respectively, to the elementary breaking of a closed or open string.
Putting $\psi = 1_a$ in the formulae above we see that this is
allowed by the Frobenius category axioms.

When we have an internal degenerate critical point, what happens, up
to time-reversal, is that a closed string is created and then joins
an existing open or closed string; this should be the same as the
trivial cobordism. Again, the unit axioms cover this.

\ifig\FIGTHIRTYSEVEN{  } {\epsfxsize2.0in\epsfbox{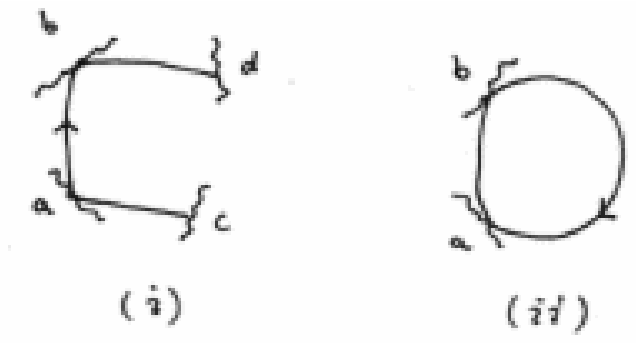}}

\ifig\FIGTHIRTYEIGHT{  } {\epsfxsize2.0in\epsfbox{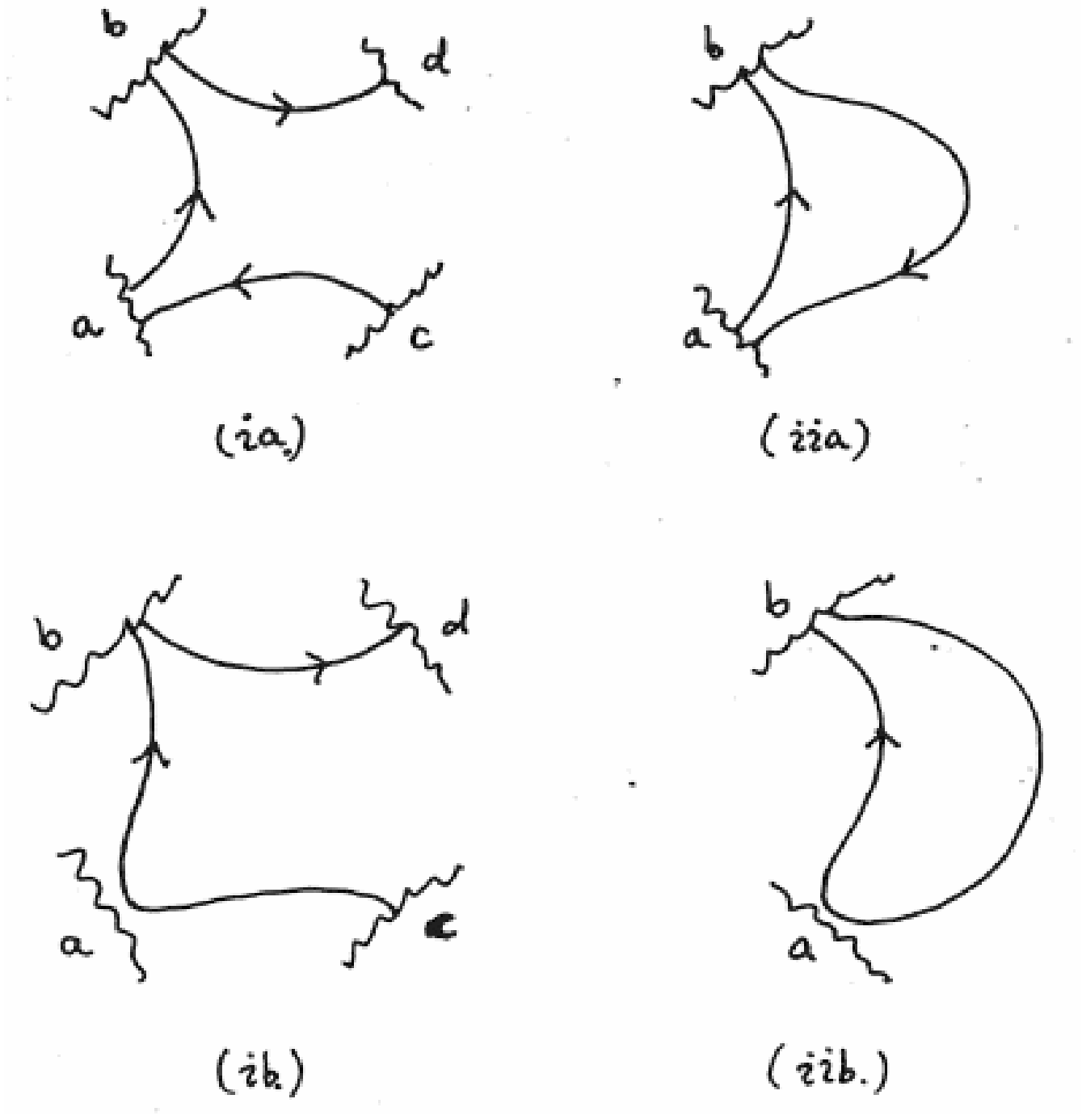}}

Finally, we have to consider what happens when two critical values
cross. They can be two boundary critical points, two internal ones,
or one of each.

If two boundary critical points are linked by a critical contour, it
has the form \FIGTHIRTYSEVEN.   These give us four cases to check,
where the contour below the critical level is \FIGTHIRTYEIGHT.

\noindent Case $(i)_a$ is accounted for by the associativity of
composition in the category $\CB$; case $(i)_b$ by the open-string
analogue of the identity \twototwo; case $(ii)_a$ by the trace axiom
$\iota^a(\psi_1\psi_2)=\iota^b(\psi_2\psi_1)$, which follows by
combining (cyclic),(center),and (adjoint); and case $(ii)_b$ by the
Cardy identity.

\ifig\FIGTHIRTYNINE{  } {\epsfxsize2.0in\epsfbox{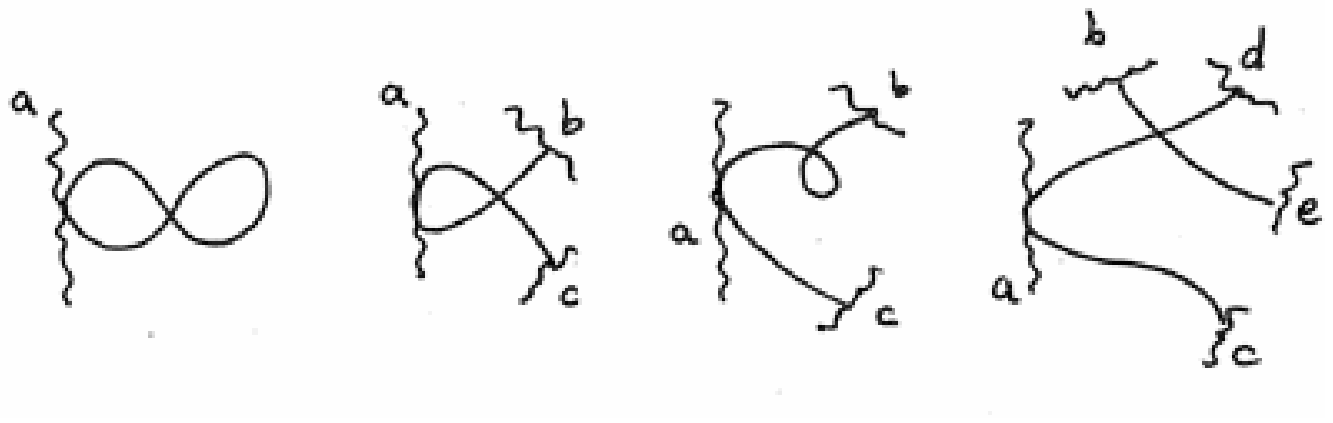}}

When we have one boundary and one internal critical point at the
same level we may as well assume the boundary point is of type $0-$
and the internal critical point is of index 1, and that they are
joined in the critical contour,which must have one of the four forms
\FIGTHIRTYNINE.

  At the boundary point either an open string becomes
closed, or else two open strings join. We shall consider each
possibility in turn. In the first case, if the boundary point is
encountered first, then at the interior point three things can
happen: the closed string can split into two closed strings, or it
can combine with another closed or open string. Thus the
possibilities are

$$o \ \to \ c \ \to \ c \ + \ c$$
$$o \ +  \ c \ \to \ c \ + \ c \ \to \ c$$
$$o \ + \ o \  \to \ c \ + \ o \  \to \ o .$$

\noindent When the internal point is encountered first there is only
one possibility in each case, and the three sequences are replaced
respectively by

$$o \ \to \ o \ + \ c \ \to \ c \ + \ c$$
$$o \ + \ c \ \to \ o \ \to \ c$$
$$o\ \ + \ o \ \to \ o \ + \ o \ \to \ o. $$

\noindent We have to check three identities. The first two reduce to
the fact that $\iota^a:\CO_{aa} \to \CC$ is a map of modules over
$\CC$. The third is the Cardy condition.

Now let us consider the case where two open strings join at the
boundary critical point. If we meet the boundary point first,  there
are again three things that can happen at the internal critical
point: the open string can emit a closed string, or else it can
interact with another closed or open string. The possibilities are

$$o \ + \ o \ \to  \ o \ \to \ o \ + \ c$$
$$o \ +  \ o \ + \ c \ \to \ o \ \ + \ c \ \to \ o$$
$$o \ + \ o \ + \ o \ \to \ o \ + \ o \ \to \ o \ + \ o.$$

\noindent In the second and third of these cases there is only one
thing that can happen when the order of the critical points is
reversed: they become

$$o \ + \ o \ + \ c \ \to \ o \ + \ o \ \to \ o$$
$$o \ + \ o \ + \ o \ \to \ o \ + \ o \ + \ o \ \to \ o \ + \ o.$$

\noindent The identities relating the corresponding algebraic maps
$\CO_{ab} \otimes \CO_{bc} \otimes \CC \to \CO_{ac}$ and $\CO_{ab}
\otimes \CO_{bc} \otimes \CO_{de} \to \CO_{ae} \otimes \CO_{dc}$ are
immediate.

\noindent The first sequence, however, can become either
$$o \ + \ o \ \to \ o \ + \ o \ + \ c \ \to \ o \ + \ c$$
or
$$o \ + \ o \ \to \ o \ + \ o \ \to \ o \ + \ c. $$
The first of these presents nothing of interest algebraically, but
to deal with the second we need to check that
$$\sum \psi\psi'\phi_i \otimes \phi^i = \sum \psi\psi^k \otimes
\iota^b(\psi'\psi_k)$$ for $\psi \in \CO_{ab}$,\ \ $\psi' \in
\CO_{bc}$, and dual bases $\phi^i,\phi_i$ of $\CC$ and
$\psi^k,\psi_k$ of $\CO_{bc},\CO_{cb}$. This relation holds because
the inner product of the left-hand side with $\psi_m \otimes \phi_j$
is $\theta_b(\psi\psi'\phi_j\psi_m)$, while the inner product of the
right-hand side with $\psi_m \otimes \phi_j$ is
\eqn\iplhs{
\eqalign{\sum_k\theta_b(\psi\psi^k\psi_m)\theta(\iota^b(\psi'\psi_k)\phi_j)
& =  \sum_k\theta_b(\psi_m\psi\psi^k)\theta_b(\psi_k\phi_j\psi')\cr
& =
\theta_b(\psi_m\psi\phi_j\psi')=\theta_b(\psi\psi'\phi_j\psi_m).\cr}
}

\ifig\FIGFORTY{  } {\epsfxsize2.0in\epsfbox{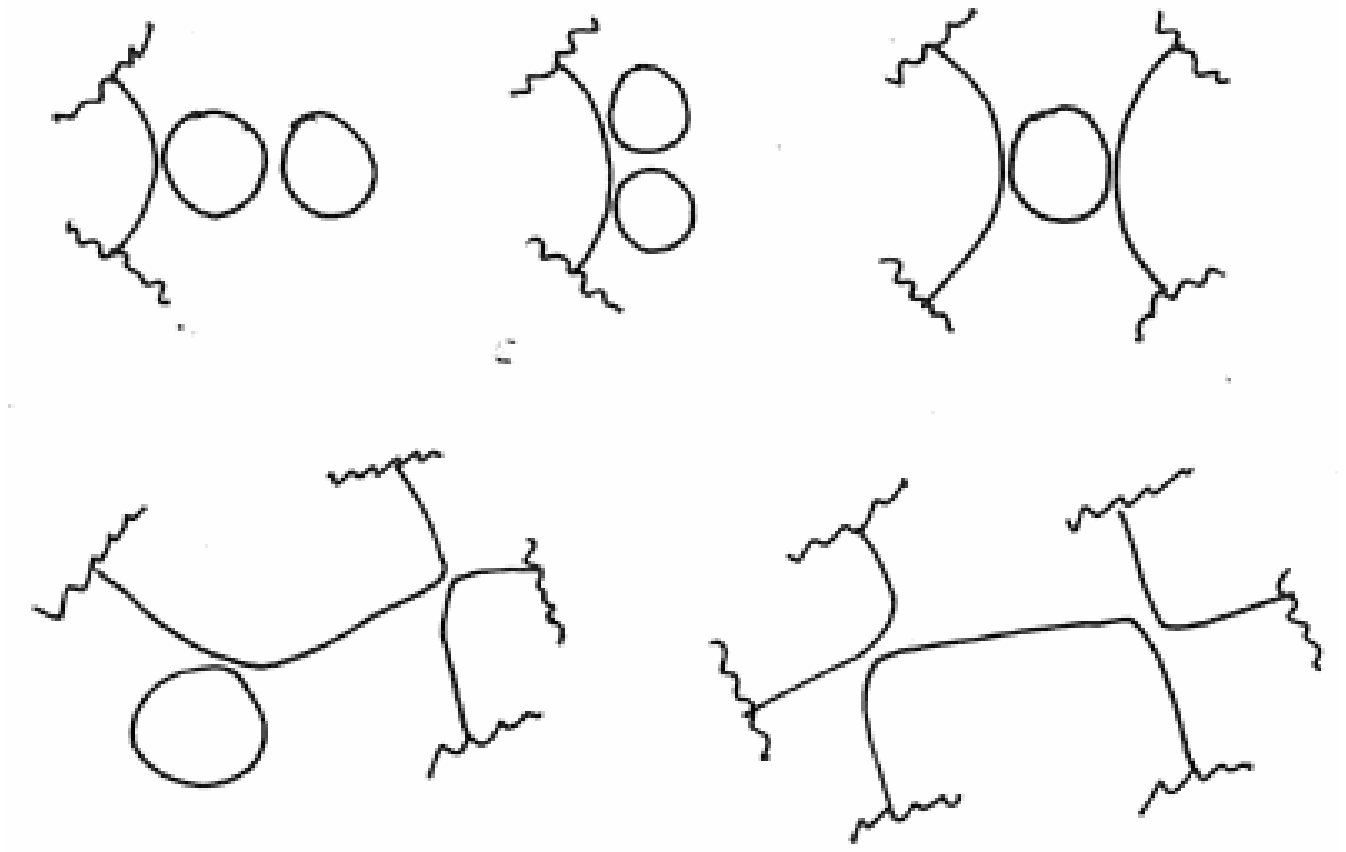}}

\ifig\FIGFORTYONE{  } {\epsfxsize2.0in\epsfbox{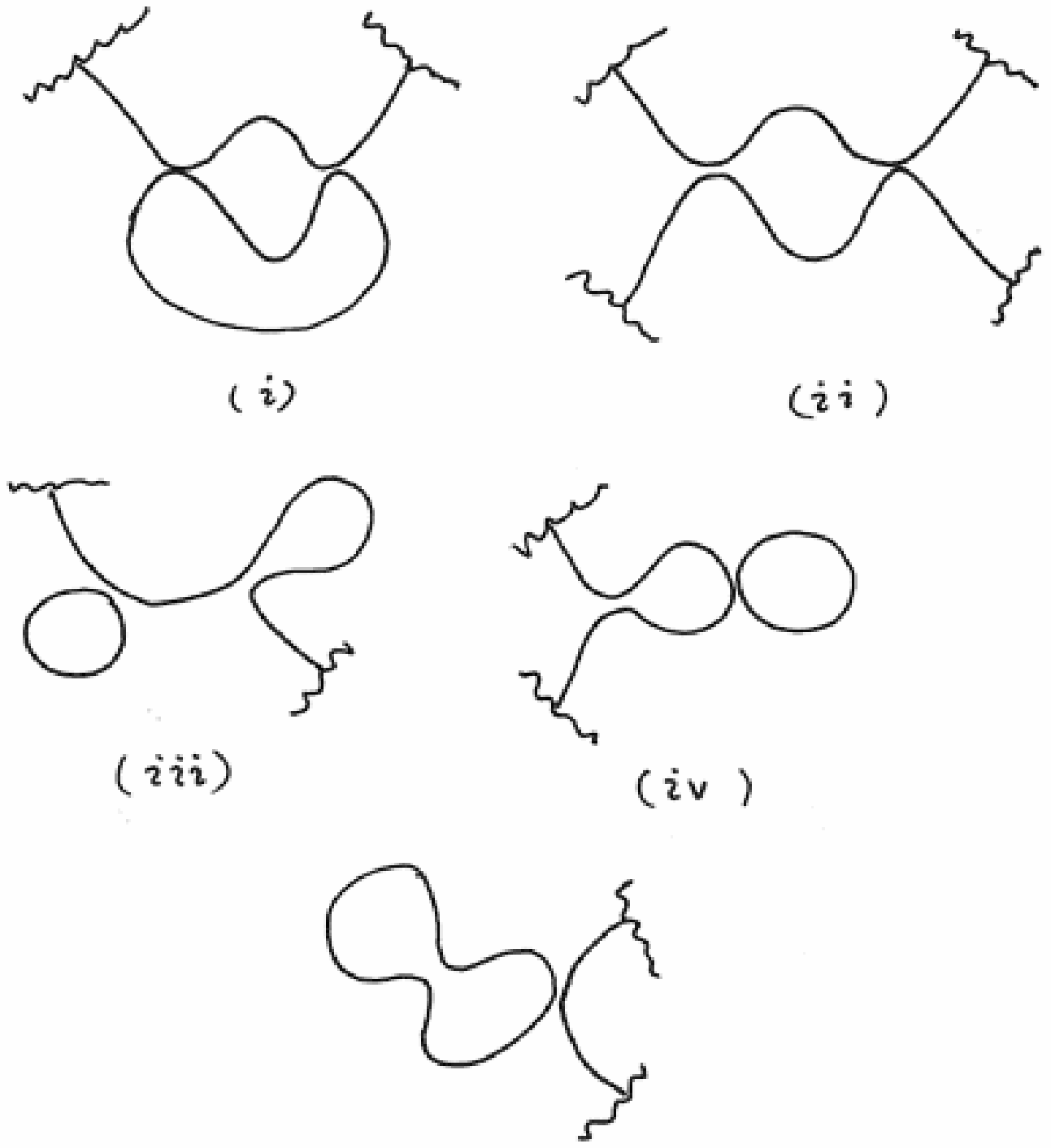}}

\ifig\FIGFORTYTWO{  } {\epsfxsize2.0in\epsfbox{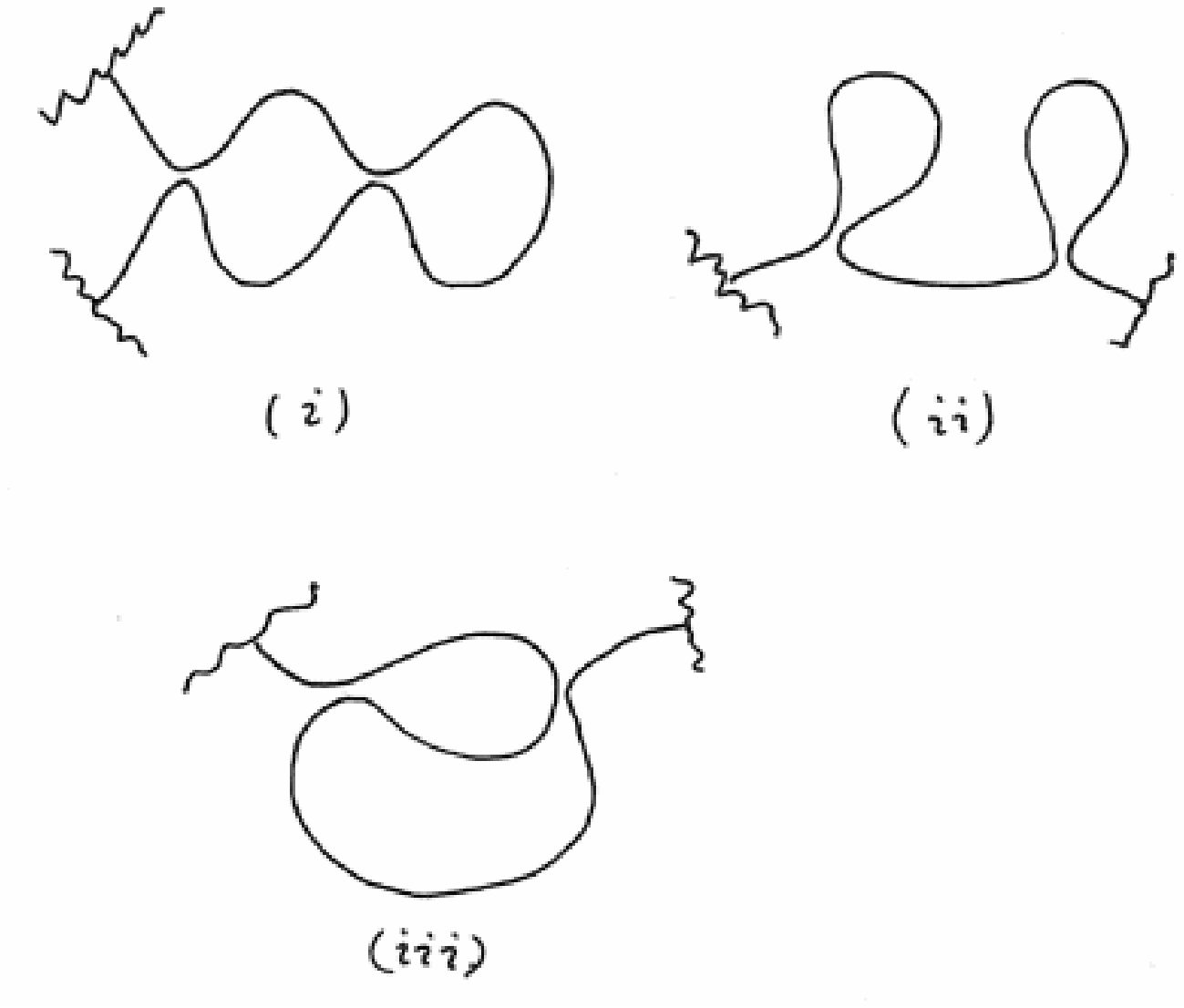}}

 Finally, we must consider what happens when there are
two internal critical points on the same level. Here we have the
possibilities which we have already discussed in the closed case,
but must also allow any or all of the strings involved to be open.
We can analyse the situation according to the number of connected
components of the part of the contour immediately below the doubly
critical level which pass close to the critical points. There must
be one, two, or three such components. If there are three they can
form five configurations (apart from the case when all three are
closed), as depicted in \FIGFORTY.  The well-definedness of the
composite map in all these cases follows immediately from the
associative law of composition in the Frobenius category.

If there are two components below the critical level then they can
again form five configurations (for either the two components meet
twice, or else they meet once, and one of them has a
self-interaction), depicted in \FIGFORTYONE.  But we have only three
cases to check, as the second is the time-reversal of one from
\FIGFORTY, and the last two are time-reversals of each other. Case
15(i) corresponds to the fact that the composition
$$\CO_{ab} \otimes \CC \to \CO_{ab} \to \CO_{ab} \otimes \CC$$
can be effected by cutting the composite cobordism in different
ways, but there is nothing to check, as there is only one possible
algebraic map.

In   \FIGFORTYONE\ case(iii), one order of the critical points gives
us the same composition
$$\CO_{ab} \otimes \CC \to \CO_{ab} \to \CO_{ab} \otimes \CC$$
as before, while the other order gives
$$\CO_{ab} \otimes \CC \to \CO_{ab}\otimes \CC \otimes \CC \to \CO_{ab}
\otimes \CC;$$ but it is very easy to check that both maps take
$\psi \otimes \phi$ to $\sum \psi\phi\phi_i\otimes \phi^i$ in the
notaion we have already used.

In  \FIGFORTYONE\ case(iii) we must again compare  compositions
$$\CO_{ab} \otimes \CC \to \CO_{ab}\otimes \CC \otimes \CC \to \CO_{ab}
\otimes \CC$$ and
$$\CO_{ab} \otimes \CC \to \CO_{ab} \to \CO_{ab} \otimes \CC.$$
This time we must check that
$$\sum \psi\phi_i \otimes \phi^i\phi = \sum
\psi\phi\phi_i\otimes\phi^i.$$  This is the same formula which we
met at the end of our discussion of closed string theories.

Finally, suppose that the contour below the critical level has only
one connected component. There are three possible configurations,
corresponding to the three ways of pairing four points on an
interval. They are \FIGFORTYTWO.  The first two of these are
time-reversals of cases we have already treated. The last one leads
--- in either order --- to a factorization
$$\CO_{ab} \to \CO_{ab} \otimes \CC \to \CO_{ab}.$$
There is only one possibility for this, so there is nothing to
check.

\bigskip

That completes the proof of the theorem about open and closed
theories.

\bigskip

\ifig\FIGFORTYTHREE{  } {\epsfxsize2.0in\epsfbox{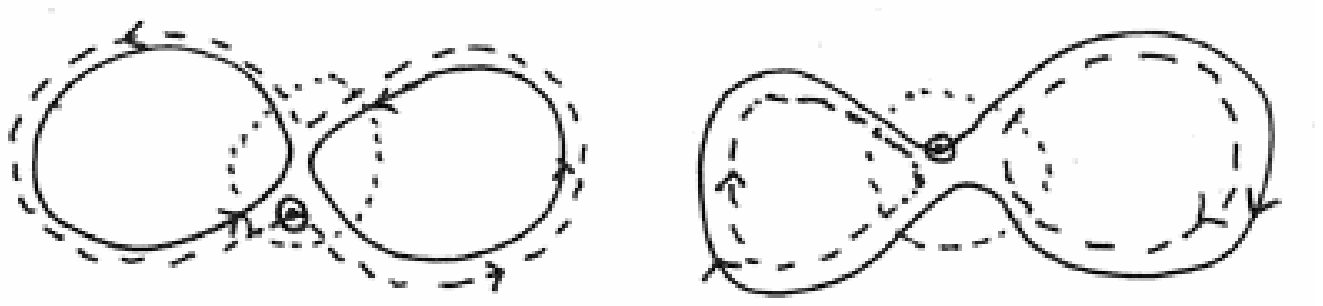}}

\ifig\FIGFORTYFOUR{  } {\epsfxsize2.0in\epsfbox{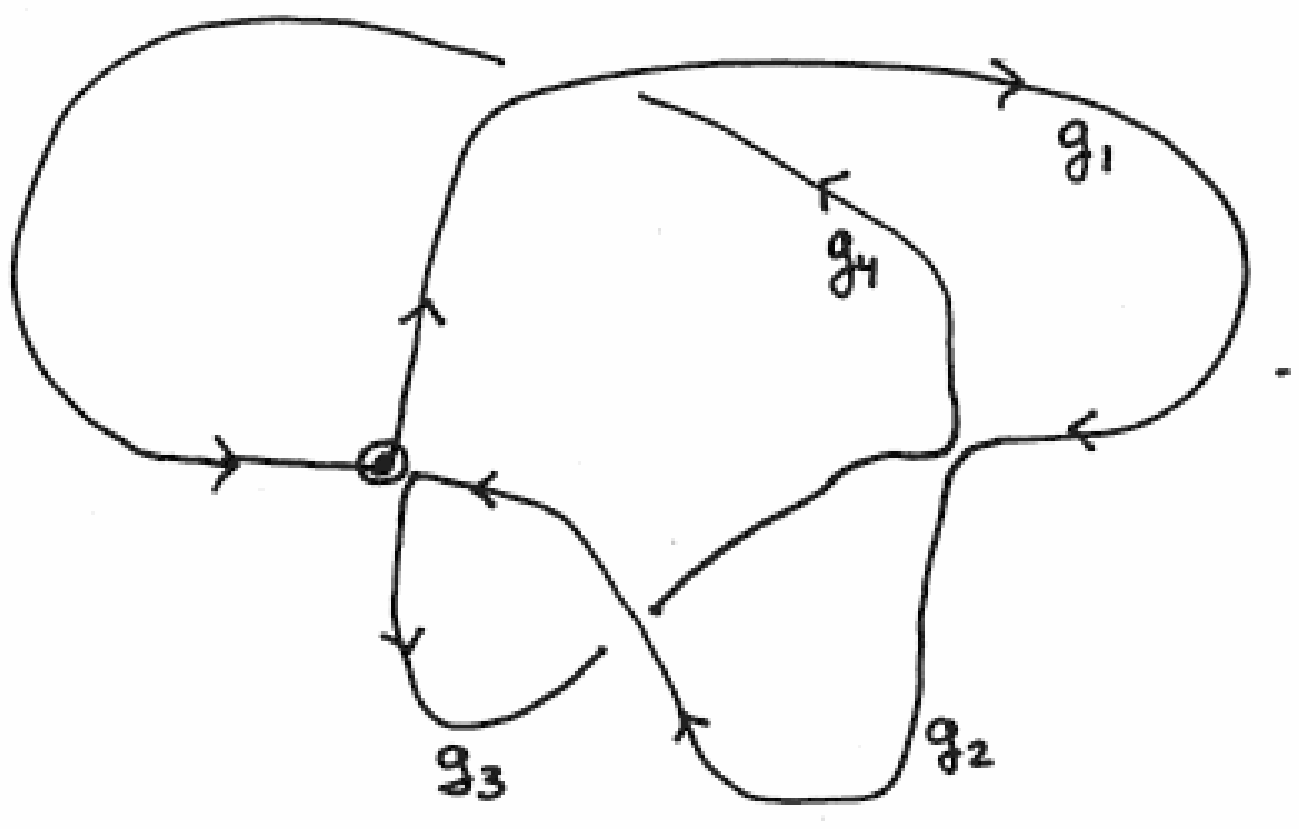}}

\ifig\FIGFORTYFIVE{  } {\epsfxsize2.0in\epsfbox{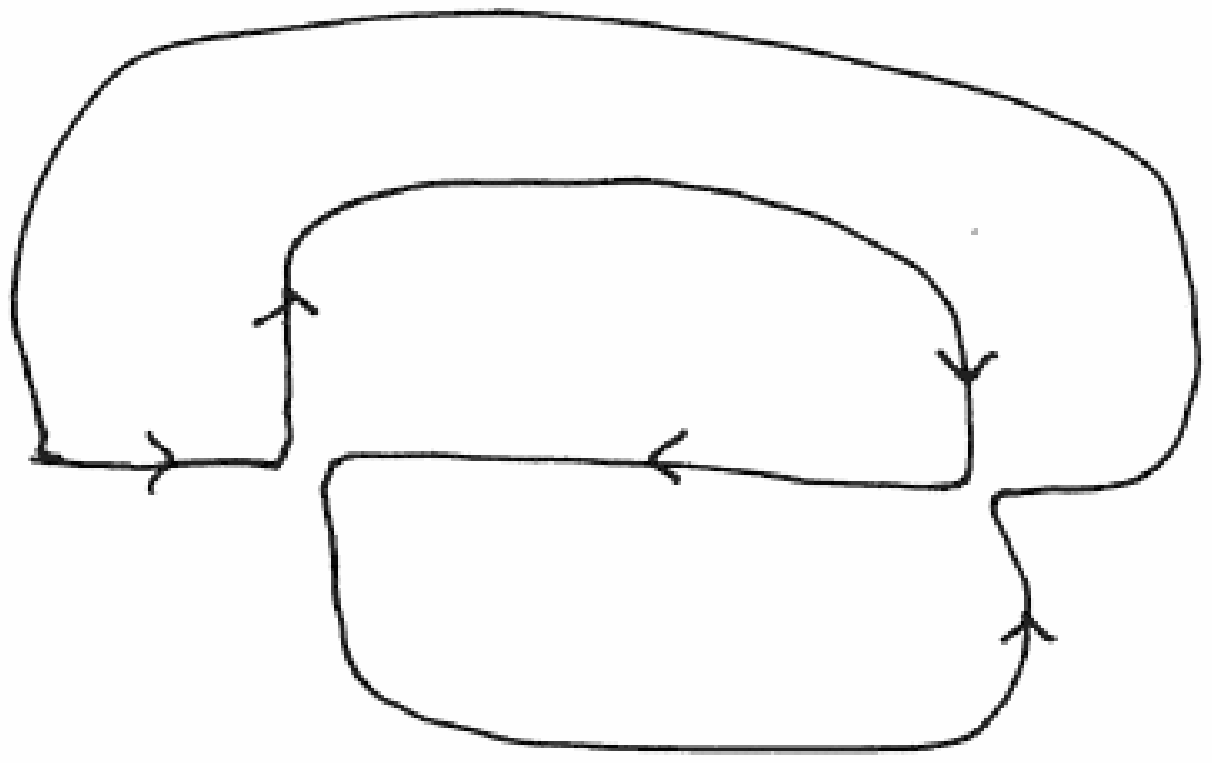}}

\ifig\FIGFORTYSIX{  } {\epsfxsize2.0in\epsfbox{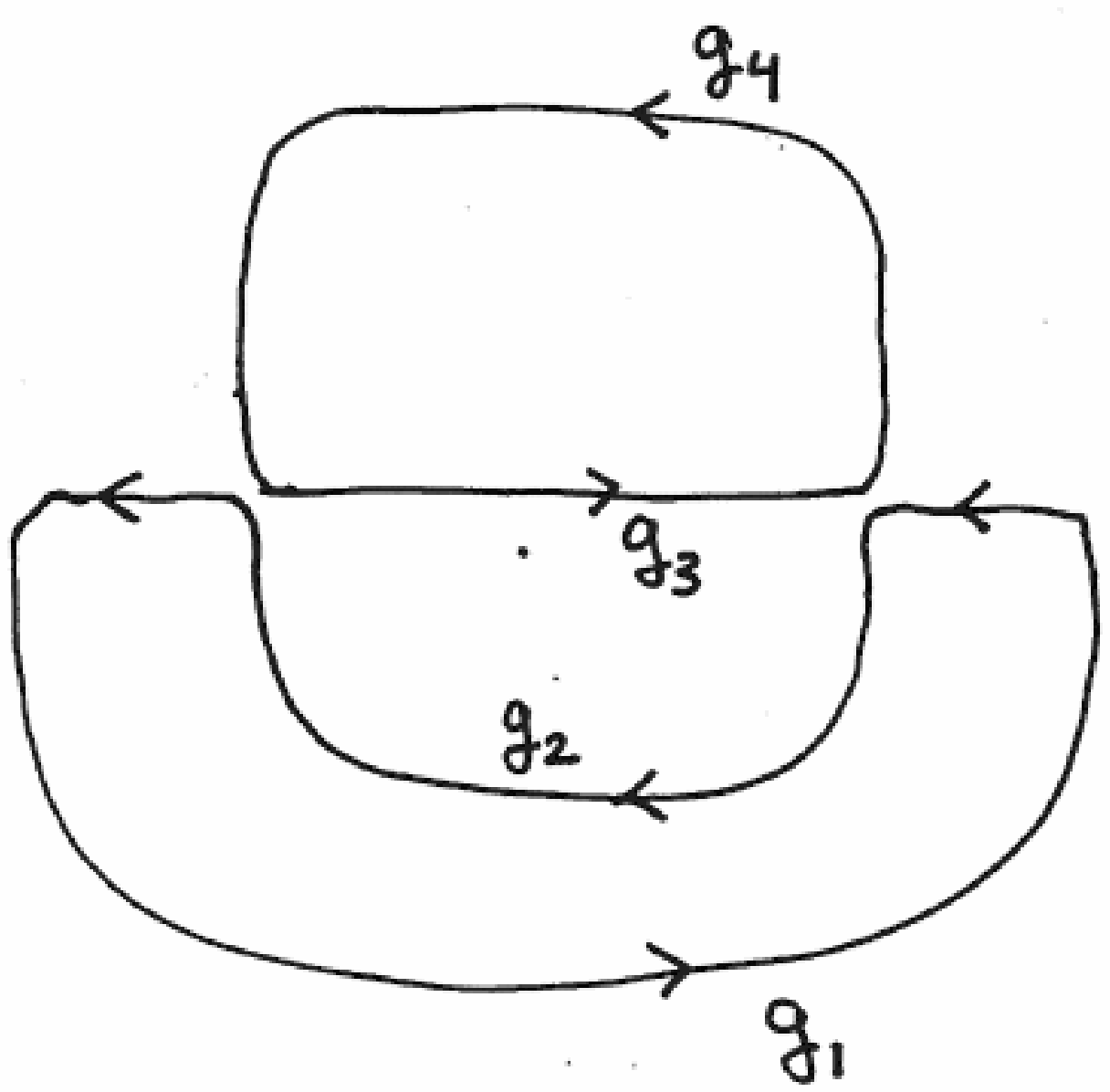}}

\subsec{Equivariant closed theories}

We must now redo the discussion in the first part of this appendix,
but for surfaces and circles equipped with a principal $G$-bundle,
where $G$ is a given finite group.

The first observation is that any circle with a bundle is isomorphic
to a standard bundle $S_g$ with holonomy $g \in G$ on the standard
circle $S^1$. Furthermore the set of morphisms from $S_g$ to
$S_{g'}$ is $\{h \in G : \ hgh^{-1}=g'\}$. In other words, the
category of bundles on $S^1$ is equivalent to the category $G//G$
formed by the group $G$ acting on itself by conjugation. An
equivariant theory therefore gives us a vector space $\CC_g$ for
each $g$, and together the $\CC_g$ form a $G$-vector-bundle on $G$.
Conversely, given the $G$-vector-bundle $\{\CC_g\}$ and a circle $S$
with a bundle $P$ on it, the theory gives us the vector space
$\CH(S,P)$ whose elements are rules which associate $\psi_{x,t}\in
\CC_{g_{x,t}}$ to each $x\in S$ and trivialization $t:P_x\to G$,
where $g_{x,t}$ is the holonomy of $P$ with base-point $(x,t)$, and
we require that
$$\psi_{x',t'}=g\psi_{x,t}$$
if $g$ is the holonomy of $P$ along the positive path from $(x,t)$
to $(x',t')$.  For this to be well-defined we need the condition
that $g_{x,t}$ acts trivially on $\CC_{g_{x,t}}$, whose necessity we
have already explained in Section 7.

Next we consider the trivial cobordism from $S_g$ to $S_{g'}$. The
possible extensions of the bundles on the ends over the cylinder
correspond to the possible holonomies from the incoming base-point
to the outgoing base-point, i.e. to the set of morphisms $\{h \in G
:\ hgh^{-1}=g'\}$ in $G//G$. Clearly these cylinders induce the
isomorphisms $\CC_g \to \CC_{g'}$ which we already know. But two
such cobordisms are to regarded as equivalent if there is a
diffeomorphism from the cylinder (with its bundle) to itself which
is the identity on the ends. The mapping-class group of the cylinder
is generated by the Dehn twist around it, so the morphism
corresponding to $h$ is equivalent to that for $hg = g'h$. This
means that $g$ must act trivially on $\CC_g$, as we already know.

Now we come to the possible bundles on the four elementary
cobordisms of diagram 1. The bundle on a cap must of course be
trivial. The pair-of-pants cobordisms that are relevant to us arise
as the regions between nearby level curves separated by a critical
level. We can draw them as in \FIGFORTYTHREE,   where the solid
contour is below the critical level, and the dashed one is above it.
We can trivialize the $G$-bundle in the neighbourhood of the
critical point (i.e. within the dotted circles), and then the bundle
on the cobordism is determined by giving the holonomies $g_1,g_2$
along the ribbons, as indicated. The operator we associate to case
(i) is the multiplication map
$$m_{g_1,g_2}:\CC_{g_1}\otimes \CC_{g_2} \to \CC_{g_1g_2}$$
of (\equivprodct). In writing it this way we are choosing an
ordering of the ribbons, i.e. a base-point on the outgoing loop. The
two orderings are related by the conjugation
$$\alpha_{g_2}:\CC_{g_1g_2}\to \CC_{g_2g_1},$$
so the consistency condition for us to have a well-defined
assignment is that
$$m_{g_2,g_1}(\psi_2\otimes \psi_1) = \alpha_{g_2}(m_{g_1,g_2}(\psi_1 \otimes \psi_2)).$$
We see that this holds in any Turaev algebra by combining
(\twistedcomm) with the facts that $G$ acts on the algebra by
algebra-automorphisms, and that $\alpha_{g_2}$ acts trivially on
$\CC_{g_2}$. As the mapping-class group of the pair of pants is
generated by the three Dehn twists parallel to its boundary circles,
there are no new conditions needed to make the assignment of the
operator to the pair of pants well-defined.

The homomorphism
$$c_{g_1,g_2}:\CC_{g_1g_2} \to \CC_{g_1} \otimes \CC_{g_2}$$
corresponding to the coordism 17(ii) is fixed by the requirement of
adjunction, bearing in mind that the dual space to $\CC_g$ is
$\CC_{g^{-1}}$. It is given by
$$c_{g_1,g_2}(\phi) = \sum \phi\phi^i \otimes \phi_i,$$
where $\{\phi_i\}$ is a basis for $\CC_{g_2}$, and $\{\phi^i\}$ is
the dual basis of $\CC_{g_2^{-1}}$.

\bigskip

Any cobordism with a bundle can be factorized by Morse theory just
as before; bundles are inherited by the elementary cobordisms. The
difficult part of the discussion is considering what happens when we
change the Morse function. But in fact the only step which presents
anything significant is the consideration of the interchange of two
critical points of index 1 on the same level, i.e. the cobordisms of
\FIGTHIRTYONE, \FIGTHIRTYTWO, \FIGTHIRTYTHREE.

Let us consider the case \FIGTHIRTYONE, where a string divides and
then rejoins
--- i.e. a torus with two holes, one incoming and one outgoing. We
draw the picture in the form \FIGFORTYFOUR.   (We do not draw it in
the apparently more perspicuous form \FIGFORTYFIVE,   as then the
neighbourhoods of the two critical points would have opposite
orientation in the plane.)

The cobordism corresponds to a map $\CC_{4321} \to \CC_{2341}$,
where, as in the following, we have abbreviated $\CC_{g_4g_3g_2g_1}$
to $\CC_{4321}$. If the left-hand critical point is encountered
first, the map we obtain is
$$\CC_{4321} \to \CC_{43} \otimes \CC_{21}
\cong \CC_{34} \otimes \CC_{12} \to \CC_{3412} \cong \CC_{2341},$$
$$\phi \mapsto \sum \phi\phi^i \otimes \phi_i
\mapsto \sum \alpha_3(\phi\phi^i) \otimes
\alpha_1(\phi_i) \mapsto \sum \alpha_3(\phi\phi^i)
\alpha_1(\phi_i) \mapsto \sum \alpha_2(\alpha_3 (\phi\phi^i)\alpha_1(\phi_i)),$$
where $\phi_i $ runs through a basis for $\CC_{21}$, and we write
$\alpha_3$ for $\alpha_{g_3}$, and so on. (The maps indicated by
$\cong$ in the previous line correspond to moving the choice of
base-point on the various strings.)

With the other order, we get
$$\CC_{4321} \cong \CC_{3214} \to \CC_{32}
\otimes \CC_{14} \cong \CC_{23} \otimes \CC_{41} \to \CC_{2341}$$
$$\phi \mapsto \alpha_4^{-1}(\phi) \mapsto
\sum\alpha_4^{-1}(\phi)\psi^i \otimes \psi_i
\mapsto \sum\alpha_2(\alpha_4^{-1}(\phi)\psi^i))
\otimes \alpha_4(\psi_i) \mapsto \sum\alpha_2(\alpha_4^{-1}(\phi)\psi^i)
\alpha_4(\psi_i),$$
where $\psi_i $ runs through a basis of $\CC_{14}$.

Thus we must prove that
$$\sum \alpha_{23}(\phi\phi^i)\phi_i =
\sum \alpha_{24^{-1}}(\phi)\alpha_2(\psi^i)\alpha_4(\psi_i).$$
We can deduce this from the axiom (newax) of \S 7, with
$h=g_2g_4^{-1}g_1^{-1}g_2^{-1}$ and $g=g_1^{-1}g_2^{-1}$, as
follows. We rewrite the right-hand side of the equation as
$$\sum \alpha_{24^{-1}}(\phi)\xi^i\alpha_g(\xi_i),$$
where $\xi^i$ is the basis $\alpha_2(\psi^i)$ of $\CC_h$, so that
$\xi_i=\alpha_2(\psi_i)$ and
$\alpha_g(\xi_i)=\alpha_1^{-1}(\psi_i)=\alpha_4(\psi_i)$. By the
axiom this equals
$$\sum\alpha_{24^{-1}}(\phi)\alpha_h(\eta^i)\eta_i
= \sum\alpha_{24^{-1}}\alpha_h(\phi^i)\phi_i.$$
Finally,
$$\alpha_{24^{-1}}(\phi)\alpha_h(\phi^i)
= \alpha_{24^{-1}}(\phi\phi^i) = \alpha_{23}(\phi\phi^i),$$
because $\phi\phi^i \in \CC_{43}$, and so
$\alpha_{24^{-1}}(\phi\phi^i)
=\alpha_{24^{-1}}\alpha_{43}(\phi\phi^i)=\alpha_{23}(\phi\phi^i).$
Thus we have dealt with the case of \FIGTHIRTYONE.

\bigskip

If fact this case is decidedly the most complicated of the set. We
shall do one more, namely case $(i)$ of \FIGTHIRTYFOUR, in which two
strings join and then split. We draw the diagram as in \FIGFORTYSIX,
corresponding to the two compositions
$$\CC_{43}\otimes \CC_{21} \to \CC_{4321}
\cong \CC_{1432} \to \CC_{14} \otimes \CC_{32} \cong \CC_{41} \otimes \CC_{23}$$
$$\CC_{43}\otimes \CC_{21} \cong \CC_{34}
\otimes \CC_{12} \to \CC_{3412} \cong \CC_{4123} \to \CC_{41} \otimes \CC_{23}.$$
The first sequence gives us
$$\psi \otimes \psi' \mapsto \psi\psi' \mapsto \alpha_1(\psi\psi') \mapsto \sum \alpha_1(\psi\psi')\phi^i \otimes \phi_i \mapsto \sum \alpha_4(\alpha_1(\psi\psi')\phi^i)\otimes \alpha_2(\phi_i),$$
where $\phi_i$ is a basis for $\CC_{32}$. The second sequence gives
$$\psi \otimes \psi' \mapsto \alpha_3(\psi)\otimes \alpha_1(\psi') \mapsto \alpha_3(\psi)\alpha_1(\psi')\mapsto \psi \alpha_{3^{-1}1}(\psi') \mapsto \sum \psi \alpha_{3^{-1}1}(\psi')\psi^i \otimes \psi_i,$$
where $\psi_i$ is a basis for $\CC_{23}$. But we can assume that
$\psi_i = \alpha_2(\phi_i)$, and hence that $\psi^i =
\alpha_2(\phi^i).$ So, noticing that $\alpha_1(\psi\psi')\phi^i \in
\CC_{14}$, and hence that
$$\alpha_4(\alpha_1(\psi\psi')\phi^i) = \alpha_{-1}(\alpha_1(\psi\psi')\phi^i),$$ what we need to prove is just that
$$\psi'\alpha_{-1}(\phi^i) =\alpha_{3^{-1}}(\alpha_1(\psi')\phi^i).$$
This is true because $\alpha_1(\psi')\phi^i \in \CC_{13^{-1}}$, and
so is fixed by $\alpha_{13^{-1}}$.

We shall leave the remaining verifications to the reader.

\ifig\FIGFORTYSEVEN{  } {\epsfxsize2.0in\epsfbox{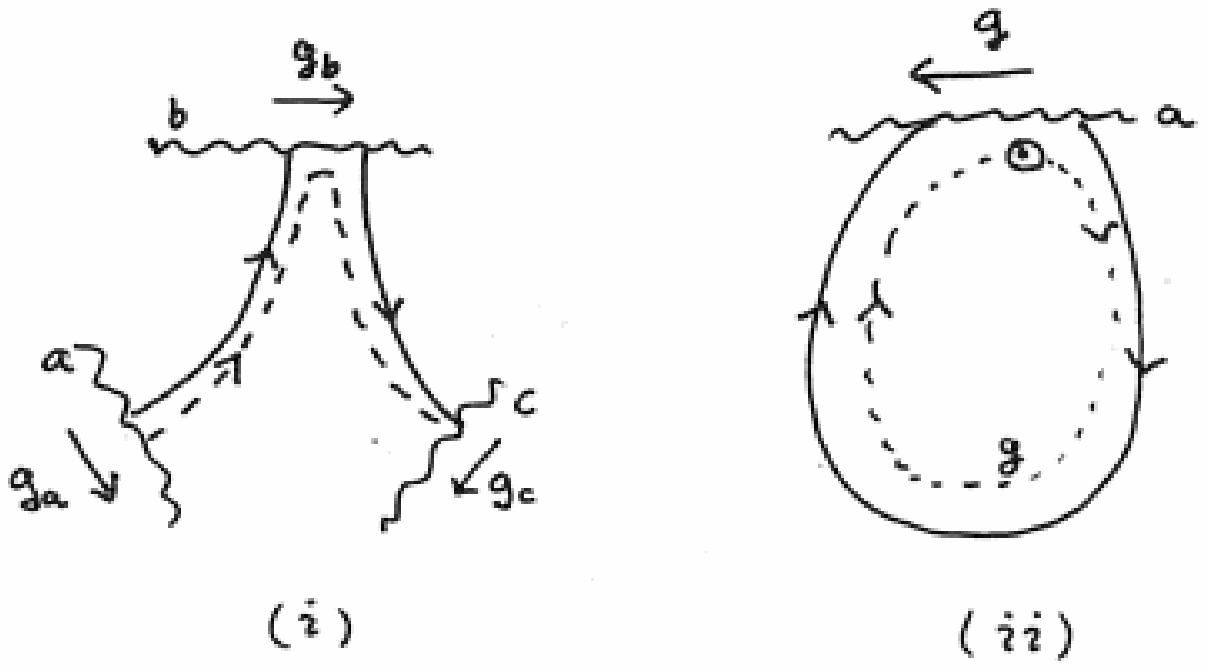}}

\ifig\FIGFORTYEIGHT{  } {\epsfxsize2.0in\epsfbox{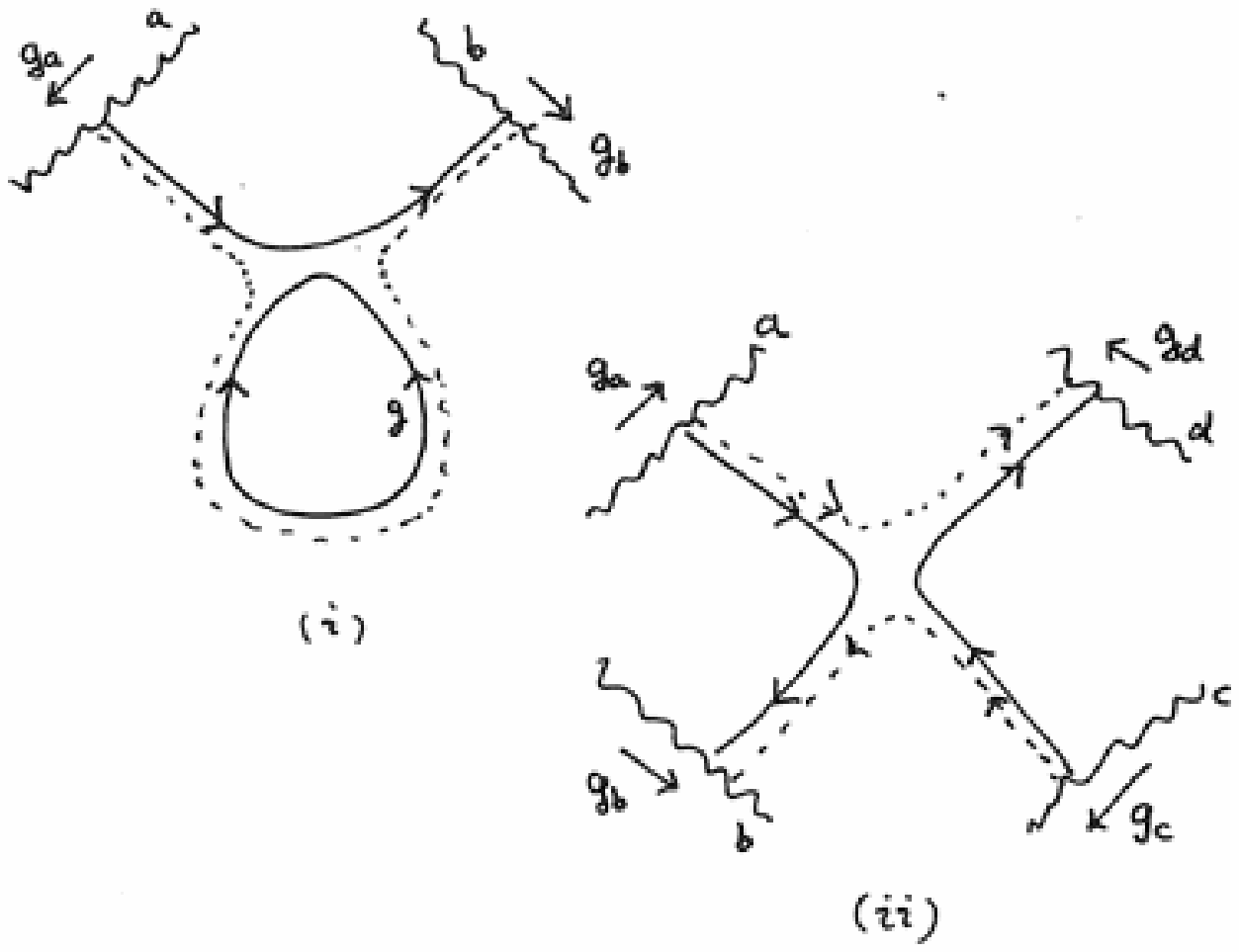}}

\subsec{Equivariant open and closed theories}

We now have  to redo the open and closed case taking account of
$G$-bundles on the cobordisms.

We assign the vector space $\CO_{ab}$ to an open string from $b$ to
$a$ equipped with a trivialization of the bundle on it. Changing the
trivialization by an element $g\in G$ corresponds to the action
$\rho_g$ of $g$ on $\CO_{ab}$, which also corresponds to the map
induced by a rectangular cobordism with holonomy $g$ along its
constrained edges.

We must consider the maps to be associated to the elementary
cobordisms corresponding to the critical points of a Morse function.
Up to time-reversal, two interesting things can happen at a boundary
critical point: either two open strings join end-to-end or an open
string becomes closed. We have the pictures of \FIGFORTYSEVEN.
  As before, the solid line is the contour below the
critical point, and the dashed line that above it. In 20(i),
$g_a,g_b,g_c$ are the holonomies between nearby points on the
respective D-branes, expressed in terms of the chosen
trivializations on the strings. (They satisfy $g_cg_b=g_a$.) The map
$\CO_{ab}\otimes \CO_{bc} \to \CO_{ac}$ that we associate to this
situation is
$$\psi \otimes \psi' \mapsto \rho_{g_a}(\psi)\rho_{g_b}(\psi').$$
The dual operation $\CO_{ac} \to \CO_{ab}\otimes \CO_{bc}$ is
$$\psi \mapsto \sum \rho_{g_1}(\psi \xi^i) \otimes \rho_{g_2}(\xi_i),$$
where $\xi_i$ and $\xi^i$ are dual bases of $\CO_{bc}$ and
$\CO_{cb}$.

In case $(ii)$ of \FIGFORTYSEVEN, the open string becomes a closed
string whose holonomy is $g$ with respect to the indicated
base-point and the trivialization coming from the beginning of the
open string. The corresponding map is $\iota^g$, with adjoint
$\iota_g$.

There are also the two kinds of operation coming from internal
critical points which involve open strings. They are illustrated in
\FIGFORTYEIGHT.

\noindent The map $\CC_g \otimes \CO_{ab} \to \CO_{ab}$
corresponding to 21(i) is $\phi \otimes \psi \mapsto
\rho_{g_a}(\iota_g(\phi)\psi))$, while the map $\CO_{ab} \otimes
\CO_{cd} \to \CO_{ad} \otimes \CO_{cb}$ corresponding to 21(ii) is
$$\psi\otimes\psi'\mapsto \sum \rho_{g_a}(\psi)\psi_i
 \otimes \rho_{g_c}(\psi')\rho_{g_bg_a^{-1}}(\psi^i),$$
where $\{\psi_i\}$ is a basis of $\CO_{bd}$.

\bigskip

We now have all the same verifications to make as in the
non-equivariant case. They are very tedious, but are in 1-1
correspondence with what we have already done, and present nothing
new. As an example of the modifications needed, let us point out
that the very frequently used formula A.1 which holds in any
Frobenius category when $\phi' \in \CO_{ab}$ and $\phi^i$ and
$\phi_i$ are dual bases for $\CO_{ab}$ and $\CO_{ba}$, generalizes
--- with the same proof --- when there is a $G$-action on the
category to
$$\sum \phi'\phi_i \otimes \alpha_g (\phi^i)
= \sum \phi_i \otimes \alpha_g(\phi^i\phi')$$
for any $g\in G$.

\bigskip

We shall say no more about the proof.

\listrefs

\end